\documentclass{elsarticle}

\usepackage{lineno,hyperref}
\usepackage{amsmath,amssymb,amsfonts}
\usepackage[T1]{fontenc}
\usepackage[normalem]{ulem}
\modulolinenumbers[5]

\usepackage{comment}

\usepackage{amsmath, upgreek}

\usepackage[utopia]{mathdesign}
% ...

\usepackage{caption}
\usepackage{subcaption}
\usepackage{multirow} 
\usepackage{bm}
\usepackage{xcolor}
\usepackage{multirow}

\makeatletter
\def\ps@pprintTitle{%
 \let\@oddhead\@empty
 \let\@evenhead\@empty
 \def\@oddfoot{\centerline{\thepage}}%
 \let\@evenfoot\@oddfoot}
\makeatother

\tnotetext[mytitlenote]{This is the accepted version of a paper published  in the Journal of the Mechanical Behavior of Biomedical Materials, 147 (2023) 106148. The final publication is available at https://doi.org/10.1016/j.jmbbm.2023.106148}

\biboptions{authoryear}

\begin{document}

\begin{frontmatter}

\title{Full-field in vivo experimental study of the strains of a breathing  human abdominal wall with intra-abdominal pressure variation}




\author[wilis]{Katarzyna Szepietowska}

\author[wilis]{Mateusz Troka}

\author[gumed]{Monika Lichodziejewska-Niemierko}
\author[gumed]{Michał Chmielewski}
\author[wilis]{Izabela Lubowiecka\corref{mycorrespondingauthor}}
\cortext[mycorrespondingauthor]{Corresponding author}
\ead{lubow@pg.edu.pl}

\address[wilis]{Faculty of Civil and Environmental Engineering, Gda\'nsk University of Technology, Gda\'nsk, Poland}

\address[gumed]{Department of Nephrology, Transplantology and Internal Medicine, Medical University of Gda\'nsk, Gda\'nsk, Poland}

\begin{abstract}

The presented study aims to assess the mechanical behaviour of the anterior abdominal wall based on an \textit{in vivo} experiment on humans. Full-field measurement of abdominal wall displacement during changes of intra-abdominal pressure is performed using a digital image correlation (DIC) system. Continuous measurement in time enables the observation of changes in the strain field during breathing. The understanding of the mechanical behaviour of a living human abdominal wall is important for the proper design of surgical meshes used for ventral hernia repair, which was also a motivation for the research presented below.

The research refers to the strain field of a loaded abdominal wall and presents the evolution of principal strains and their directions in the case of 12 subjects, 8 male and 4 female. Peritoneal dialysis procedure allows for the measurement of intra-abdominal pressure after fluid introduction. 

High variability among patients is observed, also in terms of principal strain direction. Subjects exhibit intra-abdominal pressure of values from 11 to 21 cmH$_2$O. However, the strain values are not strongly correlated with the pressure value, indicated variability of material properties.

\end{abstract}


\end{frontmatter}


\section{Introduction}

Understanding the mechanical behaviour of the abdominal wall can lead to improvements in surgery, such as ventral hernia repair \citep{junge2001elasticity}, abdominal wall closure \citep{le2020differences} or in deciding the stoma location \citep{tuset2022virtual}. It is known that mismatches between deformation behaviour of the native tissue and of the implanted soft biomedical materials can lead to short and long term complications \citep{mazza2015mechanical}. For instance, in the case of a hernia, the knowledge of  human living abdominal wall deformation characteristics may help  to design surgical meshes mechanically compatible with the tissue and consequently help to improve the repair efficiency \citep{mueller2022mesh}. Another problem is that testing and comparing mechanical properties of available surgical meshes is an important  issue that has not yet been standardised \citep[see protocol propsal by][]{civilini2023reliable}. This  was addressed by \cite{tomaszewska2022combined} where the importance of appropriate test choice and its influence on the identified material properties was shown. Data on the deformation field of the abdominal wall may help to design physical and computational experiments allowing to replicate the physiological loading that the implant will undergo.

Current knowledge on the constitutive behaviour  of  abdominal wall  single components and the mechanical behaviour of the whole abdominal wall has been mainly gained by \textit{ex vivo} studies, showing the anisotropic and hyperelastic behaviour of the abdominal wall depending on the anatomical location \citep[see review by][]{deeken2017mechanical}. \textit{Ex vivo} studies on abdominal muscles mainly focus on passive behaviour \citep{calvo2014determination}, but some research on active muscle behaviour has also been  performed \citep{grasa2016active}. However, it is not clear to what extent the limitations of the \textit{ex vivo} tests restrict capability of such studies to reflect the real behaviour of the living human abdominal wall.

Medical imaging is one of the solutions for collecting data on the  \textit{in vivo} performance of the abdominal wall. \cite{tran2016abdominal} used shear wave elastography to assess the elasticity of the abdominal wall together with  measurements of local stiffness under a low external load. \cite{jourdan2022dynamic} employed dynamic-MRI to study deformation of the abdominal wall muscles during forced breathing, coughing and the Valsalva manoeuvre.

Optical measurements have also been  employed to study the external surface of the abdominal wall in a noninvasive \textit{in vivo} way.   
\cite{szymczak2012investigation} investigated strains on the external surface of the abdominal wall during activities such as bending, stretching and expiration.  Elongation between tacks connecting the surgical mesh to the abdominal wall was obtained in a similar study using X-ray images of subjects in a standing position and  bending to one side, giving information about \textit{in vivo} performance of the surgical mesh \citep{lubowiecka2020vivo}. \cite{breier2017evaluation} used a digital image correlation (DIC) system to investigate strain on the abdominal wall during various movements.
Laser scanning of external abdominal walls was also performed to compare muscle contraction with relaxation    \citep{todros20193d}.  Although the aforementioned studies provide valuable data on the \textit{in vivo} performance of abdominal walls, it is difficult to relate the deformation  with a specific loading state. \cite{song2006elasticity} tracked markers on the human abdominal wall during the measurement of gas inflation pressure in laparosocpic surgery in order to investigate the passive behaviour of the abdominal wall. A similar approach was developed by \cite{simon2015developing} who used photogrametry to investigate a rabbit's abdominal wall. Knowing the deformation  and loading state allows for the identification of the material parameters of the abdominal wall by inverse analysis \cite{simon2017towards}. Nevertheless, not much is yet known about real strain field on the human living abdominal wall. 

In our previous work \citep{in_vivo_abdomen} photogrametric measurements were performed to asses strains on the abdominal wall of subjects undergoing peritoneal dialysis (PD), when intra-abdominal pressure can also be measured. The study was based on photos taken in two states: drained abdominal wall and abdominal wall filled with dialysis fluid, when the abdominal wall is under higher pressure. The drawback of this approach was that only four images were taken from different angles in sequence, which did not allow for a full-field view at a single moment.  In both reference and loaded states, photos were intended to be taken during the exhalation phase. Nevertheless, as shown by  \cite{mikolajowski2022automated} in an elastographic study, breathing  may have an influence on the measurements of the abdominal wall muscles, which are also respiratory muscles. Therefore, in the current study, the DIC system is used to measure deformation of the  abdominal wall during the entire process of dialysis fluid introduction, which enables  capturing the effect of breathing. What is more, DIC allows for a higher resolution strain field  as well as faster measurements and data processing. 

The aim of this study is   \textit{in vivo} investigation of abdominal wall deformation. The presented approach provides novel data on the strain field of the  external  living human abdominal wall surface during   intra-abdominal changes pressure caused by dialysis fluid. What is more, active breathing is included in the analysis.

\section{Materials and Methods}

\subsection{\textit{In vivo} experiments and Digital Image Correlation}

\begin{figure}[ht]\centering
    \includegraphics[width=80mm]{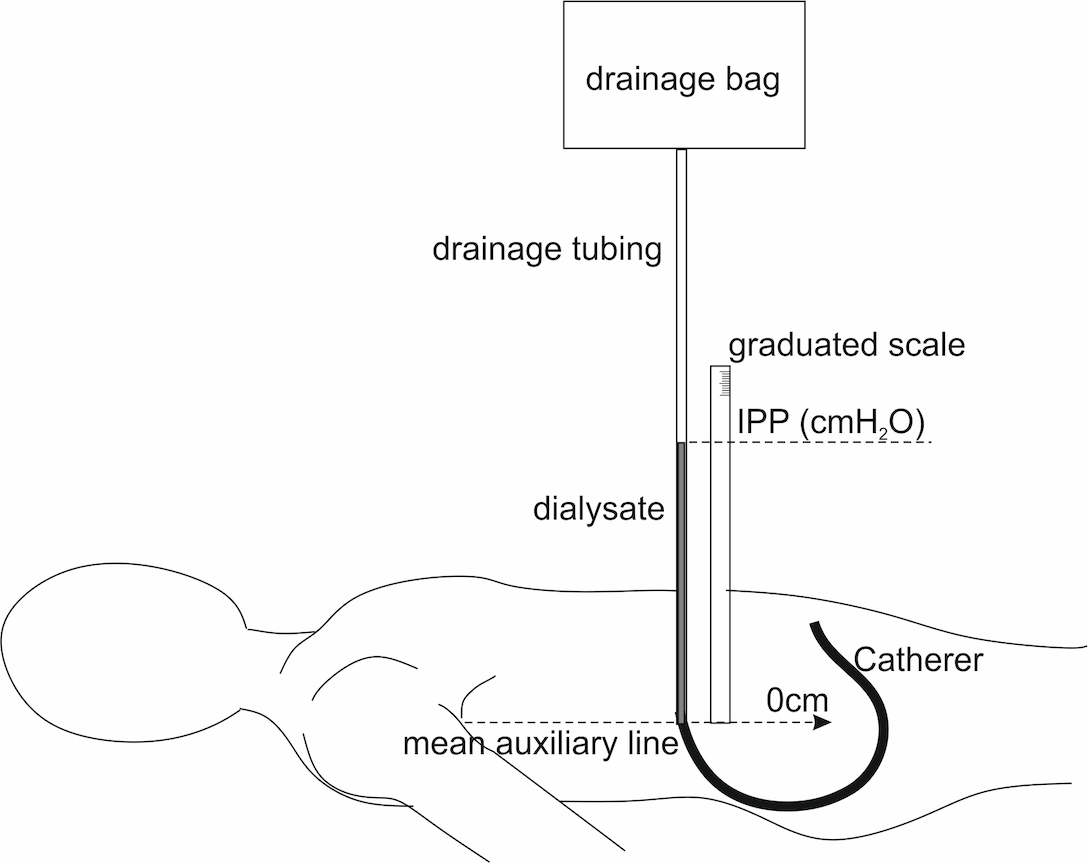}
    \caption{IPP measurement in PD}
    \label{fig_PD}
\end{figure}

\begin{figure}[ht]\centering
    \includegraphics[width=80mm]{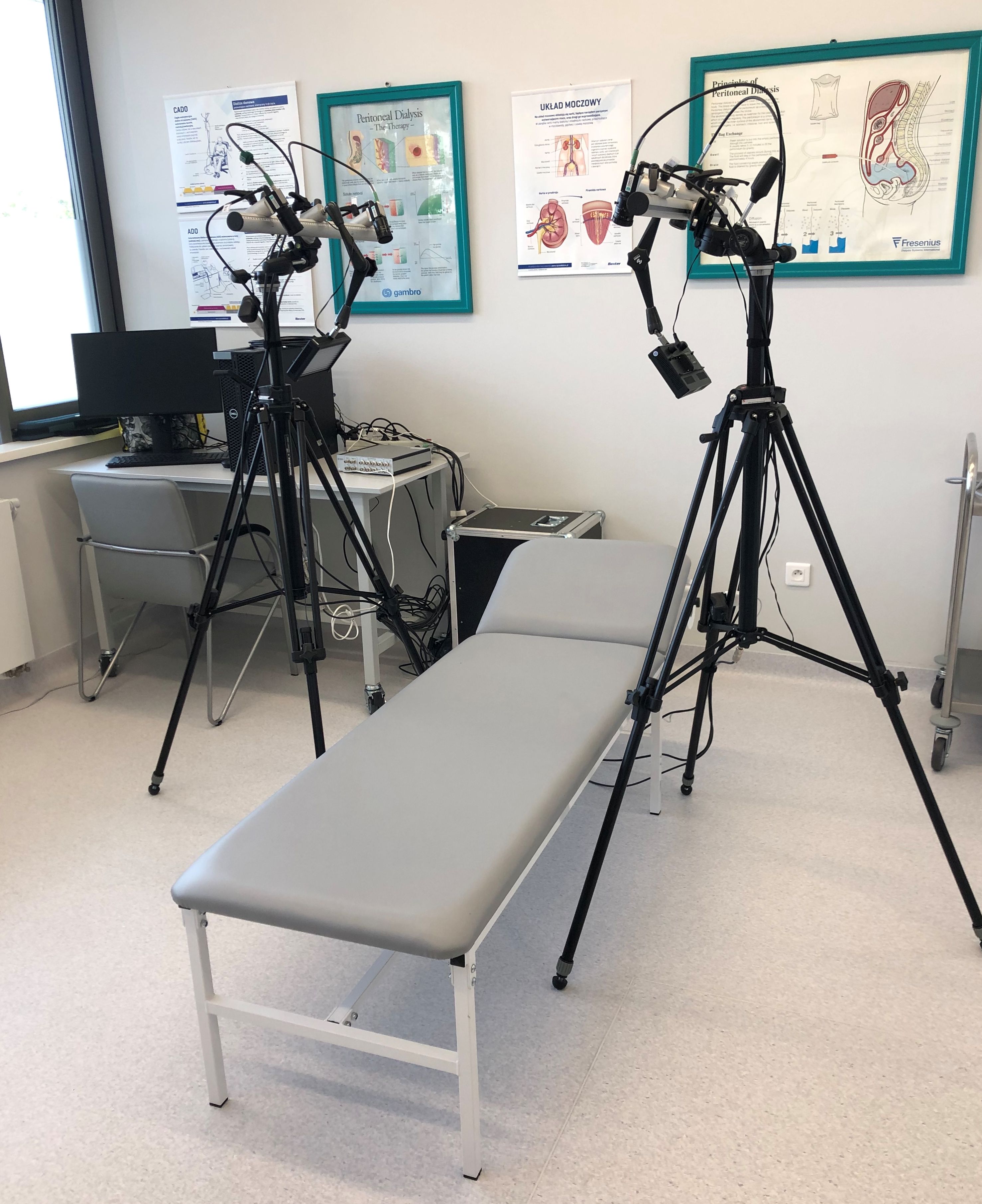}
    \caption{Digital Image Correlation - experimental stand}
    \label{fig_stand}
\end{figure}

\begin{figure}[ht]\centering
    \includegraphics[width=\textwidth]{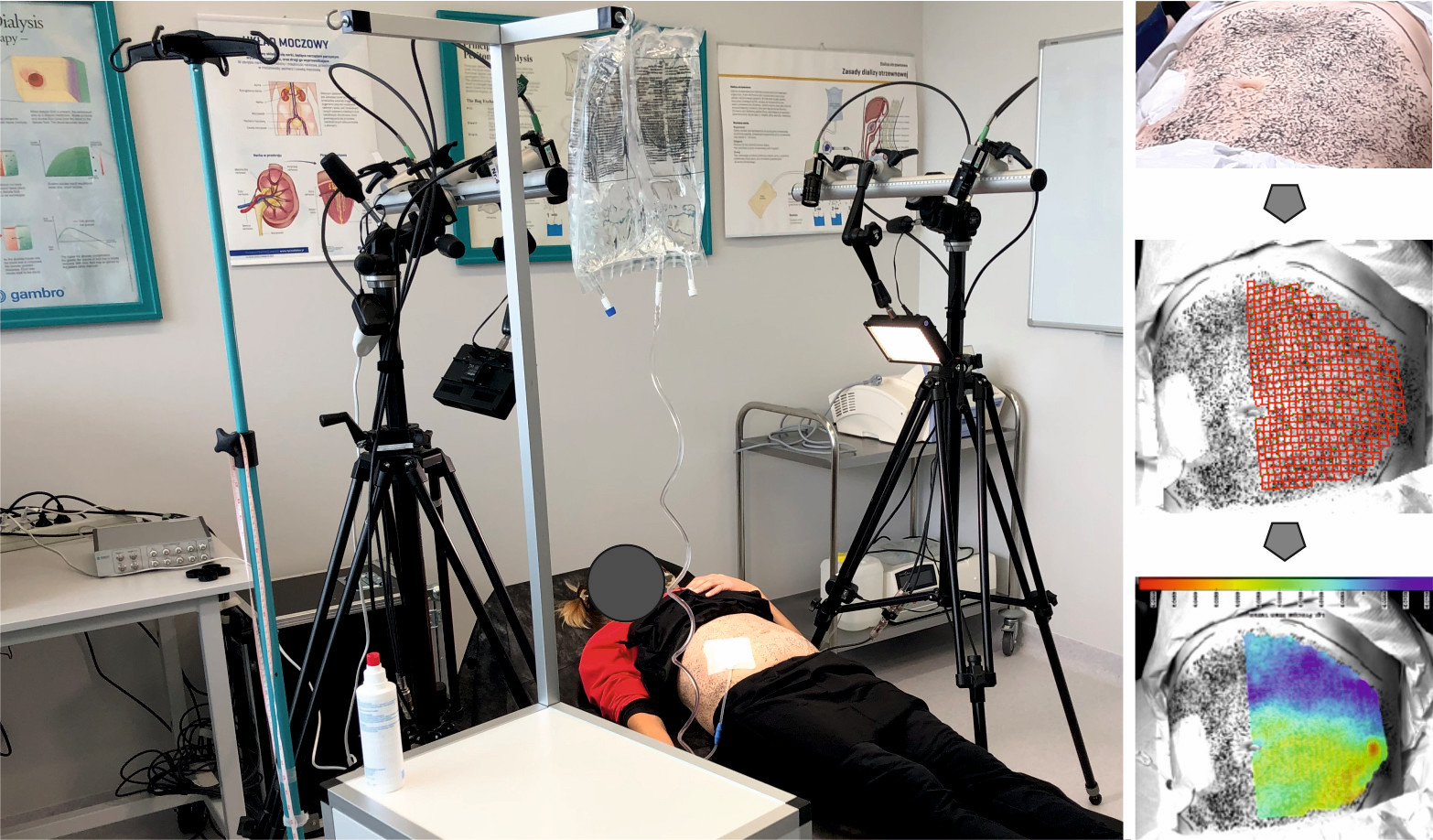}
    \caption{Experiment: subject  lying  under the DIC setup (left), speckle pattern on abdominal wall (top right),  DIC grid (middle right), and the outcome map resulting from image correlation (bottom right). }
    \label{fig_test}
\end{figure}

Twelve subjects, eight male and four female were tested during Peritoneal Dialysis fluid exchange (PD), Figure (\ref{fig_PD}); CAPD-continuous ambulatory peritoneal dialysis - 4 exchanges of PD solution per day or APD -A utomatic peritoneal dialysis - nightly PD performed by cycler with or without fluid left for the day. Their characteristics are included in Table \ref{Table_patient}. Although the patients did not receive any professional physiotherapy/rehabilitation exercises prior to the study, they were informed and encouraged to maintain physical activity and exercise. One can assume that their physical activity was appropriate to their age and gender agreed upon by average resident of Poland/Europe. The subjects suffer from end-stage kidney disease and regularly undergo PD. Thus, the experiments were performed during the dialysis procedure (see \cite{in_vivo_abdomen}). A Digital Image Correlation (DIC) system was applied in the study to register the motion of their abdominal walls.
DIC is an optically-based technique used to measure the evolving full-field 2D or 3D coordinates on the surface of a test piece throughout a mechanical test \citep {DICpractice}. The measured data can be used to calculate, e.g., displacements, strains and other quantities based on changes in the registered geometry of the objects undergoing movement. The system can map the motion and deformation of the tested piece by taking series of images that can replicate the motion of the speckle pattern applied to the specimen. It can  also be used to reconstruct the abdominal wall geometry. A four-camera DIC system Dantec Q-400 was used. This included 4 digital VCXU-23M cameras  with a 2.3 Mpx matrix (resolution: 1920 x 1200 px) and  VS-1620HV lenses(16 mm f/2.0–16) (Figure \ref{fig_stand}).
The stand was situated above the area of interest (front abdominal wall) in a way that enables the correlation of images of the whole abdominal wall (Figure \ref{fig_test}). A random speckle pattern was printed on a 3D printer with the use of a flexible filament in the form of a stamp. The pattern was applied to the abdomen of each subject using  approved  skin colouring paint (Figure \ref{fig_test}). A chequerboard calibration plate (35mm, 9x9 Marked White Plastic) was used to capture the relative positions of the four cameras mounted on two tripods and thus calibrate the system before the  measurement.

\begin{table}[ht]
\begin{tabular}
{p{1cm}p{1cm}p{1cm}p{1.2cm}p{1.2cm}p{1.2cm} p{3.5cm}}

\hline
No & sex & age & height  &weight & BMI  & hernia  \\
 &  & & [m]  &[kg] & [kg/m$^2$] &  \\
\hline
D1          & M            & 78           & 1.63             & 80  & 30.1  & no  \\
D2           & F            & 48           & 1.75             & 66    &21.6    & no  \\
D3           & F            & 73           & 1.60             & 68 & 26.6      & no \\
D4          & M            & 70           & 1.70             & 80.7      &27.9    & no    \\
D5           & M            & 74           & 1.67             & 84     &30.1  & no   \\
D6         & F            & 65           & 1.60             & 67   &26.2   & no   \\

D7         & M            & 88           & 1.72             & 89   &30.1    & no  \\
D8         & M            & 61           & 1.68             & 58   &20.5     & no \\
D9         & M            & 46           & 1.76             & 85   &27.4     & no  \\
D10         & F            & 72           & 1.54             & 61   &25.7    & no  \\
D11         & M            & 36           & 1.76             & 86   &27.8      & no\\
D12         & M            & 56           & 1.74             & 78    &25.8   &  an umbilical hernia repaired with an implant 2 years before the test\\
\hline
\end{tabular}
\caption{Characteristics of the patients } \label{Table_patient}
\end{table} 

\begin{table}[ht]
\begin{tabular}
{cccccc}
\hline
No & faceted size  & grid size & approx. grid spacing & frame rate & calibration residuum  \\
 & [px] &  [px] & [mm] &  [Hz] &  [px]    \\

\hline
D1          & 33           & 22      &   8   & 2  & 0.11 \\
D2           & 25            & 19      & 8      & 5  & 0.09   \\
D3           & 29            & 22      &   10    & 5  & 0.09    \\
D4           & 25            & 19      &   8    & 5  & 0.08    \\
D5            & 25            & 19    &    8     & 5  & 0.08   \\
D6         & 29            & 22      &   9    & 5  & 0.09   \\
D7         & 25            & 19      &   9    & 5   & 0.09   \\
D8         & 25            & 19     & 8        & 5  & 0.09  \\
D9         & 25            & 19     &  8    & 5  & 0.09    \\
D10         & 25            & 19    &  8      & 5   & 0.08   \\
D11         & 29            & 22    &   9     & 5   & 0.08  \\
D12         & 29            & 22    &   9     & 5   & 0.08   \\
\hline
\end{tabular}
\caption{DIC test parameters } \label{Table_dic_parameters}
\end{table}

The motion of the abdomen was registered in  pictures taken throughout the PD procedure, starting with a drained abdominal cavity and finishing with it being filled with two litres  of dialysis fluid \citep {Durand1996}. Then intraperitoneal pressure (IPP) was measured, (Figure  \ref{fig_PD}) with the use of manometer connected to the dialysis bag, as in \citep{PerezDiaz2017}. Due to the ethics issues the authors followed the standard procedure of peritoneal dialysis without any additional actions. That is why the IPP was measured only once during the procedure when the abdominal cavity was filled with the fluid. Additional measurements would imply connecting the subject to a new dialysis bag every time when measured, which might have risen the risk of infection during the procedure. We maintained the same procedure throughout the tests for the safety of the patients. It should be noted that the subjects visit the dialysis centre once a while for a control visit only. The regular peritoneal dialysis is done by the patients at home. The examination of each subject was performed only once within this study.

The collected images were processed using the commercial correlation software Istra 4.7.6.580 to determine the three-dimensional displacement of the  surface of each tested abdominal wall. For  selected parameters used in  the image analysis, see Table \ref{Table_dic_parameters}. DIC measurements have an approximated error radius (\ref{Error_equation})
\begin{equation}
    Err=\sqrt{1/3(V(x)+V(y)+V(z))},
    \label{Error_equation}
\end{equation}
which estimates the uncertainty of the 3D coordinates ($x$, $y$, $z$), where $V$ is the variance.

The experiments were fully non-invasive and the measurement was contactless. All participants  consented to participate in the study under a protocol approved by the Independent Ethics Committee for Scientific Research at the Medical University of Gdańsk N\textsuperscript{\underline{o}} NKBBN 314/2018.

\subsection{Strain field of abdominal wall outer surface as the basis for mechanical analysis }

The DIC method can measure strains along the surface object where a normal vector $\mathbf{n}$ is determined for each measurement point. This vector defines the local tangent plane. In a standard case, the local x direction lies at the intersection of the z tangent plane with the xz plane. The local y direction is perpendicular to x and $\mathbf{n}$. All x, y values (deformation, strain) always refer to the coordinate system used in each measurement. The strain in software is calculated based on the deformation gradient and the “original” strain calculated from it, thus it is referred to as a Green-Lagrange strain measure. What is more, the deformation at a grid node is a virtual projection of deformations observed in the images of individual cameras following a given node \cite{reu2018dic}.

The postprocessing software of the used DIC system, Istra 4.7.6.58, allows to calculate the displacements of the abdominal wall surface. The system generates a grid of points (see Figure \ref{fig_test}) on the recorded surface and in every grid point the displacements and the strains are calculated. A local Green-Lagrange strain tensor represented by formula  (\ref{GL_strain_tensor})

\begin{equation}
    \boldsymbol{\mathrm{\varepsilon}}=\frac{1}{2}\left( \bm{F}^\top 
    \bm{F} - \bm{I}\right),
    \label{GL_strain_tensor}
\end{equation}
where $\bm{F}$ is the deformation gradient, was used. The eigenevalues and eigenvectors of $\boldsymbol{\mathrm{\varepsilon}}$, i.e. principal strains ($\varepsilon_1$ and $\varepsilon_2$) and directions ($\alpha$), are the quantities on which the analysis is based. $\alpha$ is the angle between transverse direction x and principal direction. Having local tensors, we obtain the local principal strain directions. This lets us to observe their changes within the pressurizing of the abdominal wall in different area of the tested surface.  More about strain calculation via DIC can be found in \cite{reu2018dic}.

\begin{figure}[ht]\centering
    \includegraphics[width=\textwidth]{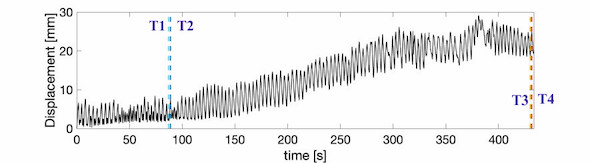 }
     \caption{Displacement of the abdominal wall during the experiment with visible breathing effect;  T1--T4 - time steps at inhalation and exhalation selected at the beginning and end of the experiment}
    \label{czas_T1T4}
\end{figure}

Due to the catheter, which covered part of the registration field, only half of the abdominal wall was considered in the study, Figure \ref{fig_test}. The presence of the catheter disturbs the strain field of the abdominal wall, as already presented in (\cite{in_vivo_abdomen}). Four time-steps representing four different deformation states of the abdominal wall were selected for analysis so that two pairs of time-steps corresponding to one inhalation of air were captured at the early and final stages of the experiment, (Figure \ref{czas_T1T4}).  T1 and T2 denote time-steps  at ca. 20\% of the experiment duration, exhalation and inhalation, respectively. T3 and T4 denote time-steps at the final stage (T3 exhalation and T4 inhalation) of the experiment, when the abdominal cavity is filled with 2l of dialysis fluid.  

\section{Results and discussion}

\subsection{Approximated error}

\begin{figure}[ht!]\centering
    \includegraphics[width=1\textwidth]{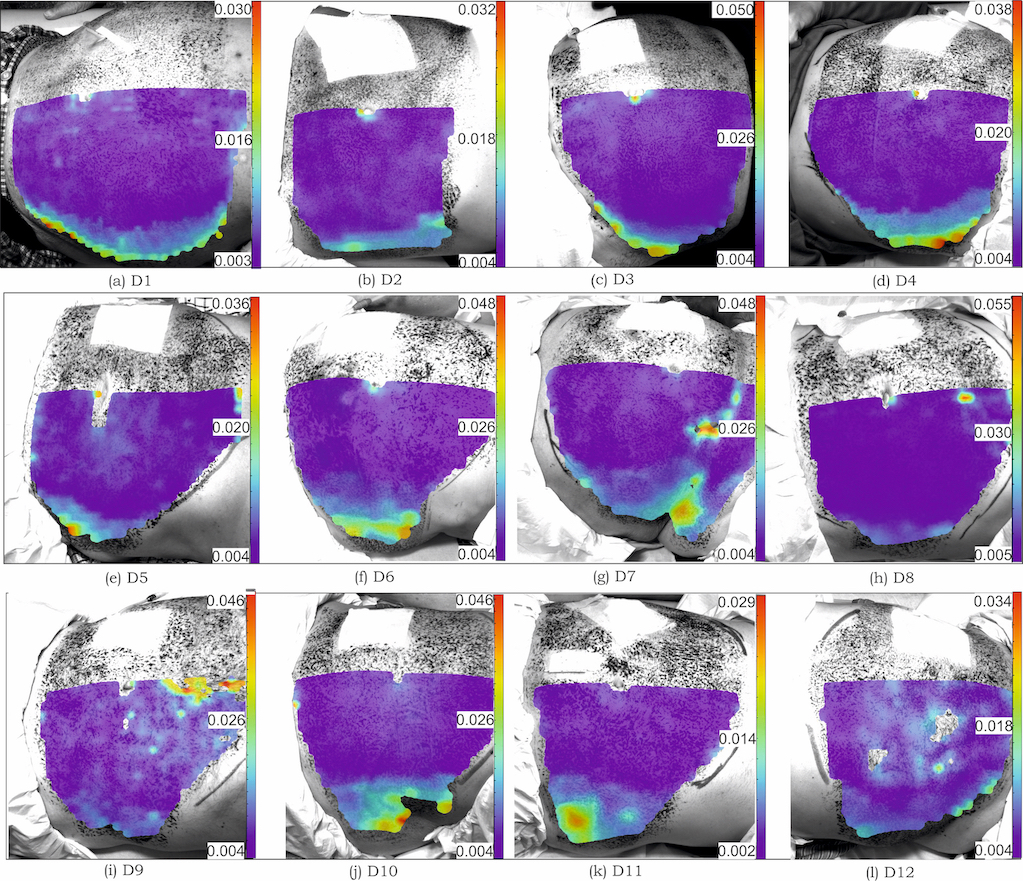}
    \caption{ Approximated error radius obtained in DIC measurements in T4  of subjects D1--D12 [mm]}
    \label{EstimatedError}
\end{figure}

The DIC system enabled obtaining deformation measurements of the abdominal wall with sufficient accuracy. 
Figure \ref{EstimatedError} shows the maps of the approximated error radius of the region of interest  as seen through one of the cameras. The maps show T4, when the deformation peaks. It may be noted that higher errors usually occur  around umbilicus and in the lateral part of the abdomen whose images in the cameras are skewed.  The highest estimated error  was 0.055 mm (D8) in one point. Intraperitoneal adhesions  of subject D7 is another area of higher error.  Subject D9 had  hairs in the mid-line that disturbed  correlation in that area.

\begin{figure}[ht!]\centering
    \includegraphics[width=\textwidth]{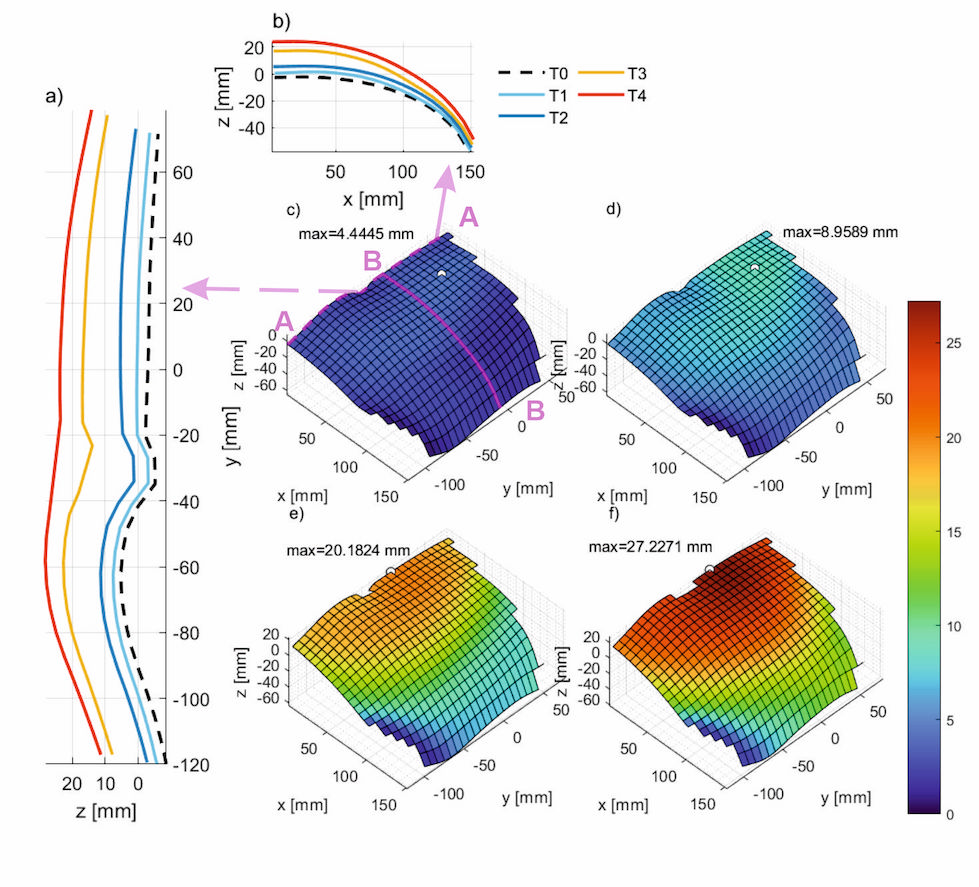}
   \caption{Shape and displacement of subject D2 in four stages T1--T4 and  reference one T0: a) profile of abdominal wall along mid-line A--A; b)  profile along transverse direction B--B; c--f) surfaces of abdominal wall with colour indicating total displacement [mm] in T1--T4, respectively with marked location of maximum displacement by a white circle; x is the mediolateral axis from right to left, y is craniocaudal axis from caudal to cranial, and z is anteriorposterior axis from anterior to posterior }
    \label{fig_dic2dis}
\end{figure}

\begin{figure}[ht!]\centering
    \includegraphics[width=\textwidth]{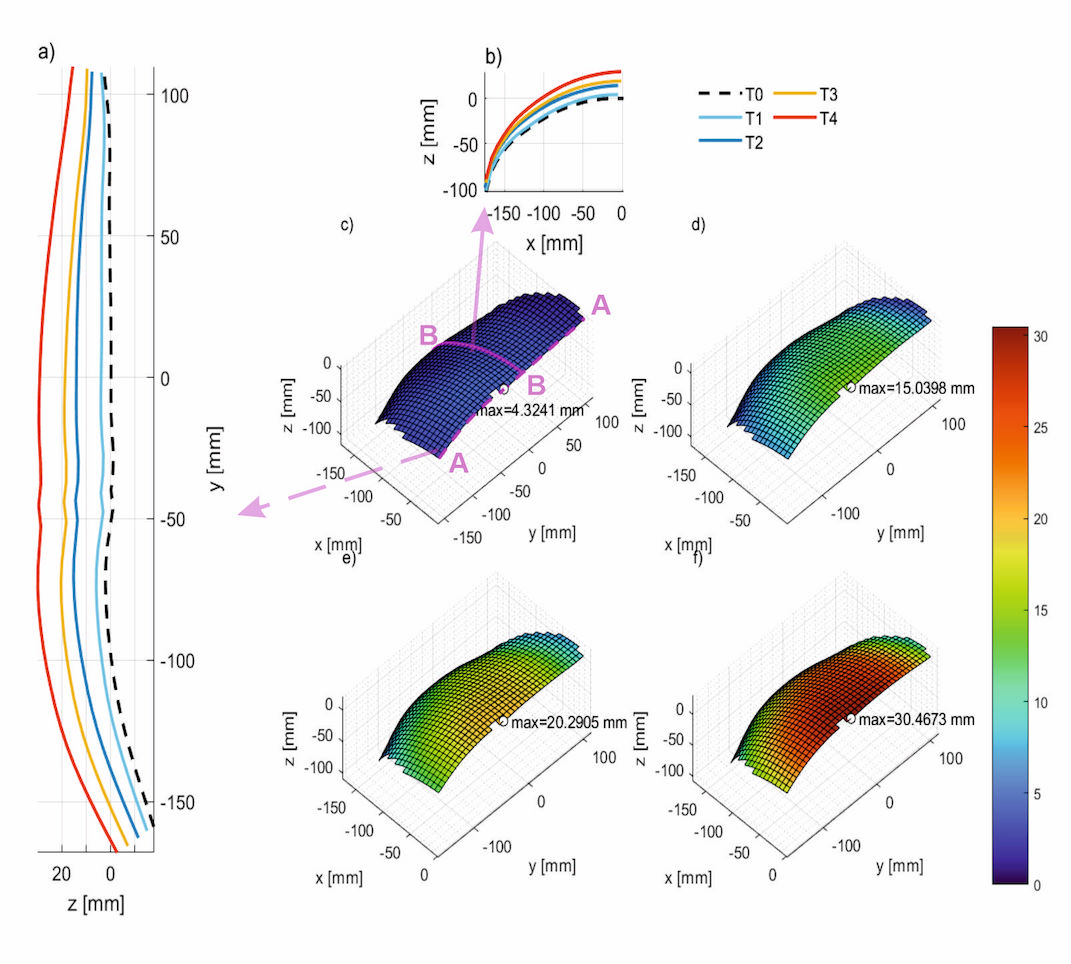}
    \caption{Shape and displacement of subject D4 in four stages T1--T4 and  reference one T0: a) profile of abdominal wall along mid-line A--A; b)  profile along transverse direction B--B; c--f) surfaces of abdominal wall with colour indicating total displacement [mm] in T1--T4, respectively with marked location of maximum displacement by a white circle; x is the mediolateral axis from right to left, y is craniocaudal axis from caudal to cranial, and z is anteriorposterior axis from anterior to posterior } 
    \label{fig_dic4dis}
\end{figure}

\begin{figure}[ht!]\centering
    \includegraphics[width=\textwidth]{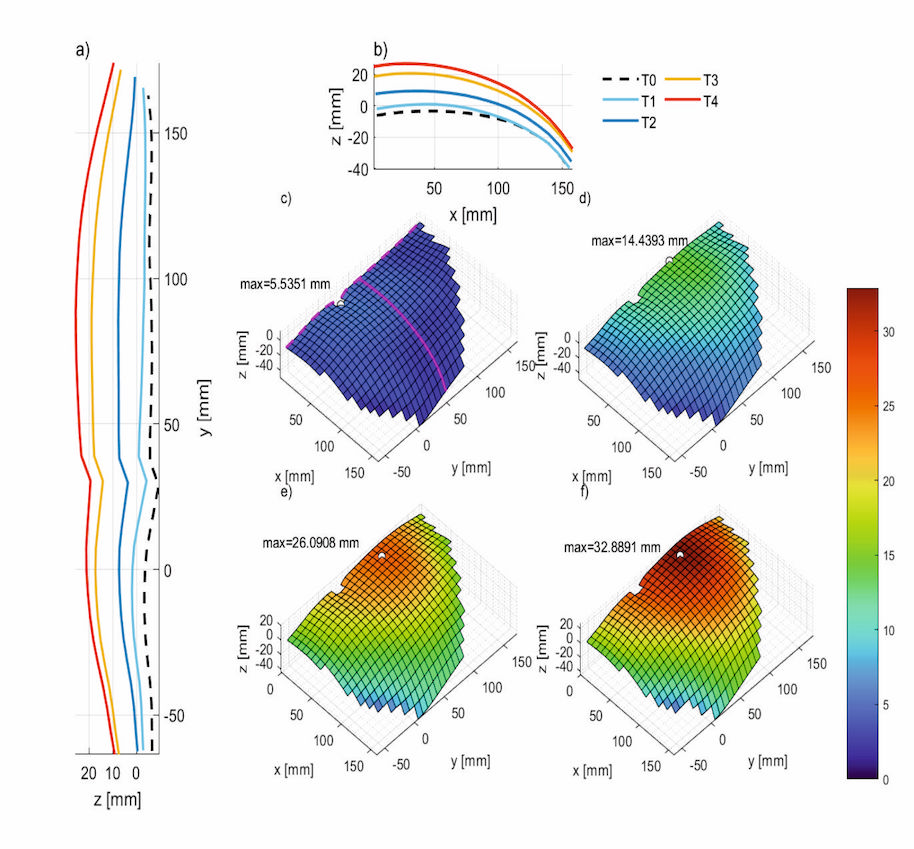}
    \caption{Shape and displacement of subject D8 in four stages T1--T4 and  reference one T0: a) profile of abdominal wall along mid-line A--A; b)  profile along transverse direction B--B; c--f) surfaces of abdominal wall with colour indicating total displacement [mm] in T1--T4, respectively with marked location of maximum displacement by a white circle; x is the mediolateral axis from right to left, y is craniocaudal axis from caudal to cranial, and z is anteriorposterior axis from anterior to posterior }
    \label{fig_dic8dis}
\end{figure}

\subsection{Displacement}

The shapes and displacements of the abdominal wall  of each subject in the four times steps in reference to the  T0 are shown in Figures \ref{fig_dic2dis},\ref{fig_dic4dis} and \ref{fig_dic8dis} (for other subjects see  \ref{appendix_d}).Origin of the Cartesian system is around 3 cm above umbilical point of each subject. The profiles of the abdominal wall along its mid-line A-A and along its transverse direction B-B show how the wall deforms under pressure. Here the specific moments T1, T2 and T3, T4 and the reference step T0 are marked with different lines. Thus the deformation resulting from breathing is also visible. Even if the same amount of dialysis fluid is introduced into the abdominal cavity, one can notice different levels of maximum displacement in different subjects. Sometimes, breathing provokes relatively high maximum abdominal wall displacement in relation to dialysis pressure, as observed in subject D4. In that case, the IPP was also quite high, 20 cmH$_2$O (Table \ref{Table_dic_result}).  Obtained pressure values are given for T4. Most of the subjects increased pressure by around 1 cmH$_2$O during inhalation (T4) comparing to exhalation (T3), both with dialysis fluid in the abdominal cavity. A similar observation was  reported by \cite{soucasse2022better} who noted that natural breathing in the supine position was in the 1--6 mmHg range. Differences between maximum displacements in corresponding stages of exhalation/inhalation are 6.23$\pm$3.83 in case of T1,T2 and 8$\pm$4.86  in case of T3, T4. Therefore, the values are smaller than observed in case of study of \cite{jourdan2022dynamic} when forced breathing were analysed. However, it should be noted that our study concerns natural breathing. Therefore lower displacement can be expected.

\begin{table}[ht]
\begin{tabular}{cccccccccc}
\hline
 &               &   & \multicolumn{3}{c}{$\varepsilon_1$ [-]}   & & \multicolumn{3}{c}{$\varepsilon_2$ [-]}    \\\cline{4-6}\cline{8-10}
subject & pressure                 & & median & \multicolumn{2}{c}{percentile}  & & median & \multicolumn{2}{c}{percentile}  \\\cline{5-6}\cline{9-10}
 &    [cmH$_2$O]              & & & 25th    & 75th   & &  & 25th    & 75th    \\
\hline
    D1     & 11    &       & 0.044 & 0.033 & 0.066 &       & 0.008 & 0.002 & 0.014 \\
    D2     & 15    &       & 0.086 & 0.07  & 0.094 &       & 0.029 & 0.022  & 0.038 \\
   D3     & 11    &       & 0.081 & 0.053 & 0.104 &       & 0.001 & -0.008 & 0.017 \\
    D4     & 21    &       & 0.064 & 0.048 & 0.081 &       & 0.027 & 0.006 & 0.043 \\
    D5     & 12    &       & 0.047 & 0.025 & 0.105 &       & 0.008 & -0.012 & 0.019 \\
    D6     & 15    &       & 0.124 & 0.094 & 0.153 &       & 0.024 & 0.007 & 0.039 \\
    D7     & 21    &       & 0.063 & 0.03  & 0.113 &       & -0.002 & -0.014  & 0.006 \\
    D8     & 16    &       & 0.079 & 0.061 & 0.104 &       & 0.04  & 0.020 & 0.048 \\
    D9     & 20    &       & 0.039 & 0.028 & 0.047 &       & 0.014 & 0.003 & 0.024 \\
    D10    & 12    &       & 0.099 & 0.062 & 0.166 &       & 0.016 & -0.020 & 0.033 \\
    D11    & 18    &       & 0.043 & 0.032 & 0.049 &       & 0.019 & 0.009 & 0.028 \\
    D12    & 10    &       & 0.086 & 0.061 & 0.115 &       & 0.043 & 0.016 & 0.06 \\
\hline
\end{tabular}
\caption{Intra-abdominal pressure and principal strains statistics for each subject (T4)  } \label{Table_dic_result}
\end{table}

\subsection{Principal strains}

\begin{figure}[ht!]\centering
    \includegraphics[width=\textwidth]{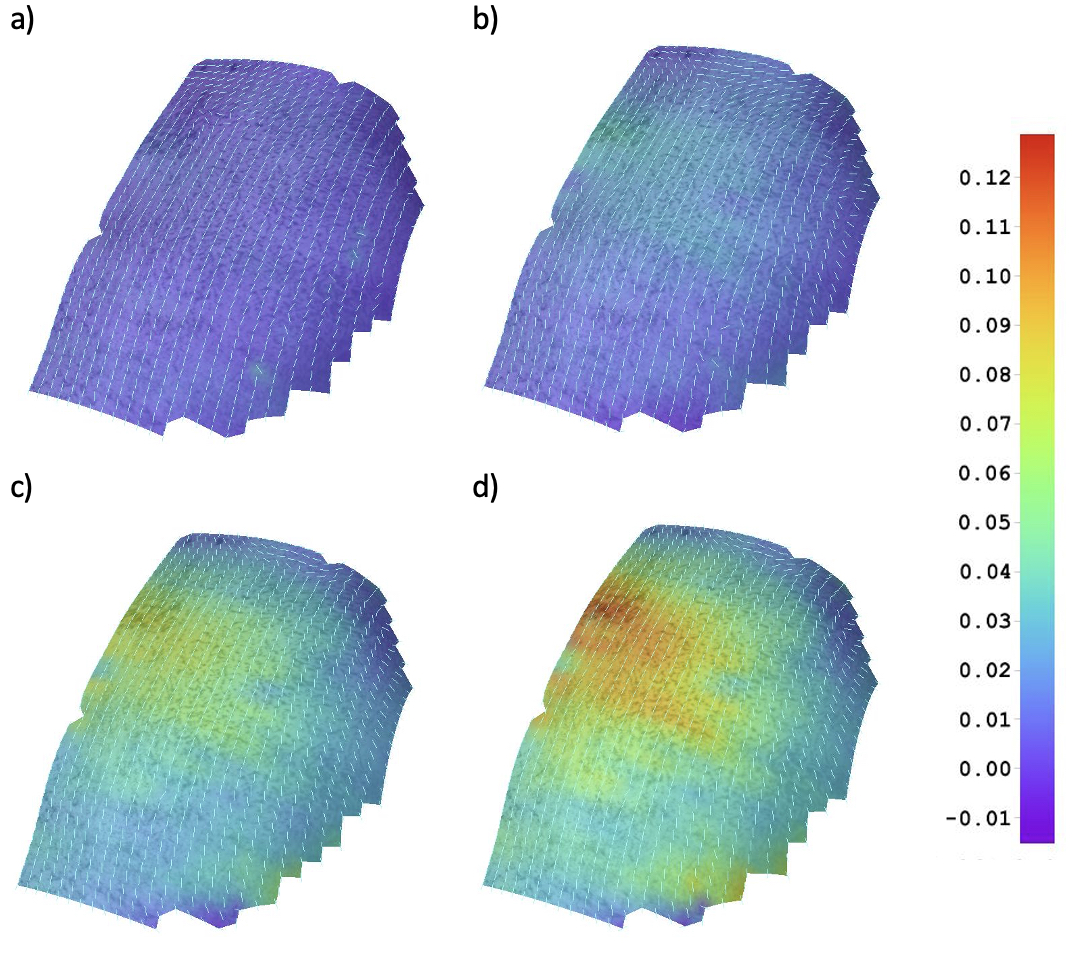}
    \caption{ Map of  principal Lagrangian strains $\varepsilon_1$ and  directions for subject D1 in stages T1 (a), T2 (b), T3 (c) and T4 (d)}
    \label{fig_dic1strain}
\end{figure}

\begin{figure}[ht!]\centering
    \includegraphics[width=0.7\textwidth]{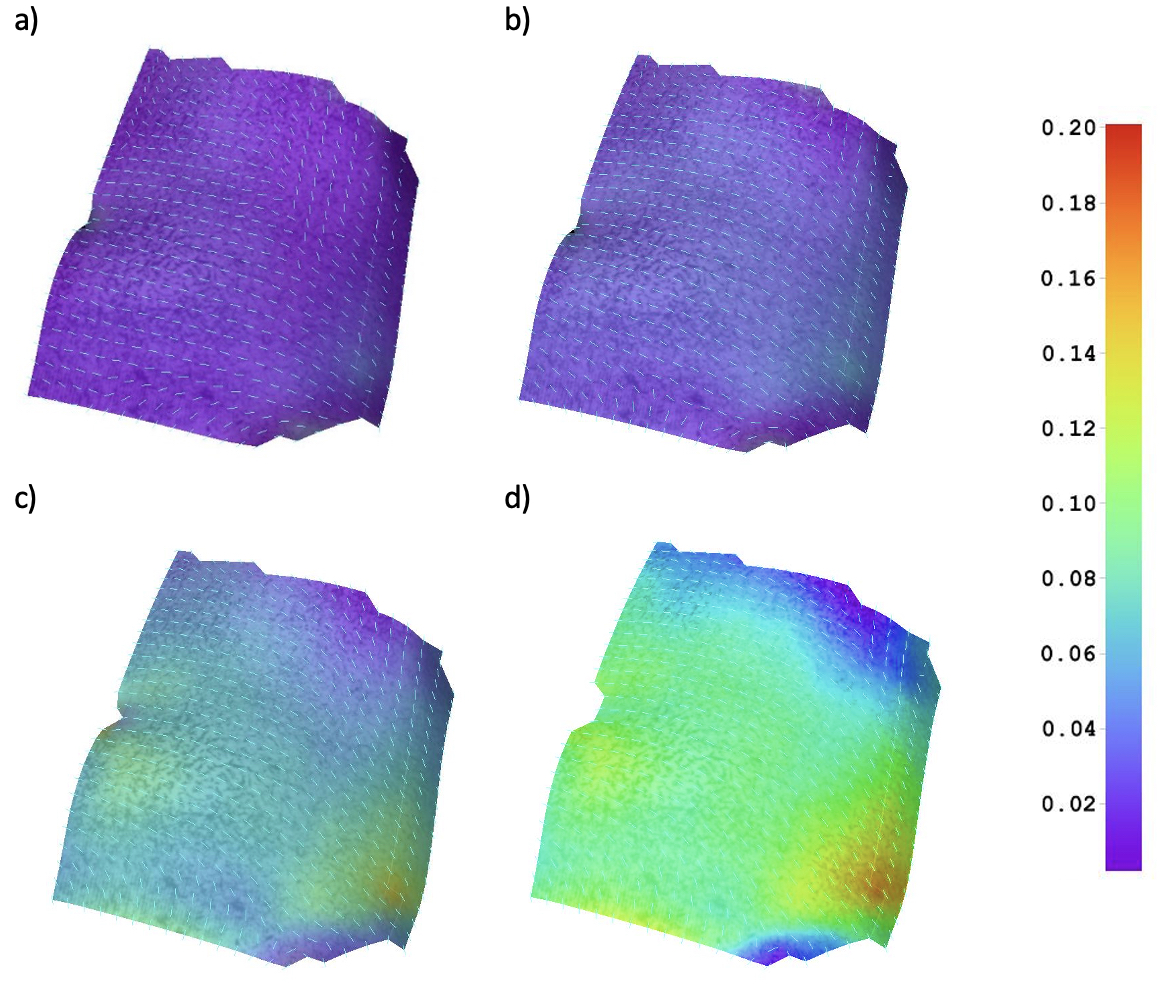}
    \caption{ Map of  principal Lagrangian strains $\varepsilon_1$ and  directions for subject D2 in stages T1 (a), T2 (b), T3 (c) and T4 (d)}
    \label{fig_dic2strain}
\end{figure}

\begin{figure}[ht!]\centering
    \includegraphics[width=0.7\textwidth]{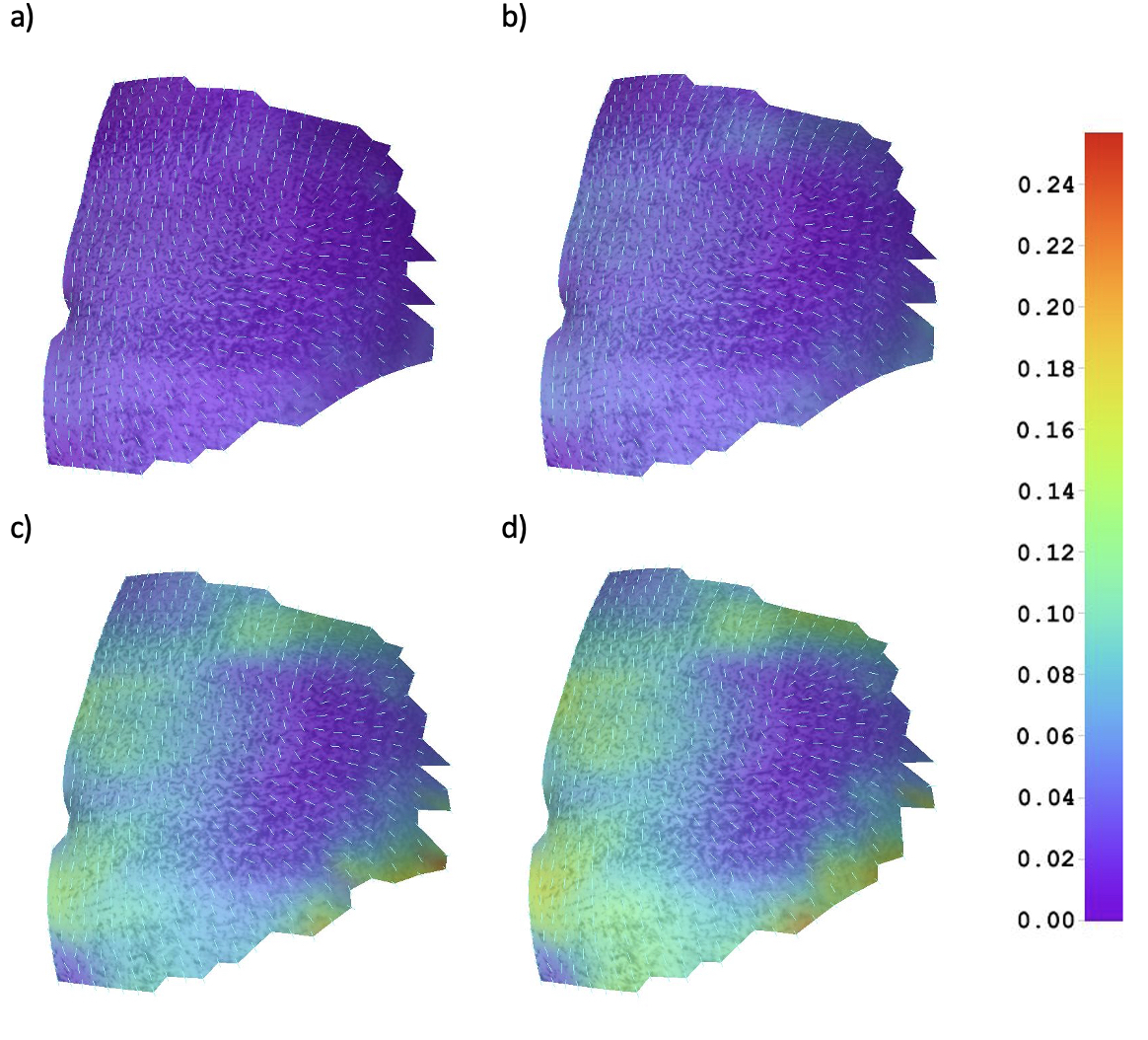}
    \caption{ Map of  principal Lagrangian strains $\varepsilon_1$ and  directions for subject D3 in stages T1 (a), T2 (b), T3 (c) and T4 (d)}
    \label{fig_dic3strain}
\end{figure}

\begin{figure}[ht!]\centering
    \includegraphics[width=0.7\textwidth]{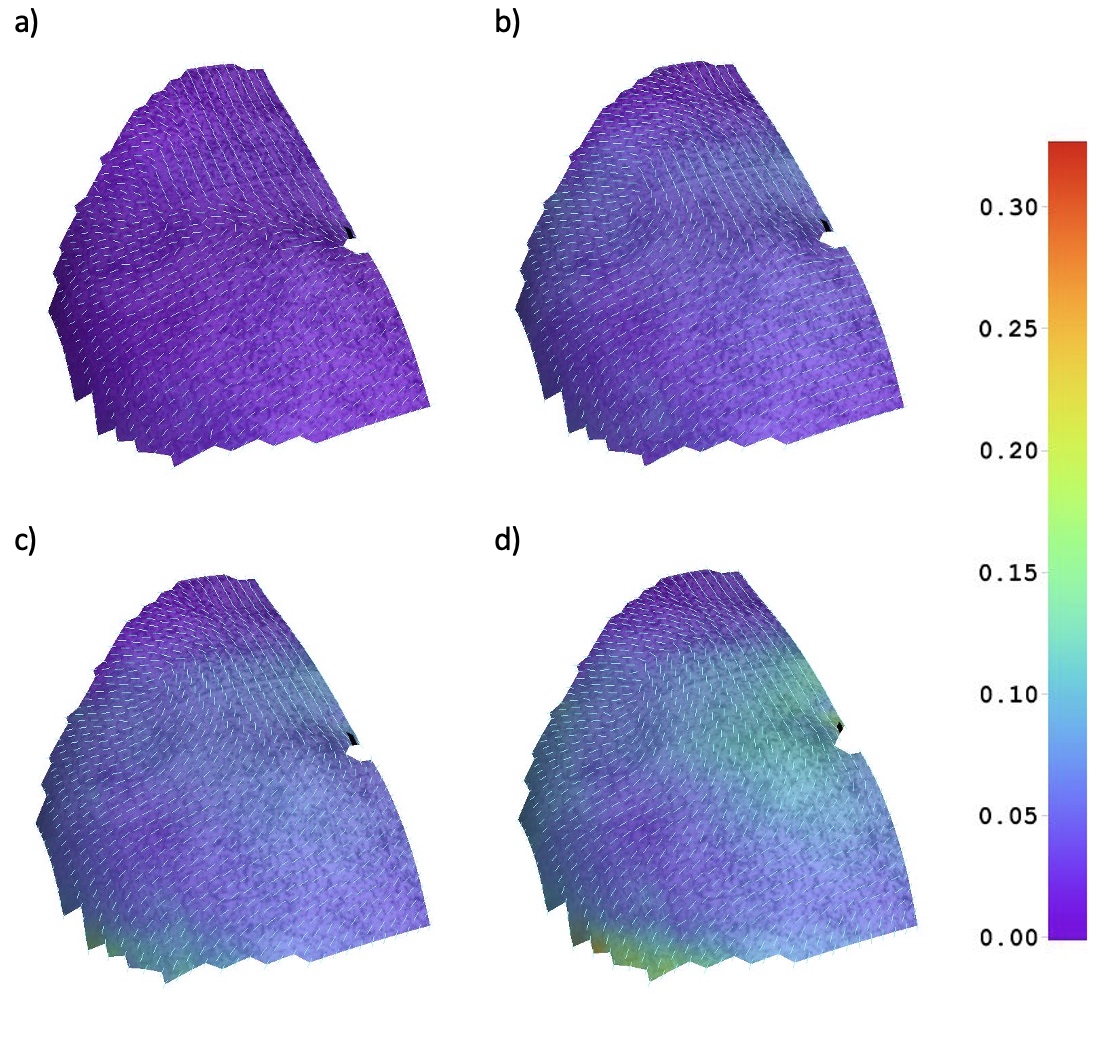}
    \caption{ Map of  principal Lagrangian strains $\varepsilon_1$ and  directions for subject D4 in stages T1 (a), T2 (b), T3 (c) and T4 (d)}
    \label{fig_dic4strain}
\end{figure}

\begin{figure}[ht!]\centering
    \includegraphics[width=0.7\textwidth]{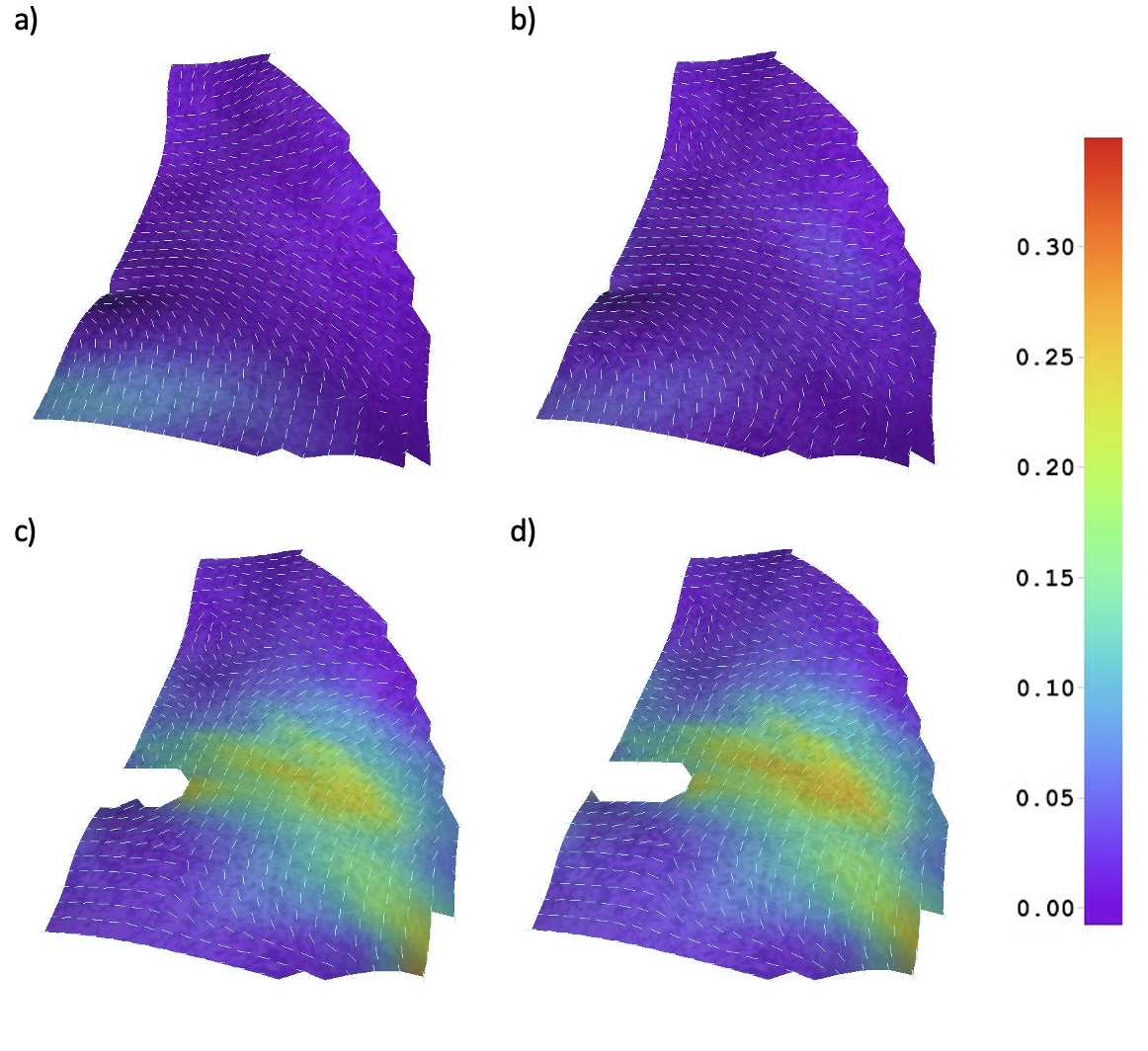}
    \caption{ Map of  principal Lagrangian strains $\varepsilon_1$ and  directions for subject D5 in stages T1 (a), T2 (b), T3 (c) and T4 (d)}
    \label{fig_dic5strain}
\end{figure}

\begin{figure}[ht!]\centering
    \includegraphics[width=0.7\textwidth]{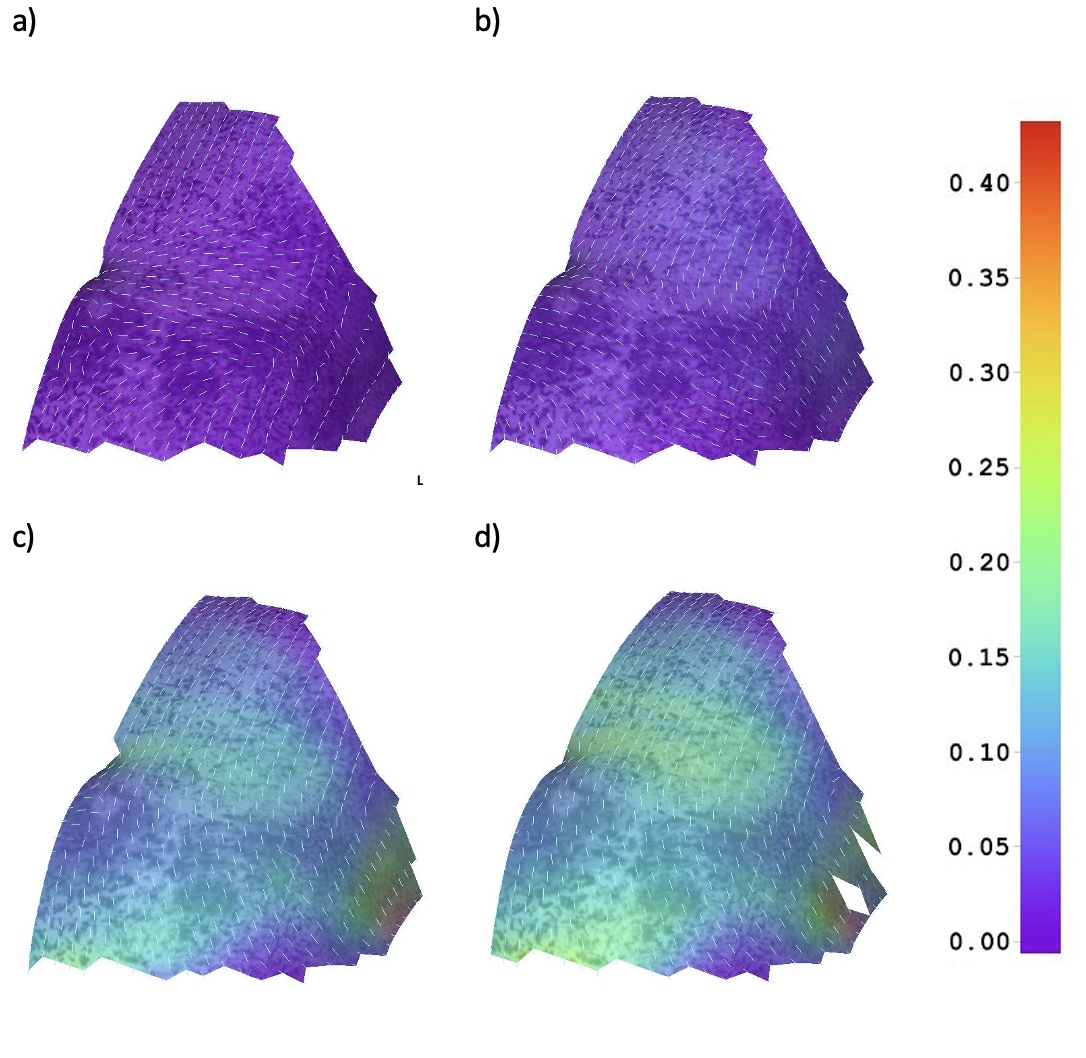}
    \caption{ Map of  principal Lagrangian strains $\varepsilon_1$ and  directions for subject D6 in stages T1 (a), T2 (b), T3 (c) and T4 (d)}
    \label{fig_dic6strain}
\end{figure}

\begin{figure}[ht!]\centering
    \includegraphics[width=0.7\textwidth]{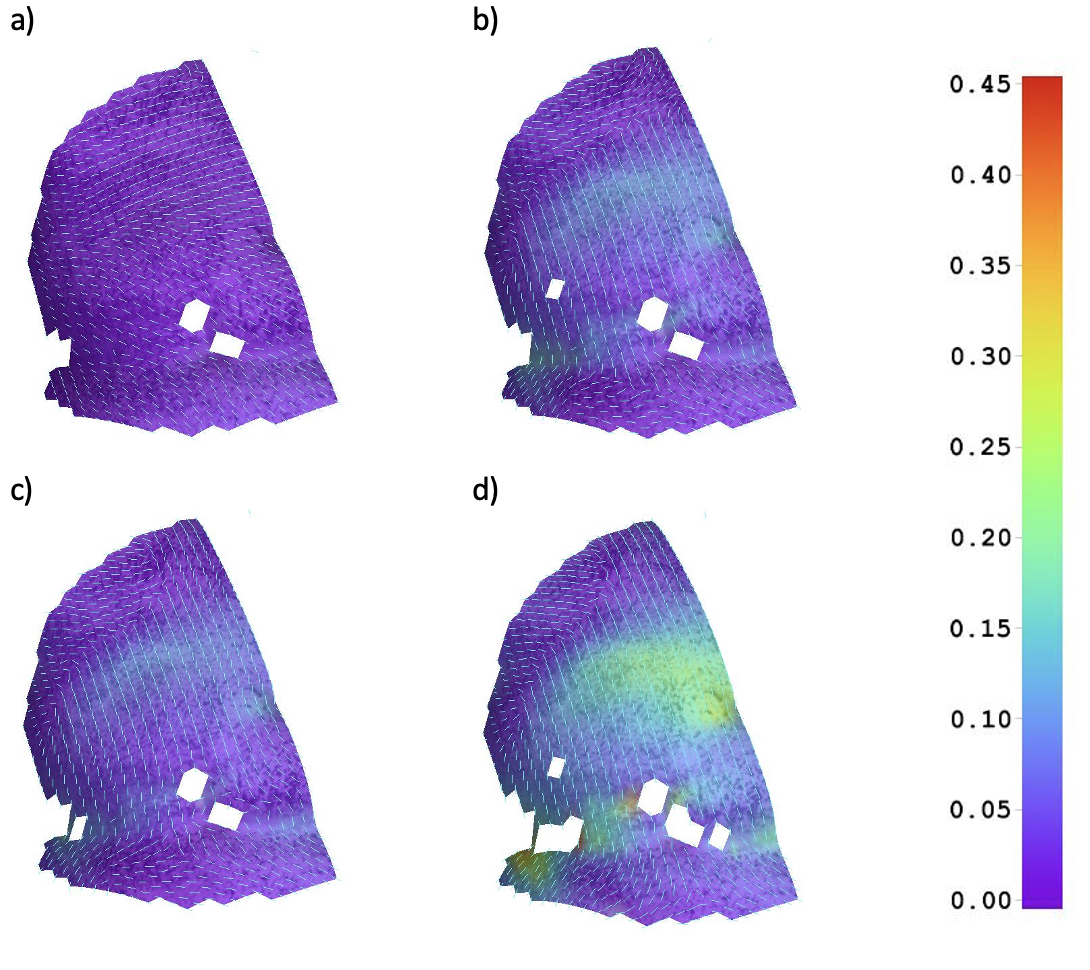}
    \caption{ Map of  principal Lagrangian strains $\varepsilon_1$ and  directions for subject D7 in stages T1 (a), T2 (b), T3 (c) and T4 (d)}
    \label{fig_dic7strain}
\end{figure}

\begin{figure}[ht!]\centering
    \includegraphics[width=0.7\textwidth]{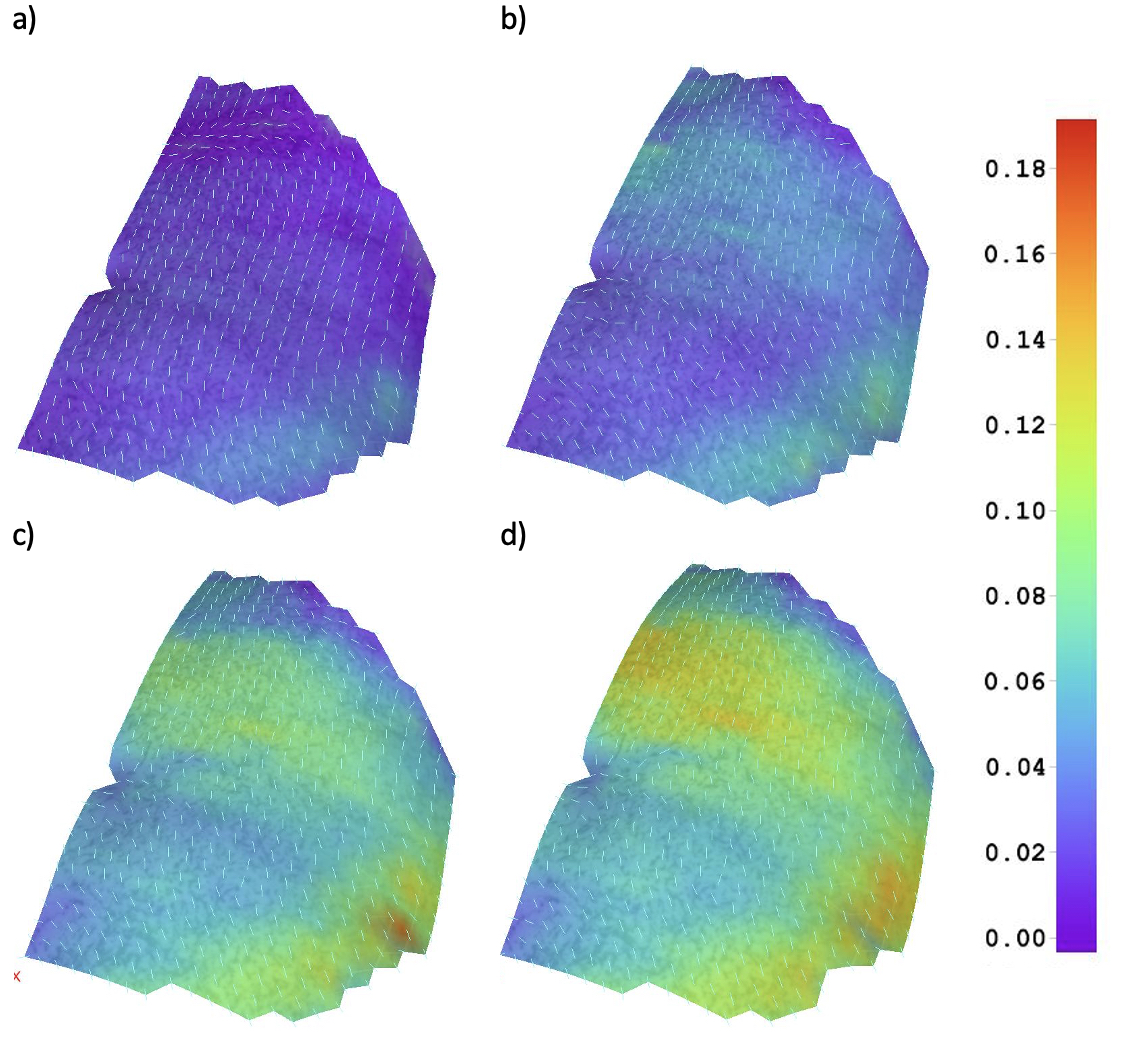}
    \caption{ Map of  principal Lagrangian strains $\varepsilon_1$ and  directions for subject D8 in stages T1 (a), T2 (b), T3 (c) and T4 (d)}
    \label{fig_dic8strain}
\end{figure}

\begin{figure}[ht!]\centering
    \includegraphics[width=0.7\textwidth]{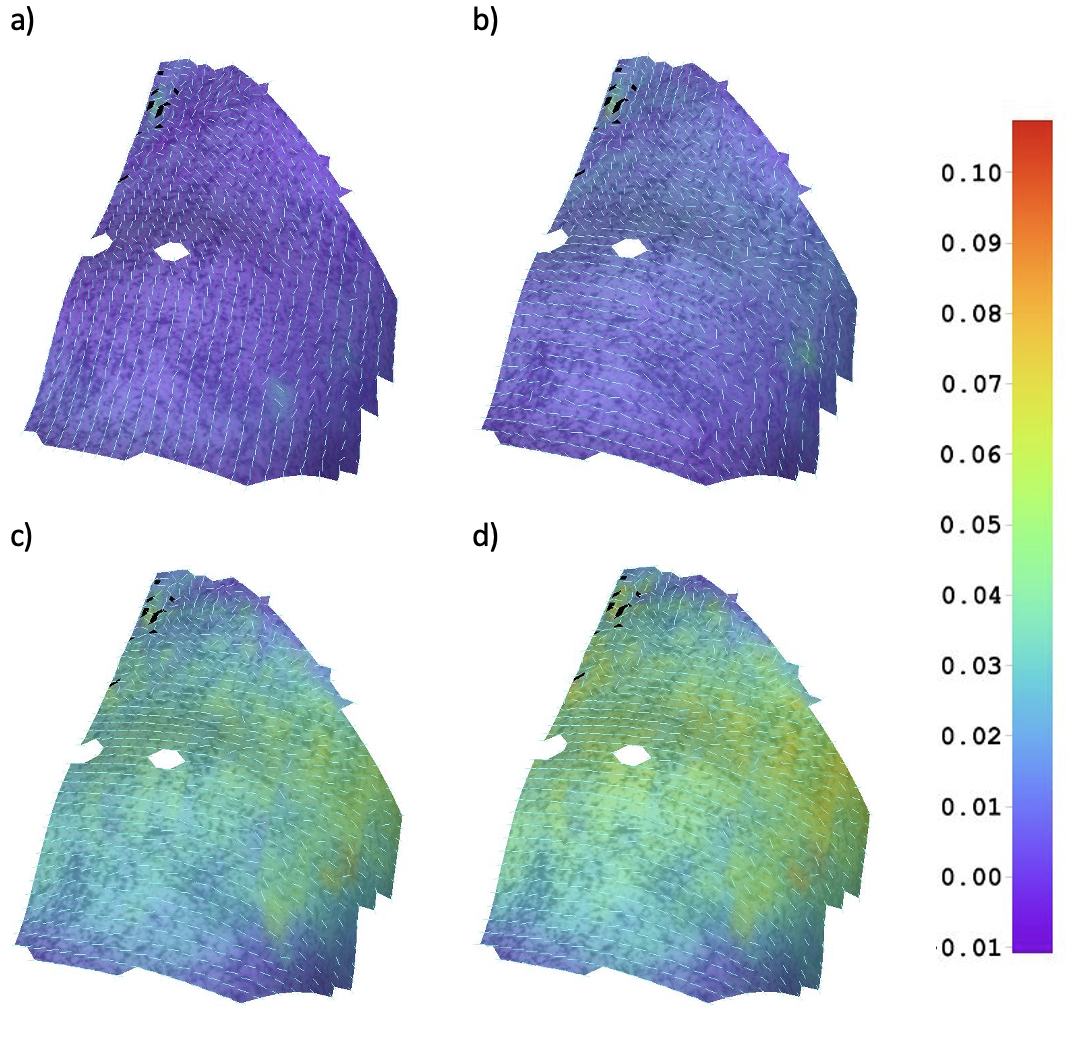}
    \caption{Map of  principal Lagrangian strains $\varepsilon_1$ and  directions for subject D9 in stages T1 (a), T2 (b), T3 (c) and T4 (d)}
    \label{fig_dic9strain}
\end{figure}

\begin{figure}[ht!]\centering
    \includegraphics[width=0.7\textwidth]{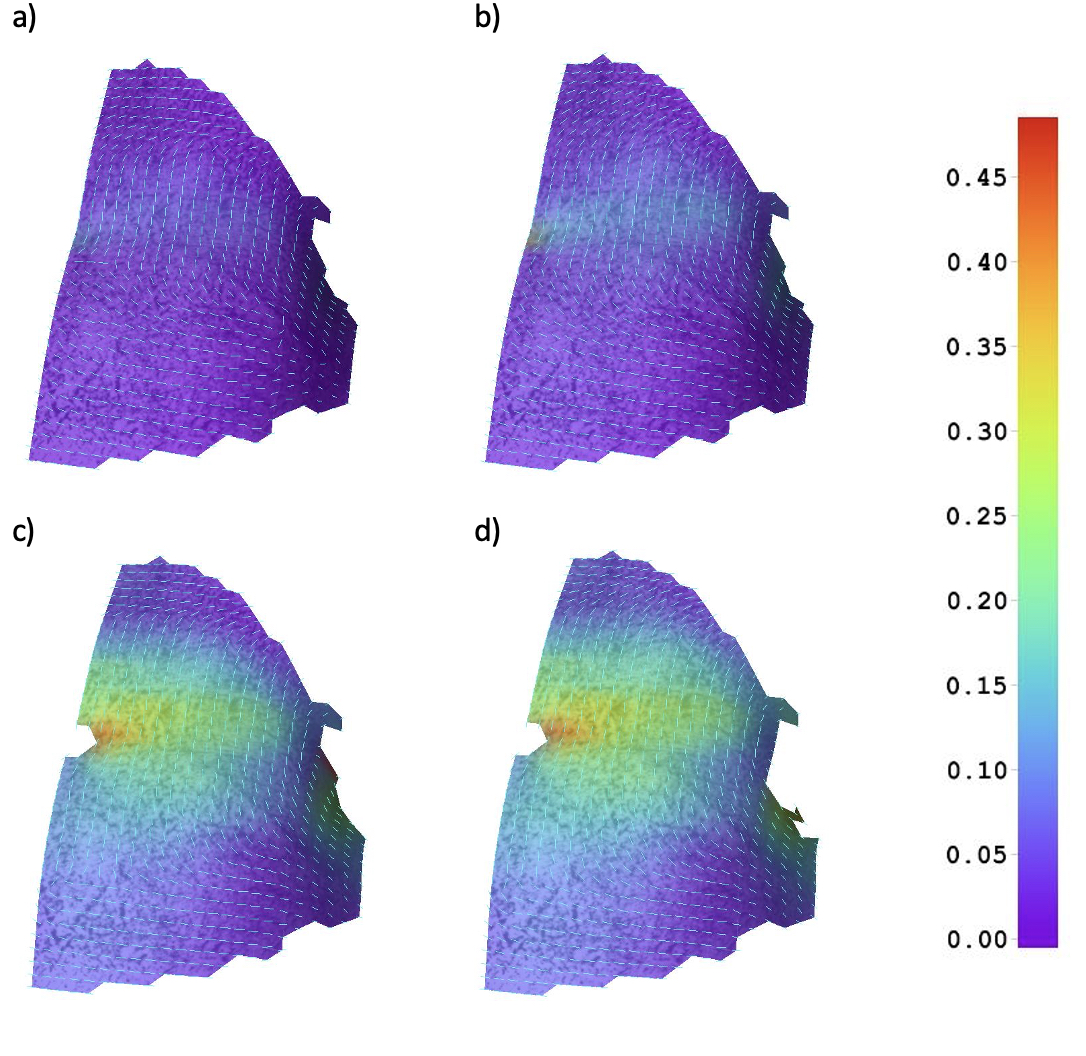}
    \caption{ Map of  principal Lagrangian strains $\varepsilon_1$ and  directions for subject D10 in stages T1 (a), T2 (b), T3 (c) and T4 (d)}
    \label{fig_dic10strain}
\end{figure}

\begin{figure}[ht!]\centering
    \includegraphics[width=0.7\textwidth]{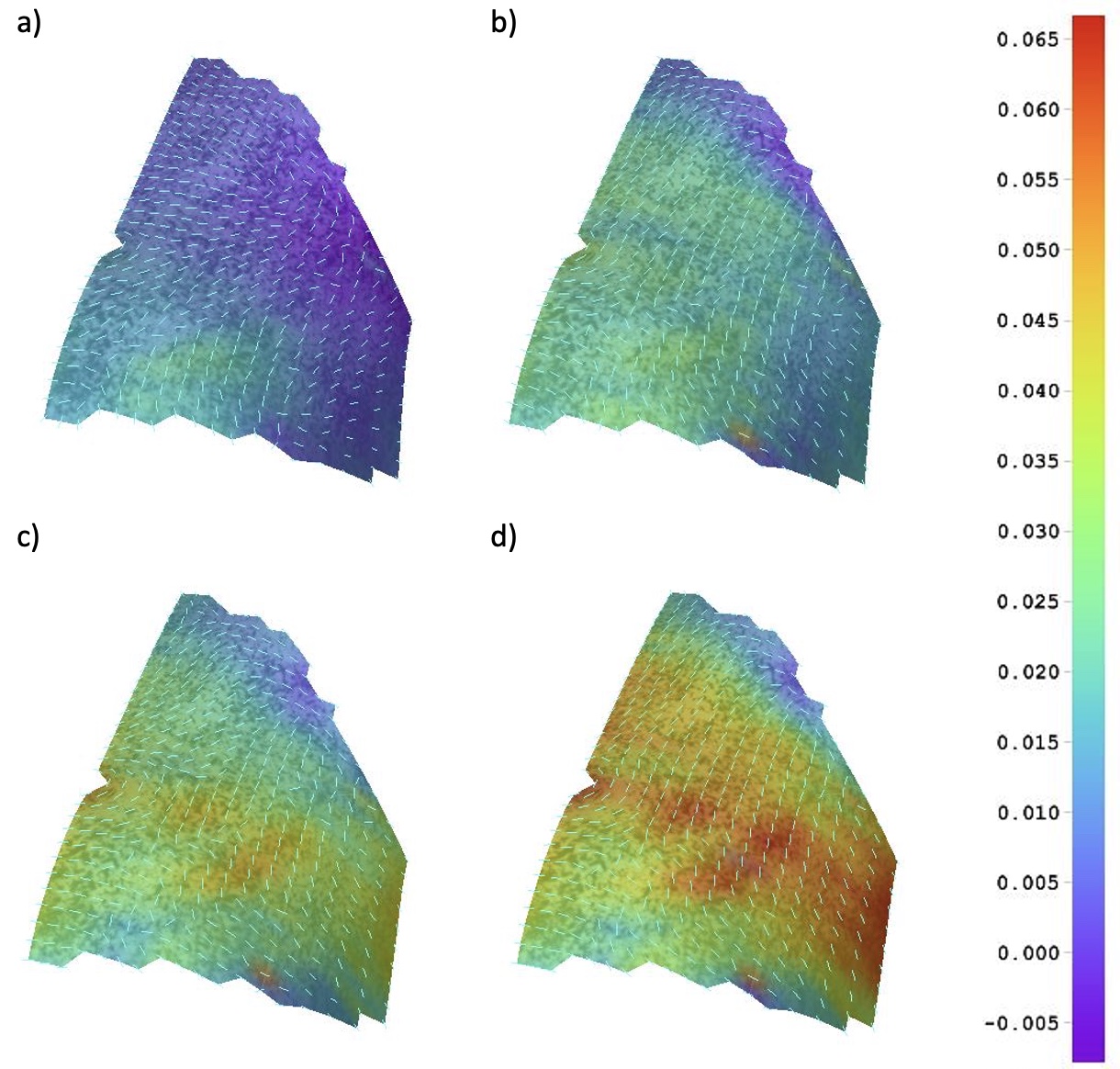}
    \caption{ Map of  principal Lagrangian strains $\varepsilon_1$ and  directions for subject D11 in stages T1 (a), T2 (b), T3 (c) and T4 (d)}
    \label{fig_dic11strain}
\end{figure}

\begin{figure}[ht!]\centering
    \includegraphics[width=0.7\textwidth]{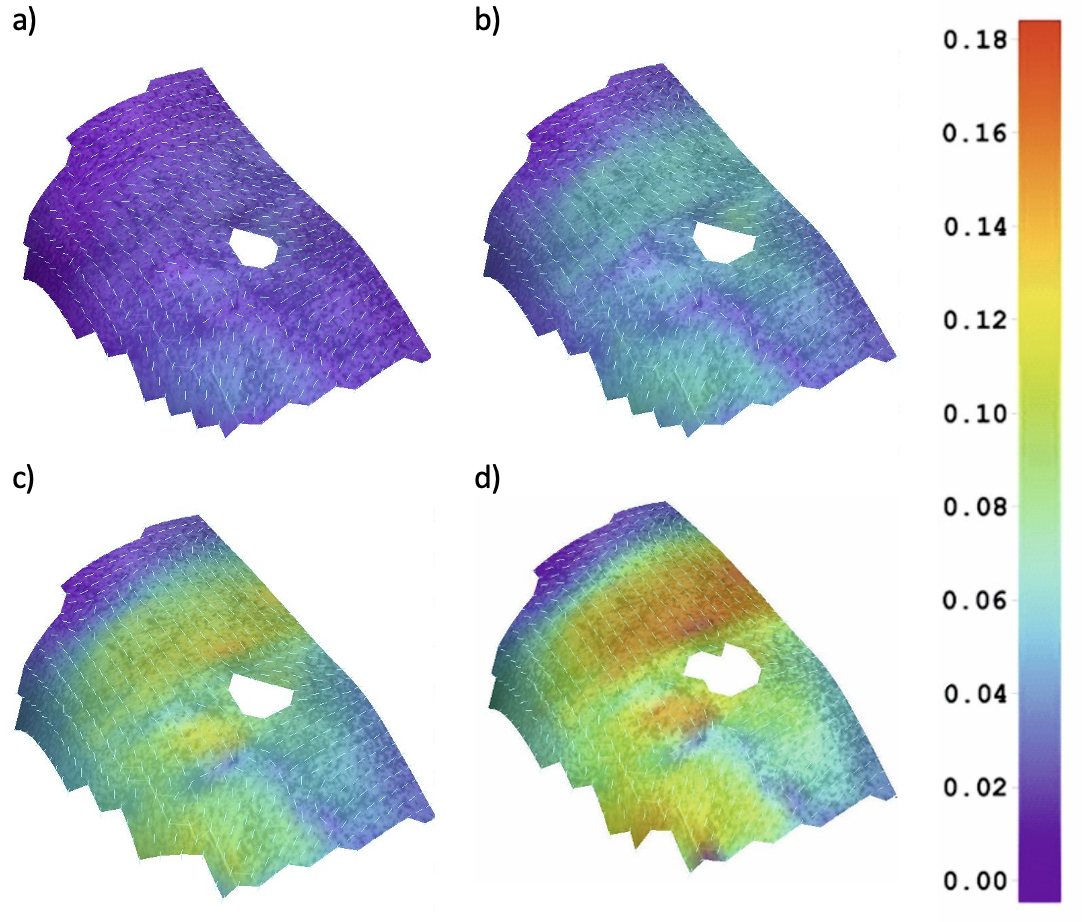}
    \caption{ Map of  principal Lagrangian strains $\varepsilon_1$ and  directions for subject D12 in stages T1 (a), T2 (b), T3 (c) and T4 (d)}
    \label{fig_dic12strain}
\end{figure}

\begin{figure}[ht!]\centering
    \includegraphics[width=\textwidth]{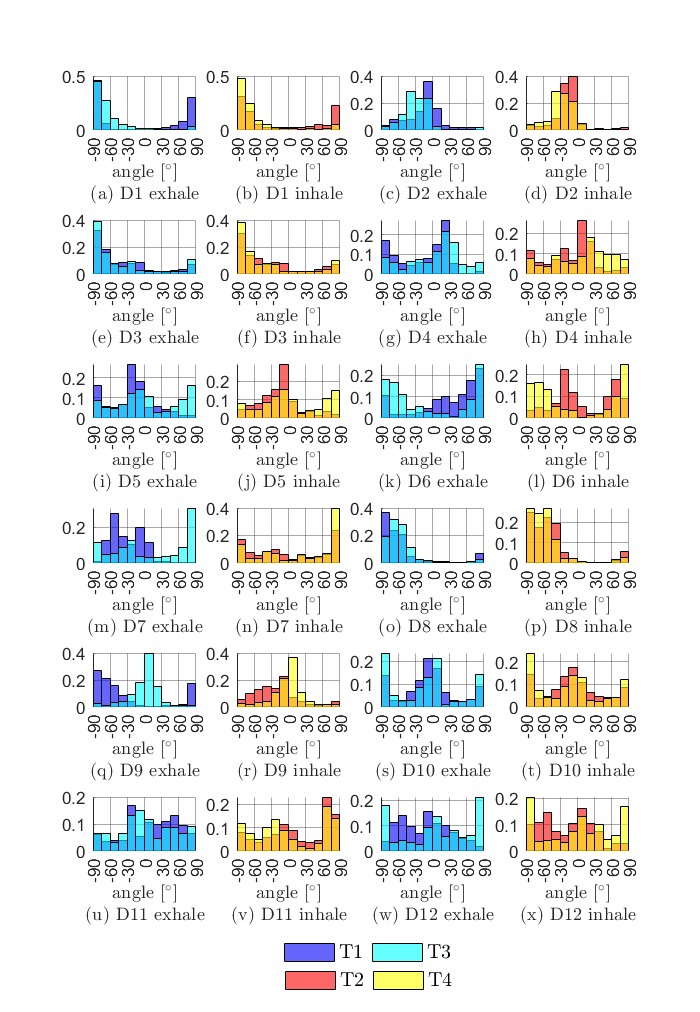}
    \caption{Normalised histograms of the angle between the first principal direction and x-axis in the early stages of fluid introduction during exhalation (T1) and inhalation (T2) and at the end of fluid introduction during exhalation (T3) and inhalation (T4) for subjects D1--D12}
    \label{fig_hist_angle}
\end{figure}

Figures \ref{fig_dic1strain}--\ref{fig_dic12strain} show maps of the principal strain $\varepsilon_1$ with its directions  for all the subjects in four time-steps (T).  Detailed information about the principal strains for  T4 is summarised in Table \ref{Table_dic_result} for each subject together with the intra-abdominal pressure value.
The principal strain directions in the case of some subjects (e.g. D8), changed while introducing the dialysis fluid. Principal directions and stains differed in various abdominal wall areas.  This  indicates different material properties dominating in various zones, which may account for the anisotropy of the complex abdominal wall structure. In most subjects,  breathing also affected the principal strain directions. In some parts of the abdominal wall, the directions changed during exhalation (e.g Figure \ref{fig_dic8strain}). It should be noted that sometimes, as in the case of subject D8, the principal direction change was less visible when the cavity was full (compare T1 and T2 vs T3 and T4). The distribution of the first principal direction angle in steps T1--T4 is presented in Figure \ref{fig_hist_angle}. Regarding values, in subjects D1 and D3, the principal direction angle $\alpha$ , close to $90\,^{\circ}$, was dominant, while in the other subjects,  greater variation of principal directions was observed. In the case of subject D9, when the abdominal cavity was filled, the dominating direction of the maximum principal strains changed by $90\,^{\circ}$  in relation to when it was  empty.

\begin{figure}[ht!]\centering
    \includegraphics[width=\textwidth]{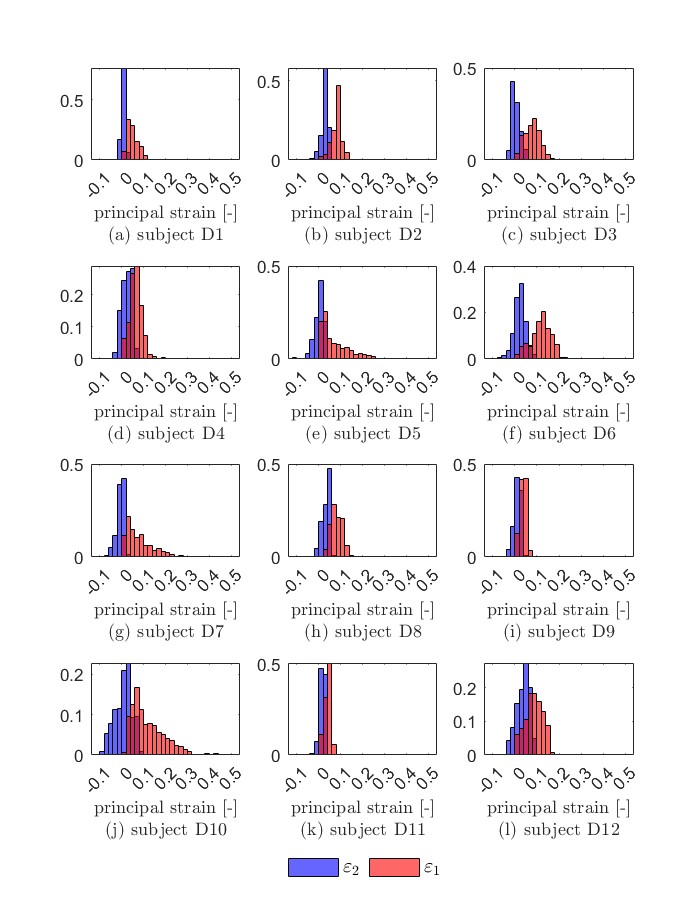}
    \caption{Normalised histograms of maximum (red) and minimum (blue) principal Lagrangian strains on the external abdominal wall surfaces of  subjects D1--D12 during inhalation  with abdominal cavities filled with dialysis fluid ( T4)}
    \label{fig_histograms}
\end{figure}

The difference in the distribution of principal strains in all subjects during inhaling when the abdominal cavity filled is presented in the Figure \ref{fig_histograms}. Table \ref{Table_dic_result} shows statistics of these values. The dominating maximum principal strains are presented in Table \ref{tab_mode}. The highest mode values are observed in subjects D2, D3, D6 (female ) and D12 (male  with operated hernia).  In D3 and D6, the values of $\varepsilon_1$ and $\varepsilon_2$ differ most. On the other hand, an opposite situation is observed in case of D1, D4, D5, D7, D9 and D11, who were male subjects. The level of   minimum and maximum strain variability differed in each subject. D9 and D11 had a narrow range of minimum and maximum principal strains, i.e. strain value variability was low. These were the youngest male subjects with similar body-mass indexes (BMI). Female subject D2,  in the same age range, had slightly higher  $\varepsilon_1$ and $\varepsilon_2$ variability.


\begin{table}[!ht]
    \centering
    \begin{tabular}{p{0.7cm}p{0.7cm}p{0.7cm}p{0.7cm}p{0.7cm}p{0.7cm}p{0.7cm}p{0.7cm}p{0.7cm}p{0.7cm}p{0.7cm}p{0.7cm}}
    \hline
         D1 & D2 & D3 & D4 & D5 & D6 & D7 & D8 & D9 & D10 & D11 & D12 \\ \hline
        0.02-0.04 & 0.08-0.1 & 0.08-0.1 & 0.06-0.08 & 0.02-0.04 & 0.12-0.14 & 0.02-0.04 & 0.06-0.08 & 0.04-0.06 & 0.06-0.08 & 0.06-0.08 & 0.08-0.1 \\ \hline

            \end{tabular}
            \caption{Most frequent range of $\varepsilon_1$ for  subjects D1--D12 in T4}\label{tab_mode}
\end{table}

\begin{figure}[ht!]\centering
    \includegraphics[width=\textwidth]{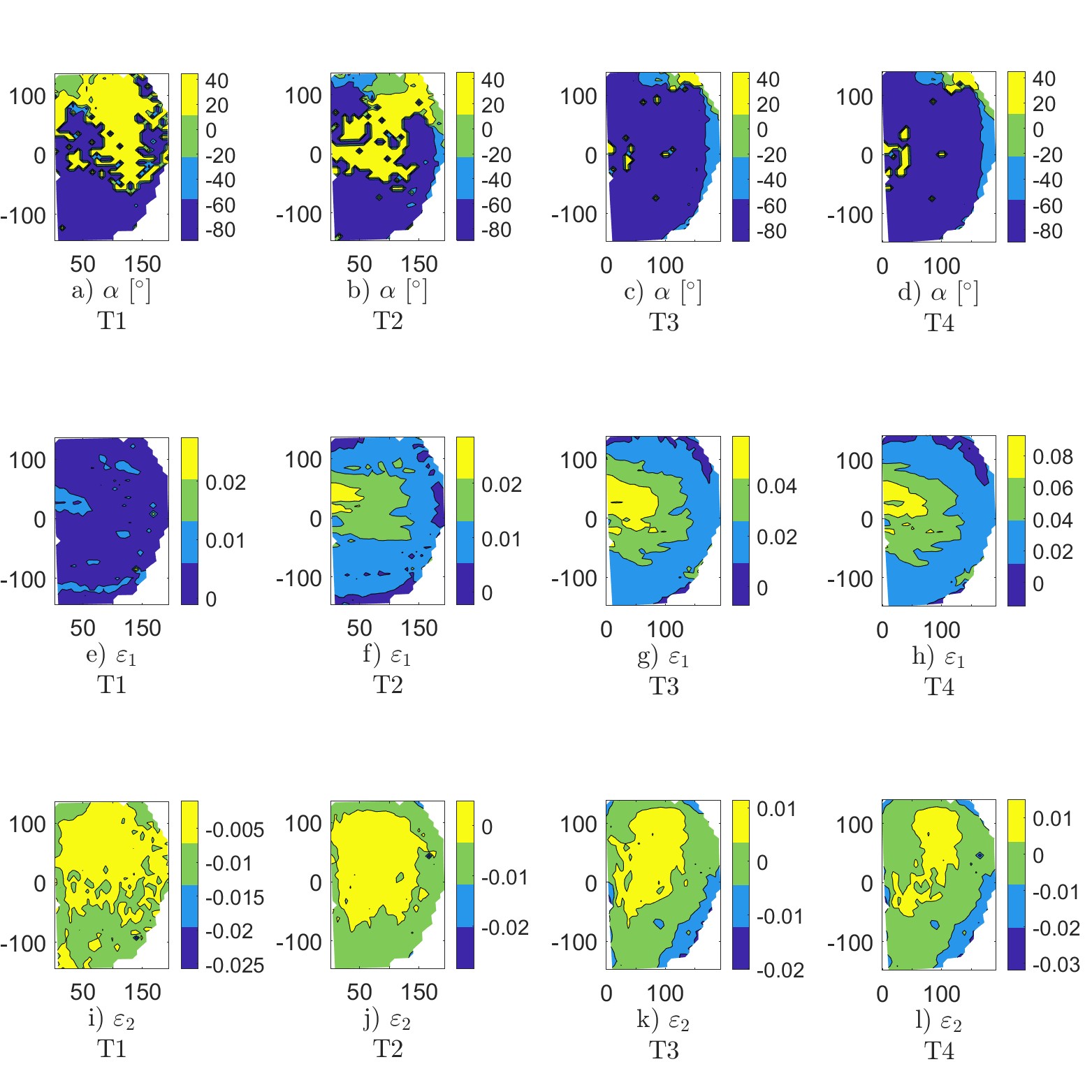}
    \caption{Contour maps with 3 levels of isolines for subject D1: principal direction angle $\alpha$ a)--d), principal strain $\varepsilon_1$ e)--h) and $\varepsilon_2$ i)--l)}
    \label{figcontourD1}
\end{figure}
\begin{figure}[ht!]\centering
    \includegraphics[width=\textwidth]{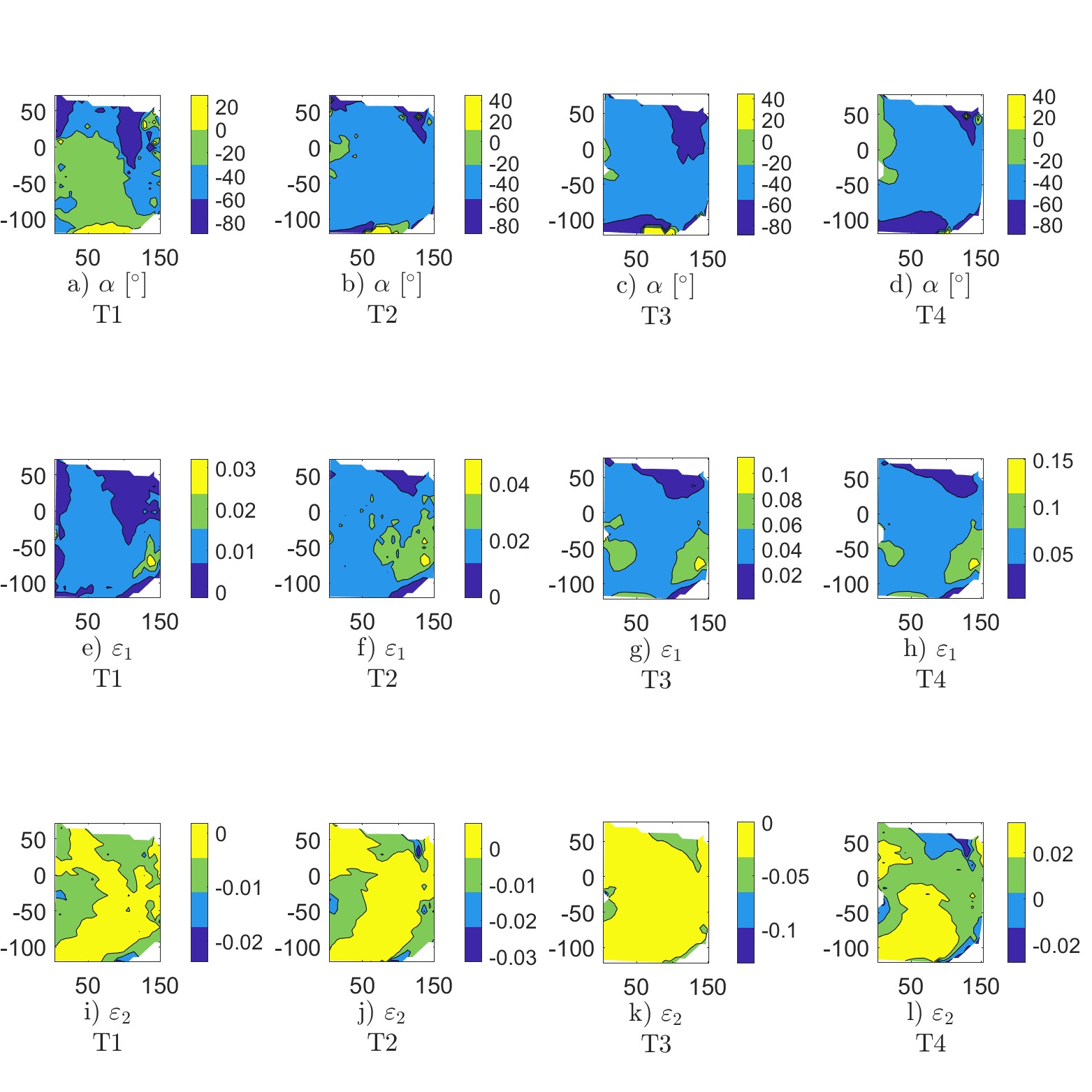}
    \caption{Contour maps with 3 levels of isolines for subject D2: principal direction angle $\alpha$ a)--d), principal strain $\varepsilon_1$ e)--h) and $\varepsilon_2$ i)--l)}
    \label{figcontourD2}
\end{figure}
\begin{figure}[ht!]\centering
    \includegraphics[width=\textwidth]{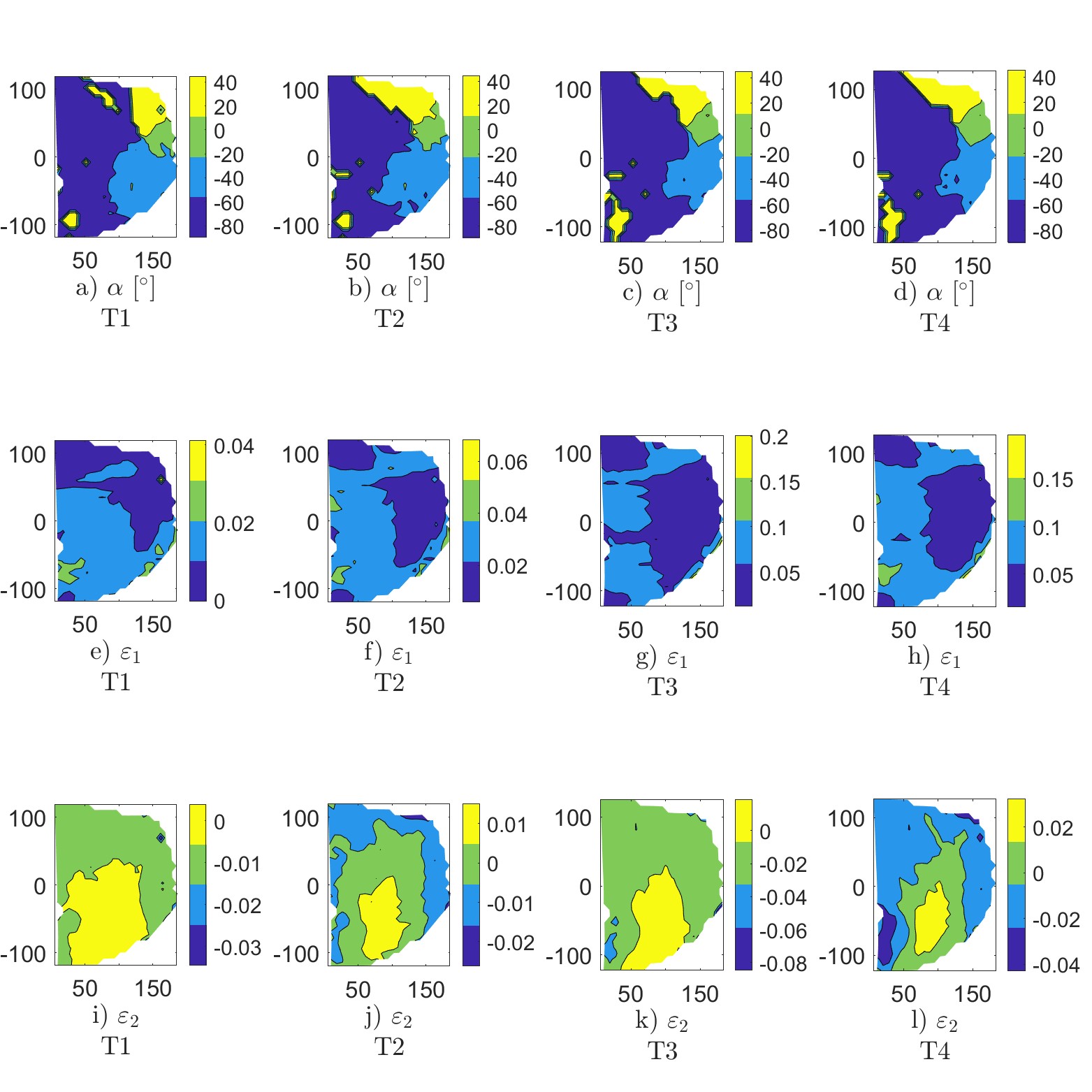}
    \caption{Contour maps with 3 levels of isolines for subject D3: principal direction angle $\alpha$ a)--d), principal strain $\varepsilon_1$ e)--h) and $\varepsilon_2$ i)--l)}
    \label{figcontourD3}
\end{figure}
\begin{figure}[ht!]\centering
    \includegraphics[width=\textwidth]{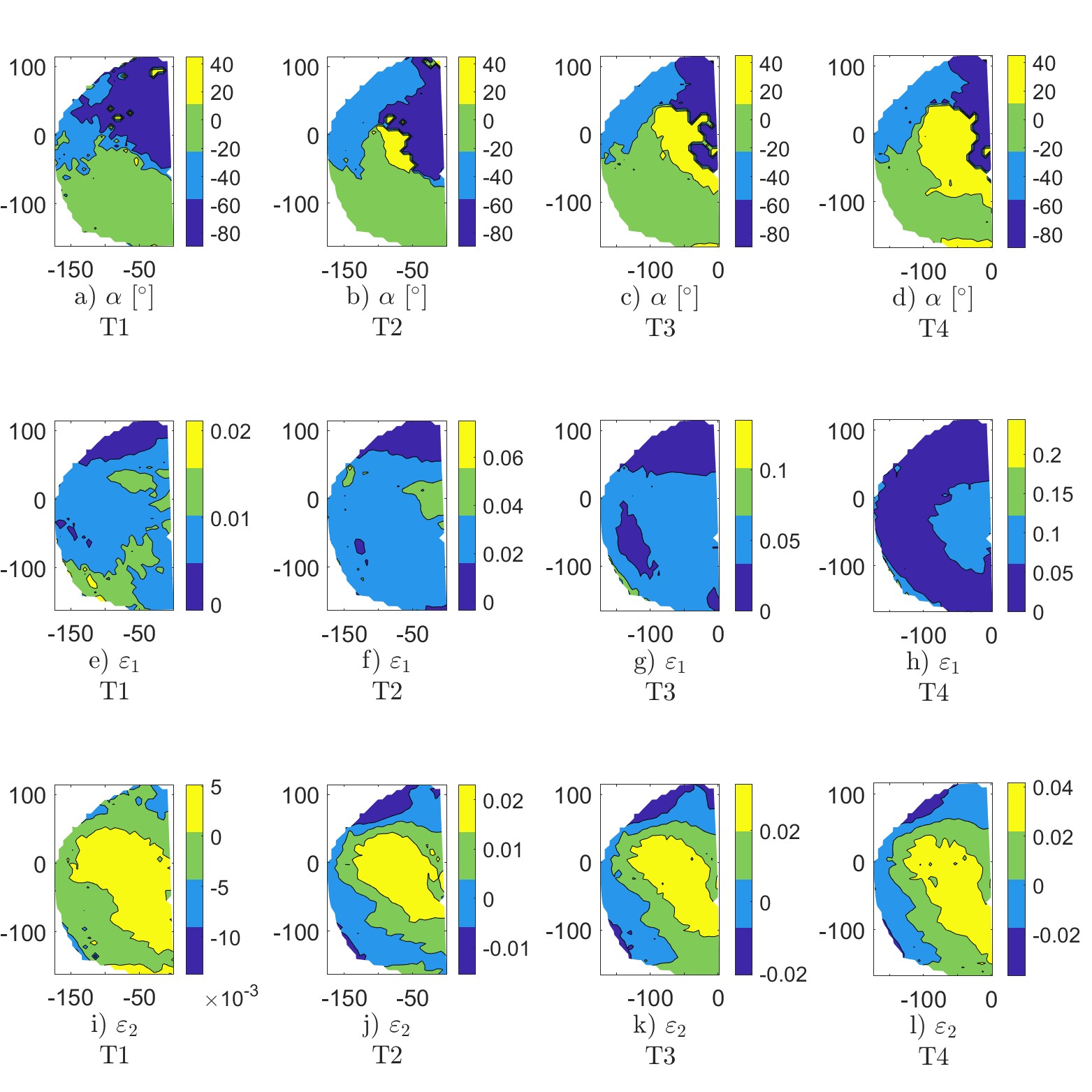}
    \caption{Contour maps with 3 levels of isolines for subject D4: principal direction angle $\alpha$ a)--d), principal strain $\varepsilon_1$ e)--h) and $\varepsilon_2$ i)--l)}
    \label{figcontourD4}
\end{figure}
\begin{figure}[ht!]\centering
    \includegraphics[width=\textwidth]{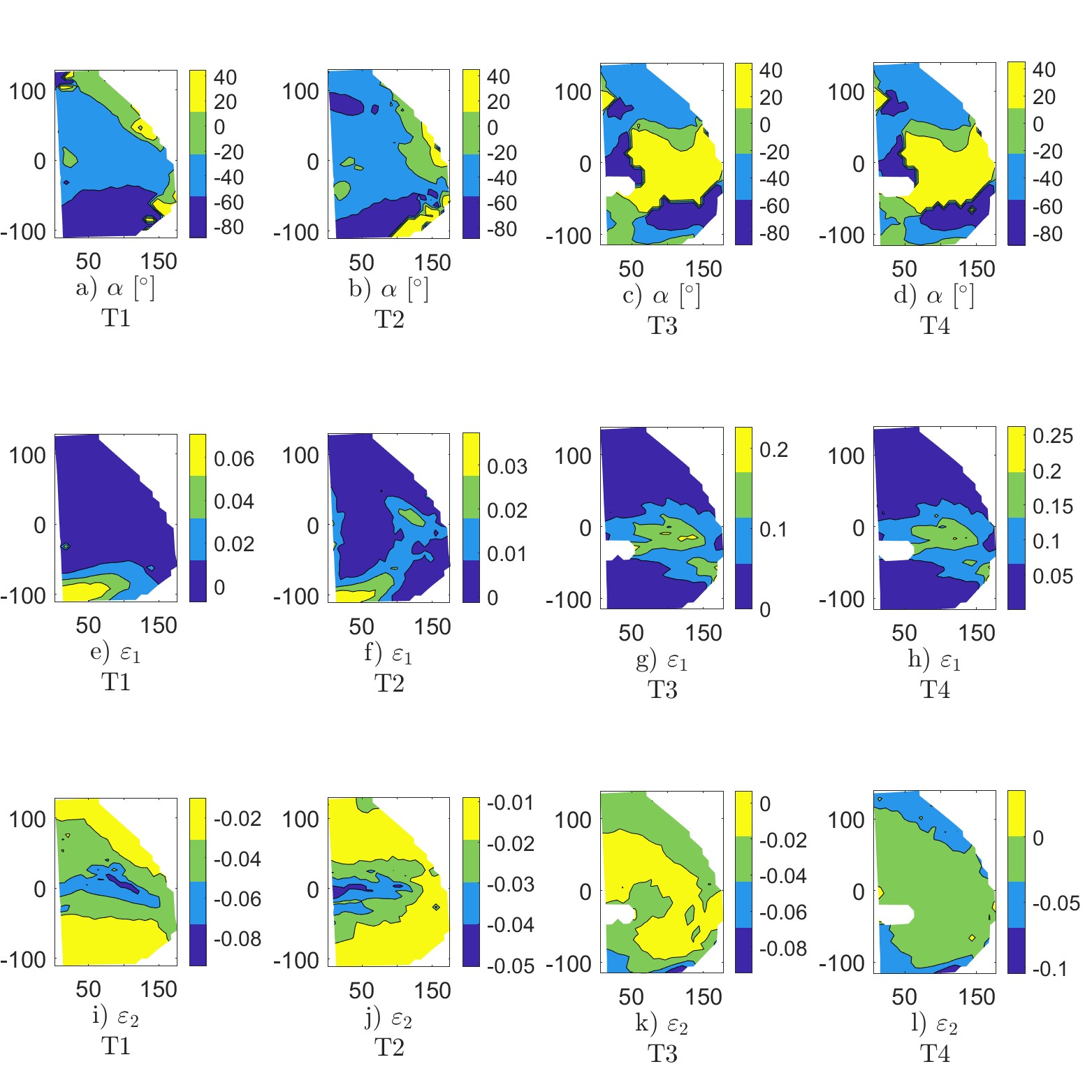}
    \caption{Contour maps with 3 levels of isolines for subject D5: principal direction angle $\alpha$ a)--d), principal strain $\varepsilon_1$ e)--h) and $\varepsilon_2$ i)--l)}
    \label{figcontourD5}
\end{figure}
\begin{figure}[ht!]\centering
    \includegraphics[width=\textwidth]{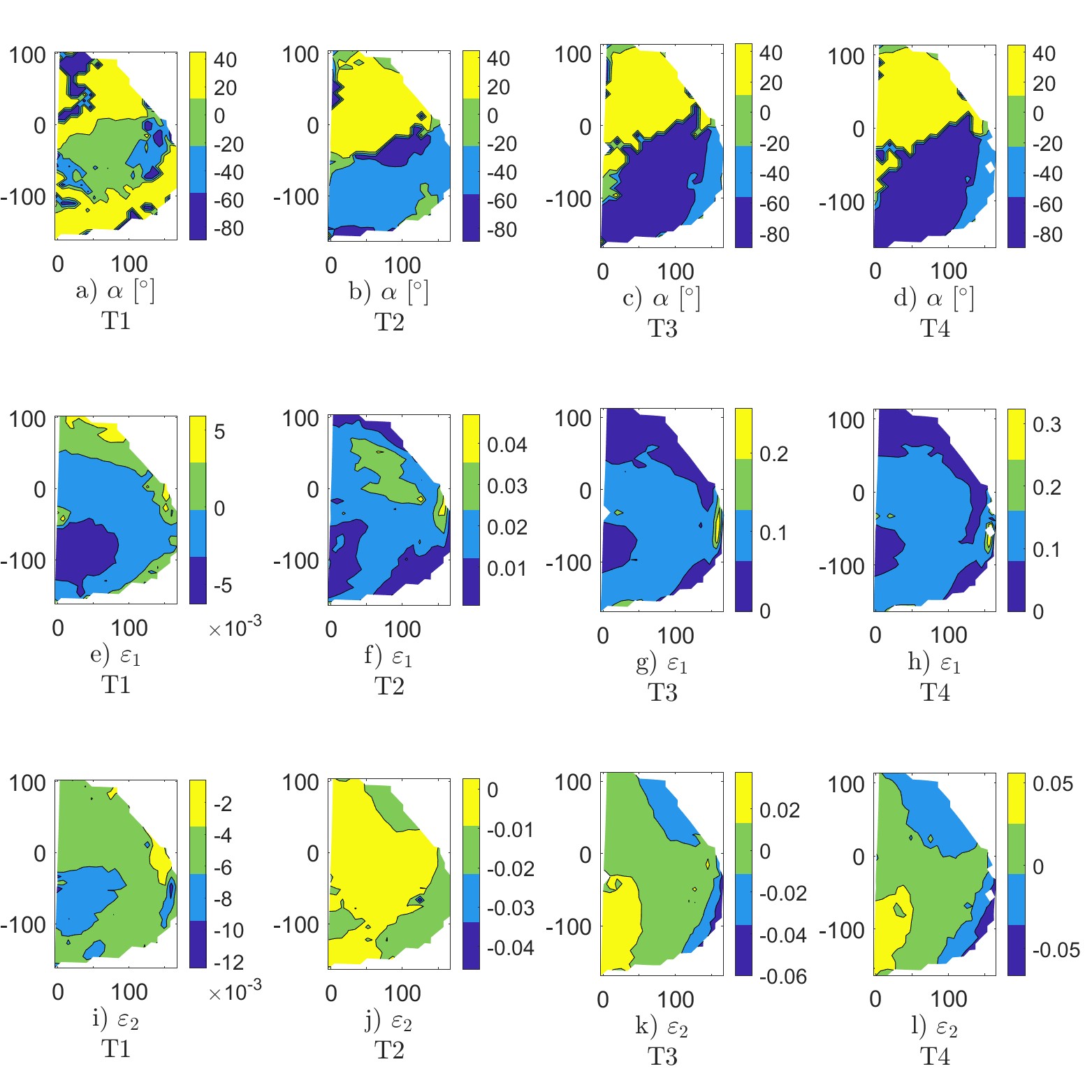}
    \caption{Contour maps with 3 levels of isolines for subject D6: principal direction angle $\alpha$ a)--d), principal strain $\varepsilon_1$ e)--h) and $\varepsilon_2$ i)--l)}
    \label{figcontourD6}
\end{figure}
\begin{figure}[ht!]\centering    \includegraphics[width=\textwidth]{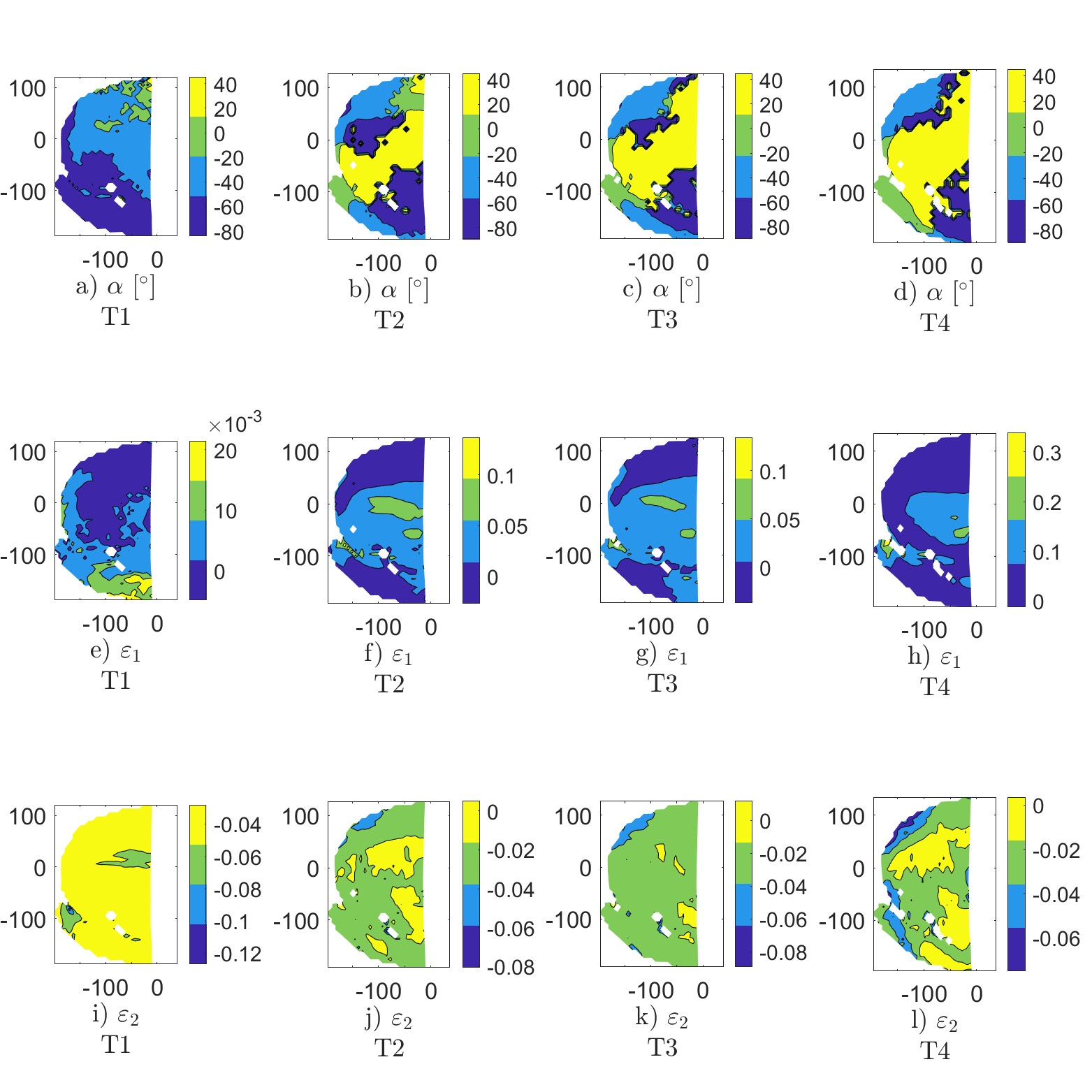}
    \caption{Contour maps with 3 levels of isolines for subject D7: principal direction angle $\alpha$ a)--d), principal strain $\varepsilon_1$ e)--h) and $\varepsilon_2$ i)--l)}
    \label{figcontourD7}
\end{figure}
\begin{figure}[ht!]\centering
    \includegraphics[width=\textwidth]{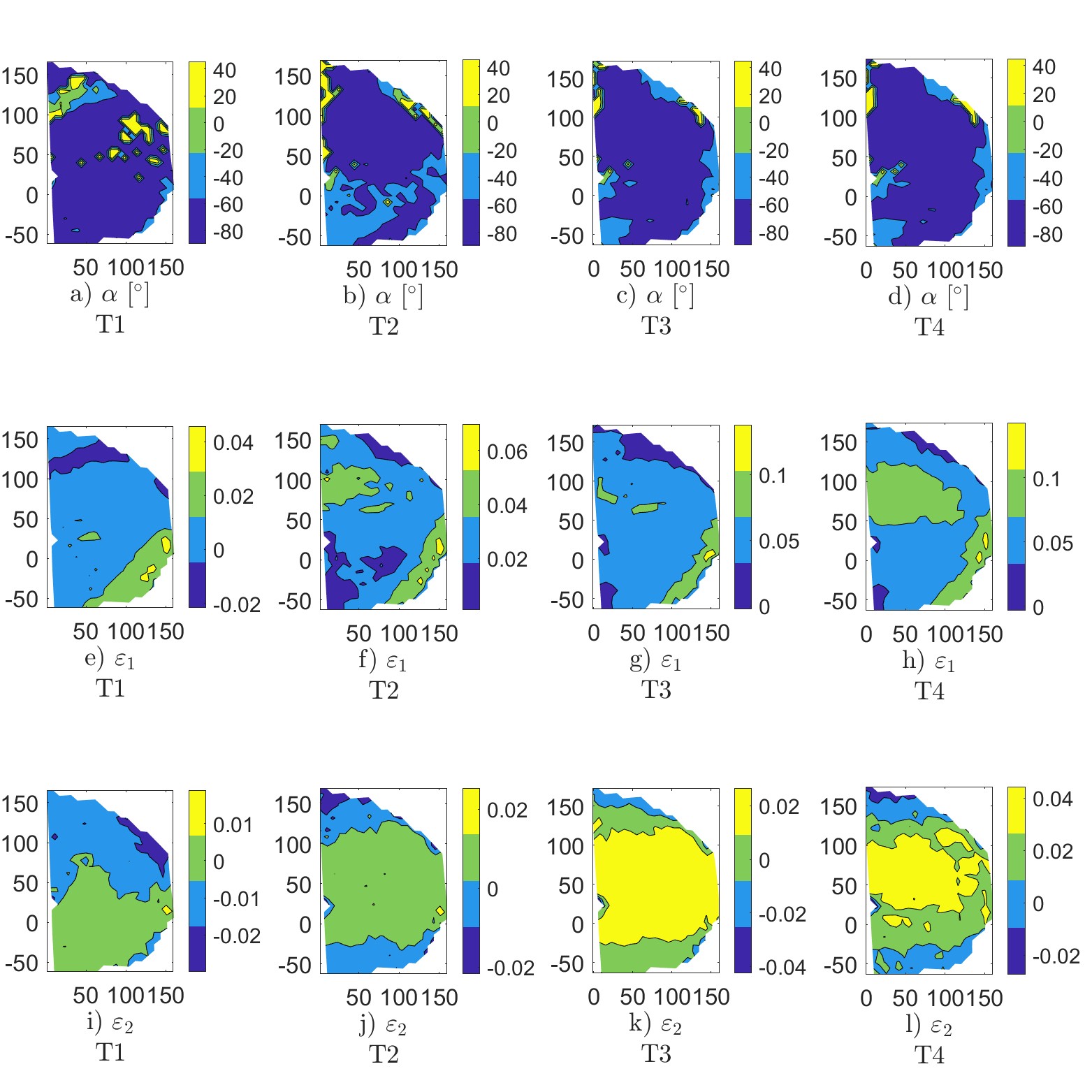}
    \caption{Contour maps with 3 levels of isolines for subject D8: principal direction angle $\alpha$ a)--d), principal strain $\varepsilon_1$ e)--h) and $\varepsilon_2$ i)--l)}
    \label{figcontourD8}
\end{figure}
\begin{figure}[ht!]\centering
    \includegraphics[width=\textwidth]{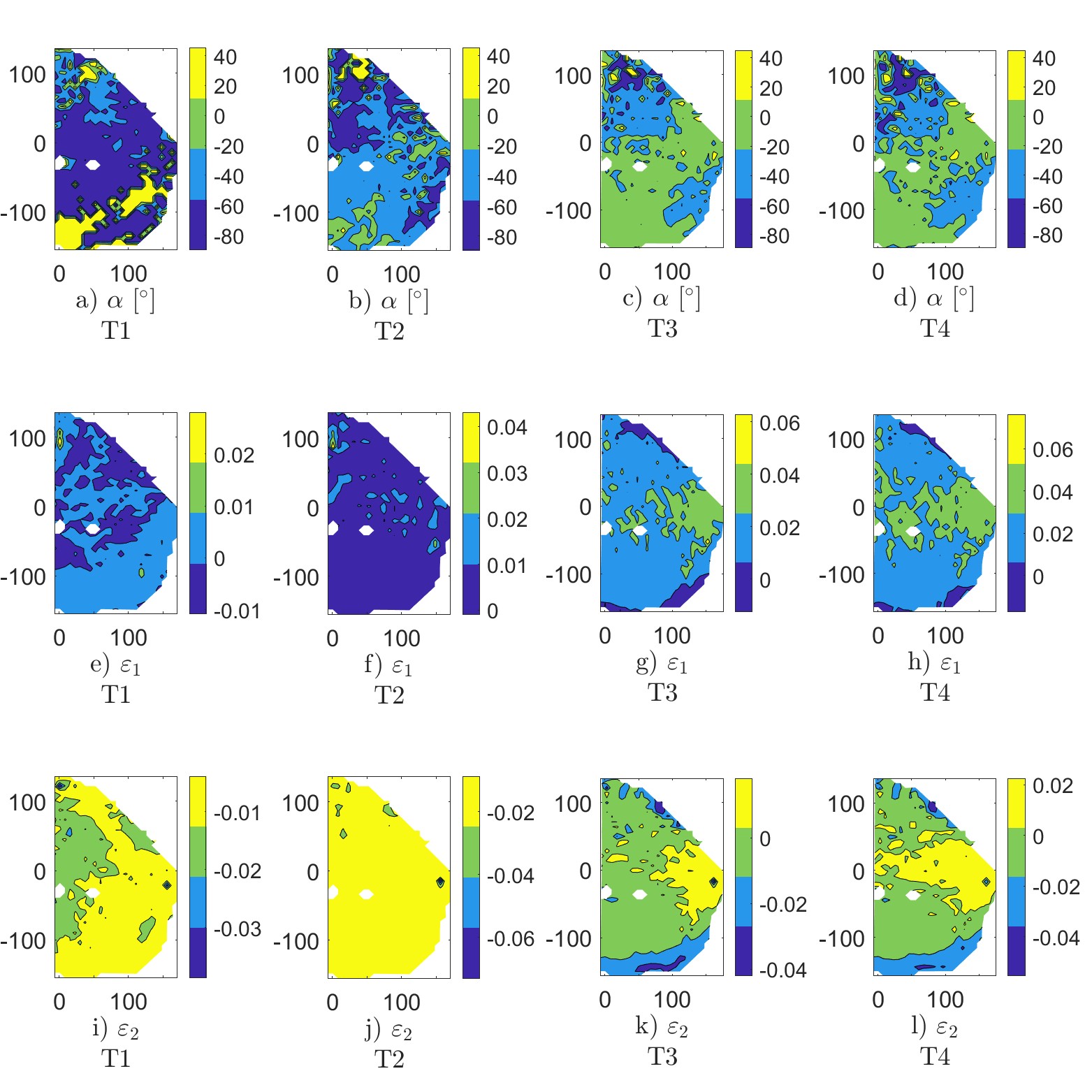}
    \caption{Contour maps with 3 levels of isolines for subject D9: principal direction angle $\alpha$ a)--d), principal strain $\varepsilon_1$ e)--h) and $\varepsilon_2$ i)--l)}
    \label{figcontourD9}
\end{figure}
\begin{figure}[ht!]\centering
    \includegraphics[width=\textwidth]{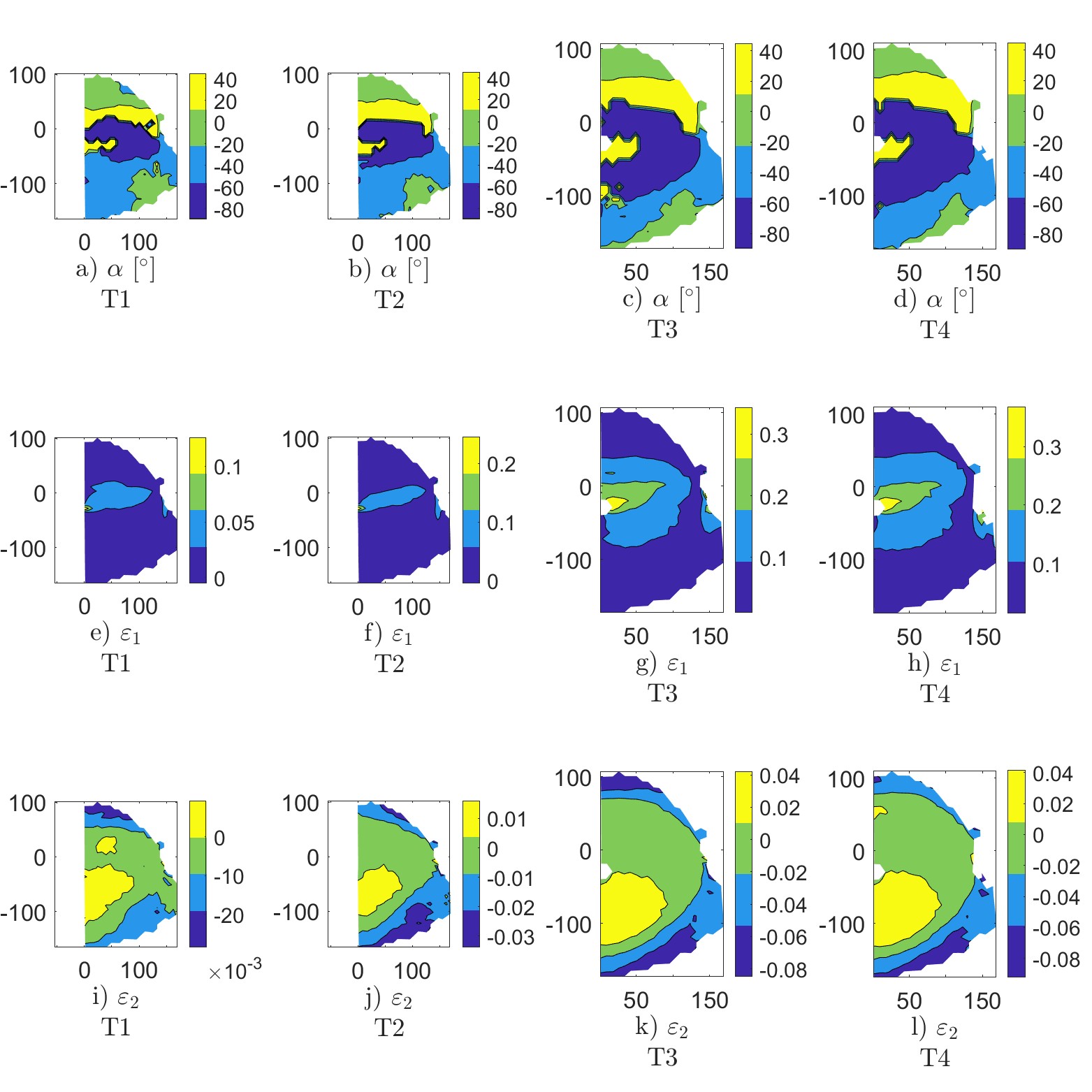}
    \caption{Contour maps with 3 levels of isolines for subject D10: principal direction angle $\alpha$ a)--d), principal strain $\varepsilon_1$ e)--h) and $\varepsilon_2$ i)--l)}
    \label{figcontourD10}
\end{figure}
\begin{figure}[ht!]\centering
    \includegraphics[width=\textwidth]{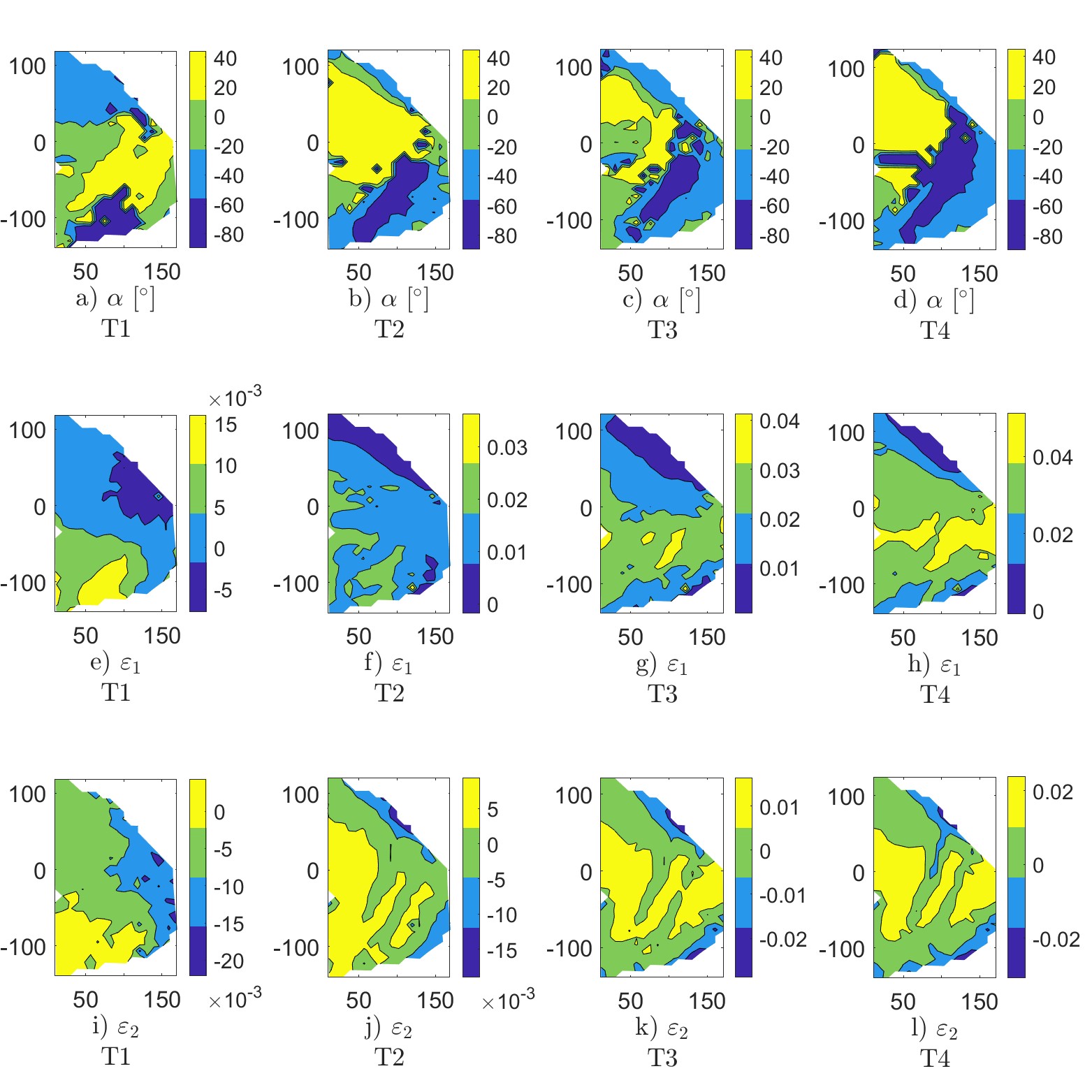}
    \caption{Contour maps with 3 levels of isolines for subject D11: principal direction angle $\alpha$ a)--d), principal strain $\varepsilon_1$ e)--h) and $\varepsilon_2$ i)--l)}
    \label{figcontourD11}
\end{figure}
\begin{figure}[ht!]\centering
    \includegraphics[width=\textwidth]{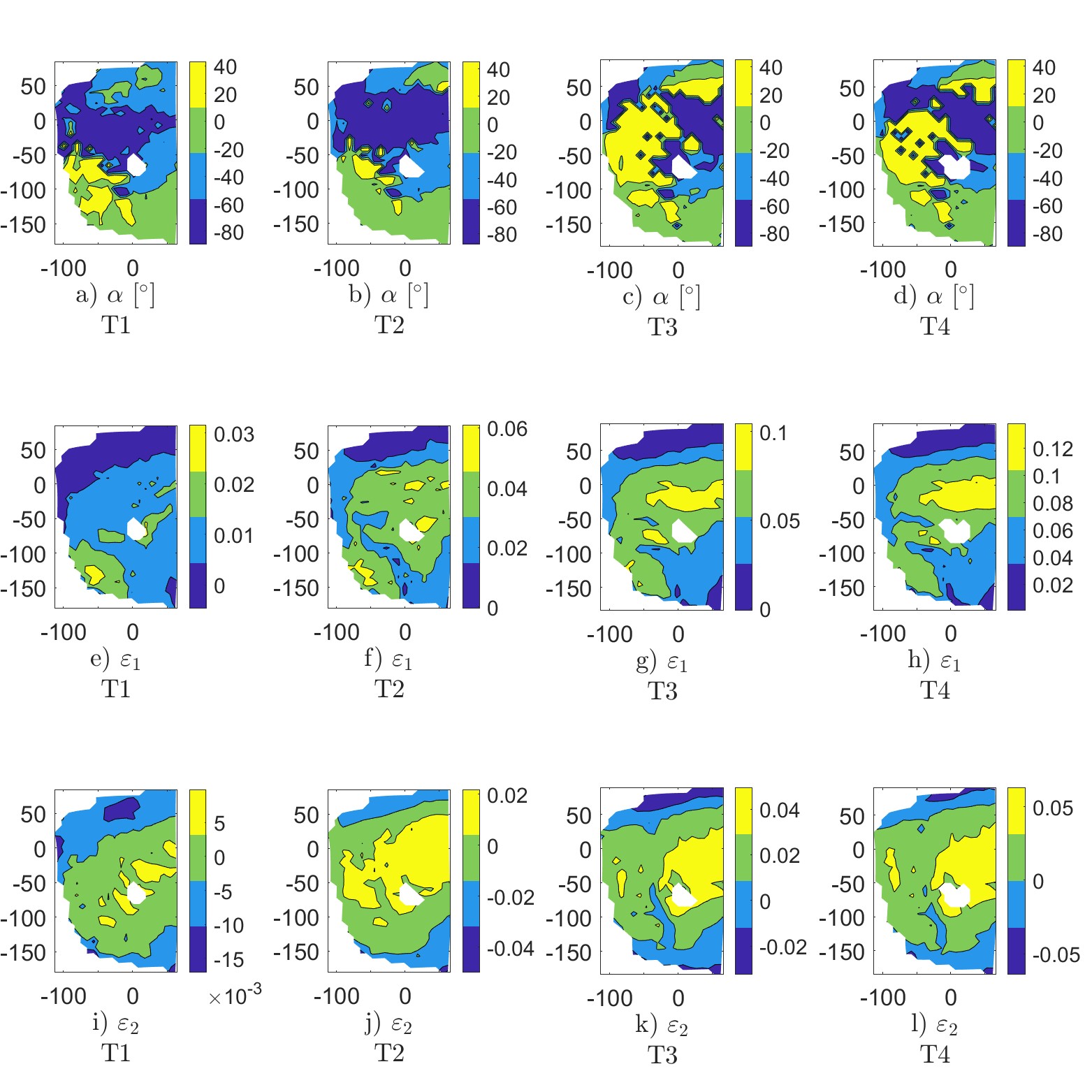}
    \caption{Contour maps with 3 levels of isolines for subject D12: principal direction angle $\alpha$ a)--d), principal strain $\varepsilon_1$ e)--h) and $\varepsilon_2$ i)--l)}
    \label{figcontourD12}
\end{figure}

It may be observed that not only distribution of strain values, but also the distribution of zones with principal directions differed between subjects. Figures \ref{figcontourD1}--\ref{figcontourD12} present isolines  to demostrate what ranges of the mechanical values are typical for various zones. The ranges of the principal Lagrangian strain $\varepsilon_1$, $\varepsilon_2$ values and the direction of their  $\alpha$ angle  on the abdominal walls of each subjects in T1--T4 are shown in the form of  contour maps in Figures \ref{figcontourD1}--\ref{figcontourD12}.  The contour maps contain isolines on the x-y plane of  half of the abdominal wall surface, showing the range of values observed in different areas of the abdominal wall for each subject.  Identifying these ranges can be used when planning abdominal hernia surgery to specific parts of the abdominal wall and selecting surgical implants with appropriate material properties, such as stiffness or anisotropy. In the latter case, the analysis of principal strain directions is also important.


The shapes of the areas separated by the isolines differ depending on the  subject and on the analysed variable. In most subjects (D5, D7, D8, D10--D12), the isolines resemble transversal regions of the abdomen  or in other cases (D1,D4, D6, D7), have a semicircular shape around the central area of the region of interest. Zones parallel to cranio-caudal axis are observed in D3.

 The maximum principal strains on the abdominal surface in subject D1 have a semicircular shape, where higher values of strains are closer to the centre of the area of interest. However, this observation refers only to T2--T4. T1 is more homogeneous. The opposite is observed for the direction of principal strains, which becomes more homogeneous with the increase of fluid pressure. 
 
Zones with specific principal strain  directions often change in size, depending on the pressure. In particular the central zone usually expands. In terms of the principal direction, D6 is divided approximately into two halves - the lower and upper part of the abdomen. The changing direction of principal strains observed on the  abdominal wall surface may indicate that different components of the abdominal wall play different roles during the various stages.

\subsection{Strains along profile lines}

\begin{figure}[ht!]\centering
    \includegraphics[width=1\textwidth]{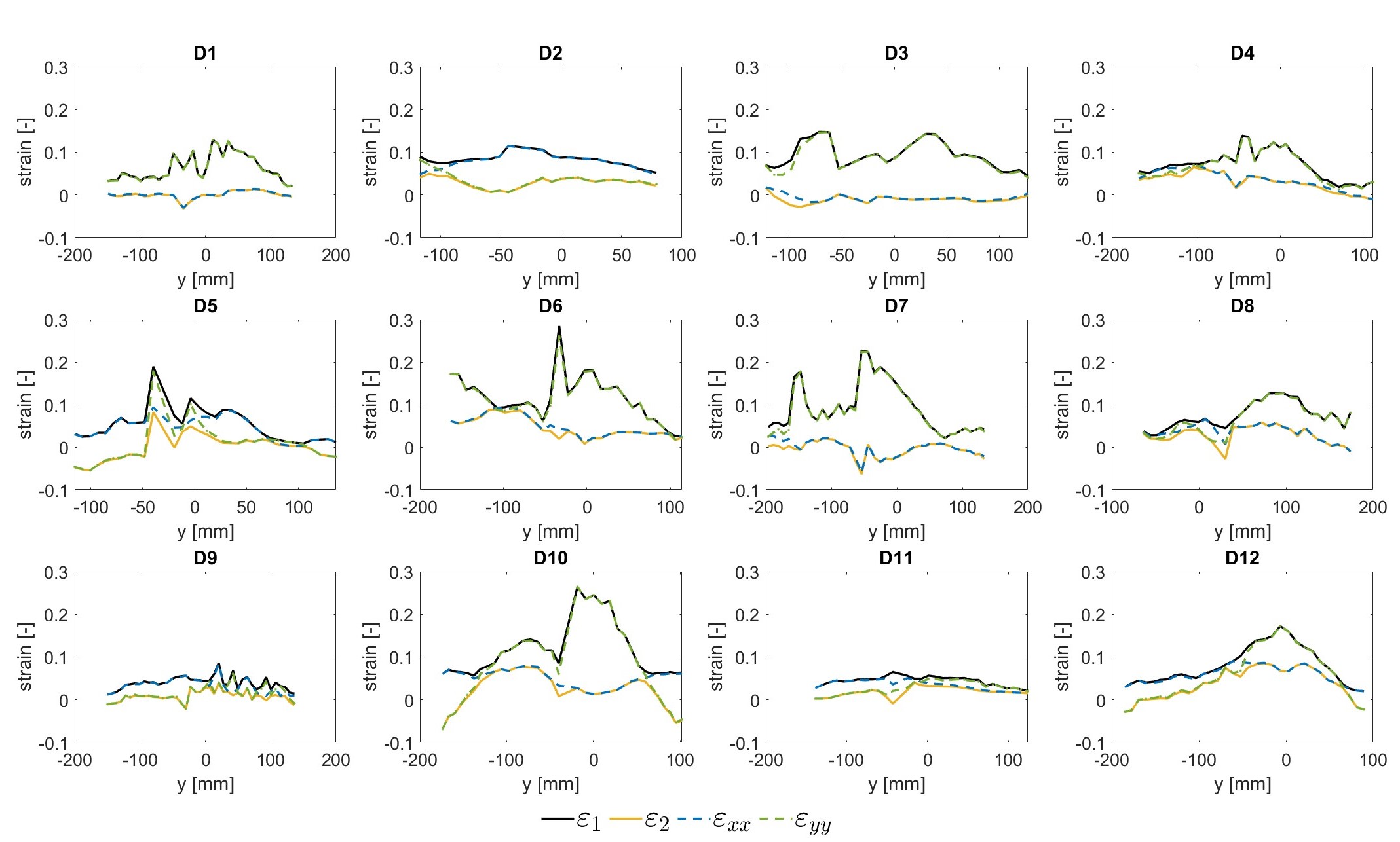}
    \caption{ Profile lines of strains  $\varepsilon_1$, $\varepsilon_2$, $\varepsilon_{xx}$ and $\varepsilon_{yy}$  along the mid-line line (line A--A) at  T4 . }
    \label{fig_profile_LA}
\end{figure}

\begin{figure}[ht!]\centering
    \includegraphics[width=1\textwidth]{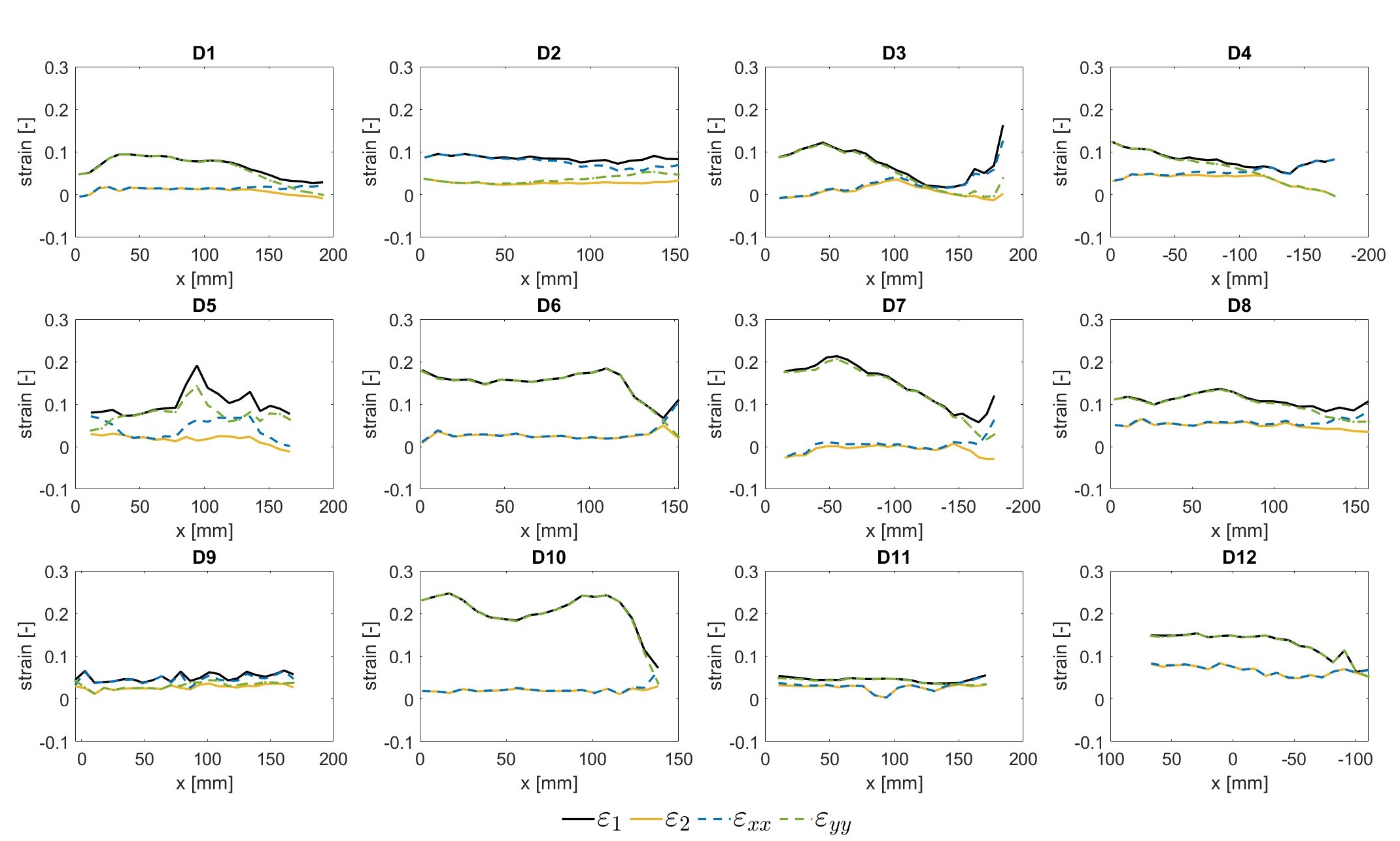}
    \caption{ Profile lines of strains  $\varepsilon_1$, $\varepsilon_2$, $\varepsilon_{xx}$ and $\varepsilon_{yy}$  along the transverse (line B--B) around 3 cm above umbilical in  T4 (the x-axes of D4, D7 and D12 are flipped due to the catheter).}
    \label{fig_profile_TA}
\end{figure}

Table \ref{Table_dic_result_lines} shows mean values of strain $\varepsilon_{xx}$ and $\varepsilon_{yy}$ obtained from grid points located in the section lines A--A and B--B (see Figure \ref{fig_dic2dis}c) of each subject. Section A-A runs along the mid-line (y-direction) and section B-B runs in a transverse direction (x-direction), around 3 cm above the umbilicus. The profile lines of the strains along these sections together with principal strain $\varepsilon_{1}$ and $\varepsilon_{2}$ are presented in Figures \ref{fig_profile_LA}--\ref{fig_profile_TA}. Here, the deformations in the individual sections of the abdominal wall can be observed. Although to clearly distinguish the anatomical zones would require additional medical imaging to visualise muscle structures in each patient, which was not covered by the standard peritoneal dialysis procedure.

The relations of $\varepsilon_{xx}$ to $\varepsilon_{yy}$  and principal strains vary between subjects and sometimes vary along a single profile. Coordinate system of each subject is depicted under strain maps in \ref{appendix_strain}. It may be observed that $\varepsilon_{yy}$ is higher than $\varepsilon_{xx}$   along a longer section of the mid-line in the case of the majority of subjects.  However, $\varepsilon_{xx}$ is higher for mid-lines of D2, D9 and D11,  who were the youngest subjects with the most muscular abdominal walls. Mean $\varepsilon_{xx}$ along the mid-line is higher in those subjects and also in the case of D5 and D12. All of the above had higher ultrafiltration (UF) than the remaining subjects. From the clinical point of view, the high UF (0.7-1L) means higher fluid pressure on the abdominal wall and may pose a risk of hernias. Predominantly longitudinal rather than  transverse higher strains  were  found in the \textit{ex vivo} study by \cite{podwojewski2014mechanical}, though in some individual cases, the opposite was observed. The predominance of the longitudinal direction, with some exceptions, is also  observed in our study. 

\begin{figure}[ht!]\centering
    \includegraphics[width=1\textwidth]{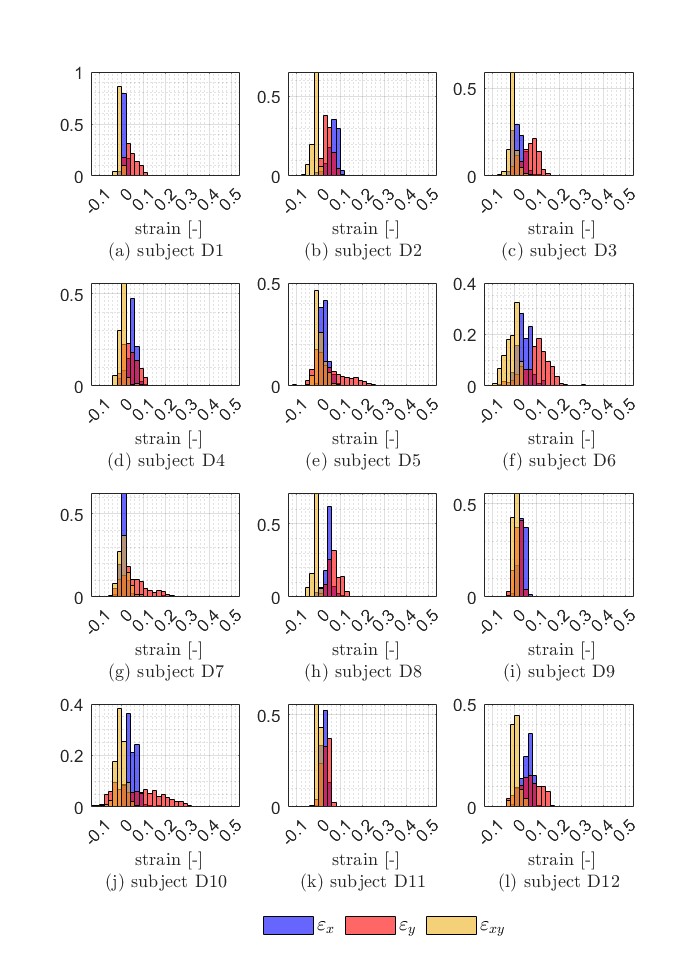}
    \caption{ Histograms of strains $\varepsilon_{xx}$,  $\varepsilon_{yy}$ and $\varepsilon_{xy}$. at  T4 .}
    \label{fig_hist_xy}
\end{figure}

The information of the average value of the deformations in both directions may be useful in case of considering surgical mesh implantation covering a large part of abdomen as well as may be important for its cognitive value in terms of the direction of maximal deformation of human abdominal wall. It can interestingly supply the discussion about in which direction (longitudinal or transversal) of the human torso, the  abdominal wall deforms more. Here x and y reflect here the both directions. It is worth of noting that the values of $\varepsilon_{xy}$ observed in the study are low comparing to other components of the local strain tensor, as one can see at Figures \ref{fig_strainxy_dic1} - \ref{fig_strainxy_dic12}. Also, the histogram at Figure \ref{fig_hist_xy} indicates their lower value when compared to $\varepsilon_{xx}$ and $\varepsilon_{yy}$. This supports the hypothesis that the abdominal wall can be modelled as a membrane structure.

\begin{table}[ht]
\begin{tabular}{ccccccc}
\hline
 & &   \multicolumn{2}{c}{transverse line }   & & \multicolumn{2}{c}{midline }    \\\cline{3-4}\cline{6-7}
subject &  &$\varepsilon_{xx}$&$\varepsilon_{yy}$ & & $\varepsilon_{xx}$& $\varepsilon_{yy}$ \\
\hline
D1&  & 0.014 & 0.061   &  & 0.001 & 0.062 \\
D2&  & 0.077 & 0.037   &  &  0.077 &  0.035\\
D3&  &  0.026 & 0.054    &  & -0.006 & 0.090 \\
D4&  &   0.055 &  0.061   &  & 0.035 &  0.064\\
D5&  &  0.042 & 0.077   &  &  0.043 & 0.007 \\
D6&  &  0.031 &  0.143  &  &  0.049 & 0.111\\
D7&  & 0.005 &  0.129   &  &  -0.001 &0.086 \\
D8&  &0.057  &  0.101  &  &0.036 &  0.067\\
D9&  & 0.046 & 0.032   &  &  0.033 & 0.014\\
D10&  & 0.023 &  0.201   &  & 0.052 &  0.083\\
D11&  &  0.030  &  0.042  &  & 0.035 &  0.028\\
D12&  & 0.068 &   0.128 &  &0.058 & 0.056 \\
\hline
\end{tabular}
\caption{Mean strain [-] results in x and y direction along  A--A and B--B at T4 } \label{Table_dic_result_lines}
\end{table}

\subsection{Strains in time}

\begin{figure}[ht!]\centering
    \includegraphics[width=1\textwidth]{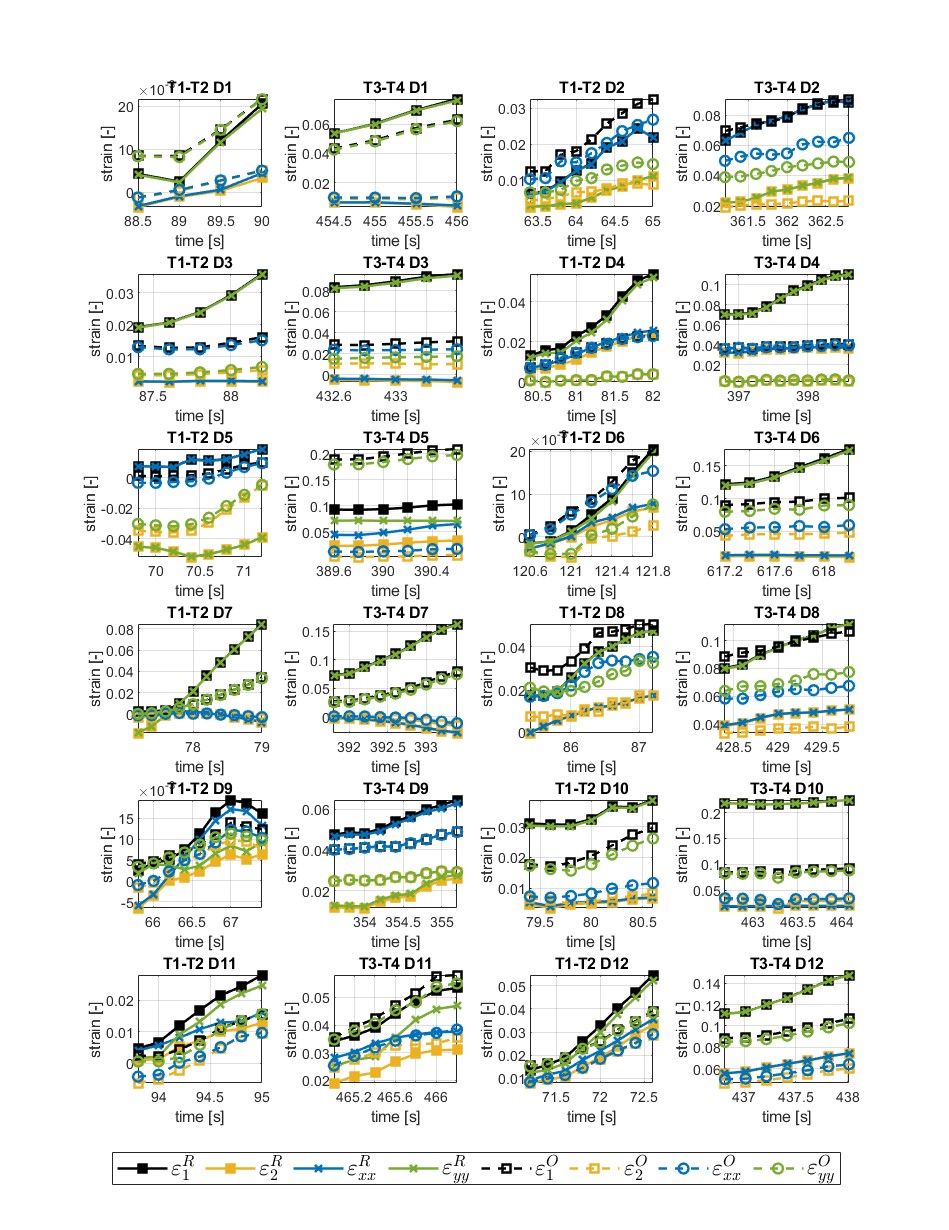}
    \caption{ Strain changes in time between  T1 and T2 and between T3 and T4, where $\varepsilon_1^R$, $\varepsilon_2^R$, $\varepsilon_{xx}^R$, $\varepsilon_{yy}^R$ denote  strains in the grid point around 3 cm above the umbilicus  and  $\varepsilon_1^O$, $\varepsilon_2^O$, $\varepsilon_{xx}^O$, $\varepsilon_{yy}^O$ denotes strain in the lateral part of the abdominal wall approximate to the oblique muscles for subjects D1--D12.}
    \label{fig_time}
\end{figure}

Figure \ref{fig_time} shows strain changes in time between T1--T2 and T3--T4, which both correspond to one breath of air. It may be noticed that in the majority of cases, strains $\varepsilon_1$, $\varepsilon_2$, $\varepsilon_{xx}$, $\varepsilon_{yy}$  increase  in the considered points above the umbilicus (R) and  in the lateral part level with the umbilicus(O) during  inhalation. The exception is subject D7, whose  $\varepsilon_{xx}$ and $\varepsilon_{2}$ decrease. However, the mechanical response of this subject may have been  influenced by strong inter-peritoneal adhesions. This was also the only subject in this study who had a negative median of $\varepsilon_2$  (Table \ref{Table_dic_result}). Similarly,  a single subject had a negative median of $\varepsilon_2$ in our previous  study \cite{in_vivo_abdomen}.

In the  majority of cases, principal strain $\varepsilon_1$ is higher in the mid-line point R than in the lateral point O. In the case of some subjects, this relation changes between stages. In the cases of D5 and D11, the strain in the lateral point O becomes higher in the T3--T4 phase. Conversely, in the cases of  D1, D2, D5, D6, strain  in point O is higher in the early stage T1-T2  . \cite{jourdan2022dynamic} using dynamic-MRI, also observed  that circumferential strains in lateral muscles  were higher than those of the rectus muscle during breathing. Although, our study refers to the external surface of abdominal wall, strains  were in the similar range. The study we present here shows that this may change when the abdominal wall undergoes the higher intra-abdominal pressure of dialysis fluid. 
The relation in the time between $\varepsilon_{xx}$ and $\varepsilon_{yy}$ differs depending on the subject. A possible explanation for this could be the differences in the  contribution of active muscles  during breathing and the passive response of the abdominal wall during the introduction of intra-abdominal pressure. 
Relating the obtained results to individual abdominal zones (\cite{jourdan2022dynamic}) would be beneficial for validating future numerical models of abdominal wall. However, this would require additional diagnostic imaging to visualise muscle structures in each patient, which was not covered by the standard peritoneal dialysis procedure.

\subsection{Discussion}

The high variability of strains may be observed  in the literature. A comparison of histograms obtained in this study (Figure \ref{fig_histograms} with those of  the previous  study on this subject \citep{in_vivo_abdomen}, reveals similar median values. However, the range of maximum values of $\varepsilon_1$ is higher in the current study, which can be explained by measurement on the inhalation phase and the higher resolution allowing for the observation of more local behaviour.  

 The abdominal wall is composed of various muscles and connective tissues with different fibre orientation. \cite{astruc2018characterization} showed that linea alba and rectus sheath are stiffer in transverse direction than the longitudinal one. An ex vivo study showed the rectus sheath to contribute significantly to passive response of the abdominal wall in \citep{tran2014contribution}. \textit{Ex vivo} studies of passive abdominal wall behaviour under pressure indicated the first principal direction to be along the cranio-caudal axis \citep{le2020differences}. \cite{szymczak2012investigation} obtained similar principal directions in the case of an \textit{in vivo} study of the body bending to one side. The transverse direction of abdominal wall was shown to exhibit lower strains in \citep{szymczak2012investigation,deeken2017mechanical}.  The current study reveals the  variability of principal directions among subjects. Although as in the aforementioned studies, our study reveals generally greater strains in the longitudinal direction, this was not the case with all the studied subjects. These variations in  principal directions may be explained by heterogeneity and variability of the mechanical properties and geometries of individual abdominal walls as well as differences in the active contribution of muscles, which requires further investigation. \cite{pavan2019effects} has already shown in a numerical investigation the importance of including the muscle activity in the mechanical response of the abdominal wall to pressure.

 Variability of the outcomes  indicates the need for a patient-specific approach to the hernia treatment. Variability in principal directions implies that the orientation of applied anisotropic surgical meshes should be personalised. Alignment of orthotropic surgical mesh was shown to be important for the sake of junction forces minimisation \citep{lubowiecka2016preliminary}. High variability can be also addressed in the simulation of abdominal wall and surgical meshes by means of  uncertainty quantification methods \citep{szepietowska2018sensitivity}.

Due to the under-representation of women compared to men, it is hard to draw specific conclusions regarding sex.

This study has several limitations. Firstly, optical measurements were performed only on the  skin  of the abdominal wall. In the case of hernia implants, the knowledge of the strain field in the interior parts of the abdominal wall may be more useful. \cite{podwojewski2014mechanical} showed via an \textit{ex vivo} study that strains on the external surface are statistically twice as high as on the internal surface. Nonetheless, the strain pattern is different.   
Secondly, exhalation and inhalation stages are assessed by the evaluation of  abdominal wall surface deformation. A more detailed approach could be achieved by additional measurements of breathing phases \citep{mikolajowski2022automated}. It should also be noted that the subjects had been undergoing regular peritoneal dialysis for some time. Only in the case of D11, was  the subject being subjected to the very first PD. Other subjects had undergone regular PD from two months to two years prior the measurements, which may have influence on their abdominal wall response.

The results have not been related to the subjects' characteristics since this is a cross-sectional study referring to mechanical behaviour of the abdominal wall under fluid pressure in a heterogenic group (not clustered by age, sex etc.) of subjects. This fact makes difficult to draw more general conclusions due to the small size of a group. However, the research is focused on assessing the variability of the mechanical response among the patients under peritoneal dialysis that obtain the same amount of dialysis fluid.   

\section{Conclusions}

Presented here has been the deformation of the human living abdominal wall subjected to intra-abdominal  dialysis fluid pressure whilst breathing.
The study concerns \textit{in vivo} tests on human subjects and shows the changes in the strain field due to loading.  The measurements performed during peritoneal dialysis  gives the possibility of linking the deformation with intra-abdominal pressure values.   The shown strain fields are not homogeneous and exhibit high variability between subjects, both in terms of strain values and principal directions during inhalation and exhalation.

The intention of this study is to advance a better understanding of living human abdominal wall mechanics. 
 The knowledge of  abdominal wall mechanics based on \textit{in vivo} experiments can support the optimisation of surgical strategies for the proper selection and design of the implants used in hernia repairs. The data reported here can be further used to identify the mechanical properties of the human abdominal wall as well as validate numerical models.
 In addition, it may indicate the character of suggested exercises in order to improve mechanical properties of the abdominal wall and reduce the risk of new and recurrent hernia formation.

The high variability of the results suggests the need for a patient specific-approach to hernia repair and other issues concerning  abdominal wall mechanics (e.g. closure). Another route of inquiry that needs to be further considered is uncertainty quantification to include this variability in  simulations. What is more, there is a need to further investigate the active behaviour of the abdominal wall. Although, the surgical mesh reinforce the abdominal wall only in a passive way, the active behaviour of the surrounding muscles may influence the physiological conditions under which the implant functions. 

\section*{Acknowledgements}

We would like to thank the staff of Peritoneal Dialysis Unit Department of Nephrology Transplantology and Internal Medicine Medical University of Gda\'nsk and Fresenius Nephrocare (dr Piotr Jagodzi\'nski, nurses Ms Gra\.zyna Szyszka and Ms Ewa Malek) for their help in accessing the patients, performing PD exchanges and measurements of the IPP.

This work was supported by the National Science Centre (Poland) [grant No. UMO-2017/27/B/ST8/02518]. Calculations were carried out partially at the Academic Computer Centre in Gdansk.

\appendix

\section{Displacement results} \label{appendix_d}

\begin{figure}[ht!]\centering
    \includegraphics[width=\textwidth]{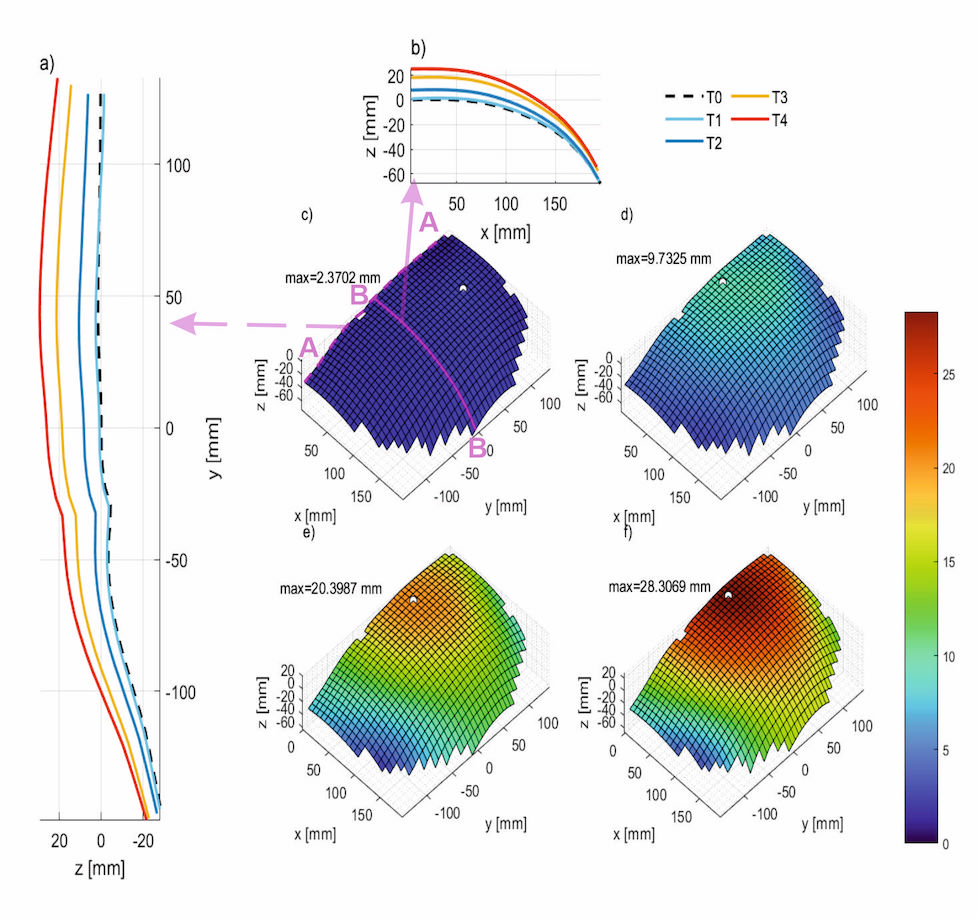}
   \caption{Shape and displacement of subject D1 in four stages T1--T4 and  reference one T0: a) profile of abdominal wall along mid-line A--A; b)  profile along transverse direction B--B; c--f) surfaces of abdominal wall with colour indicating total displacement [mm] in T1--T4, respectively with marked location of maximum displacement by a white circle ; x is the mediolateral axis from right to left, y is craniocaudal axis from caudal to cranial, and z is anteriorposterior axis from anterior to posterior }
    \label{fig_dic1dis}
\end{figure}

\begin{figure}[ht!]\centering
    \includegraphics[width=\textwidth]{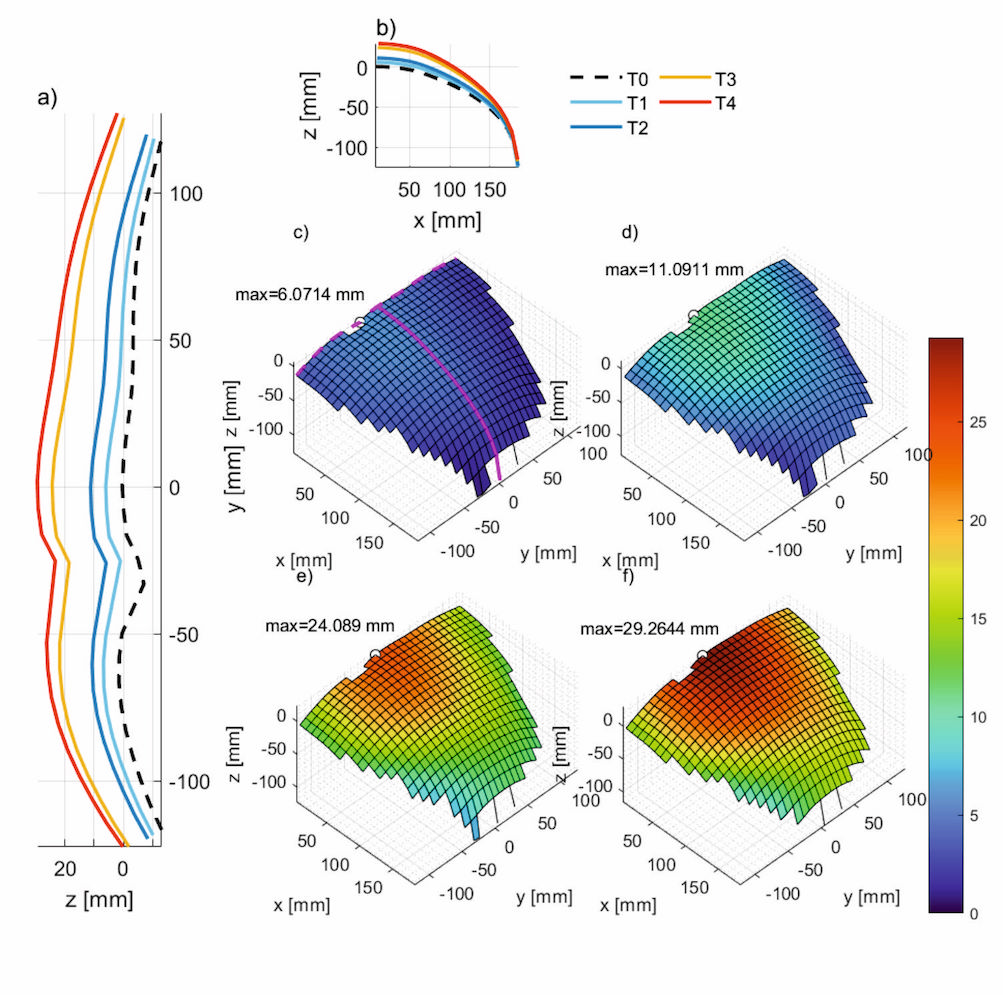}
   \caption{Shape and displacement of subject D3 in four stages T1--T4 and  reference one T0: a) profile of abdominal wall along mid-line A--A; b)  profile along transverse direction B--B; c--f) surfaces of abdominal wall with colour indicating total displacement [mm] in T1--T4, respectively with marked location of maximum displacement by a white circle; x is the mediolateral axis from right to left, y is craniocaudal axis from caudal to cranial, and z is anteriorposterior axis from anterior to posterior }
    \label{fig_dic3dis}
\end{figure}

\begin{figure}[ht!]\centering
    \includegraphics[width=\textwidth]{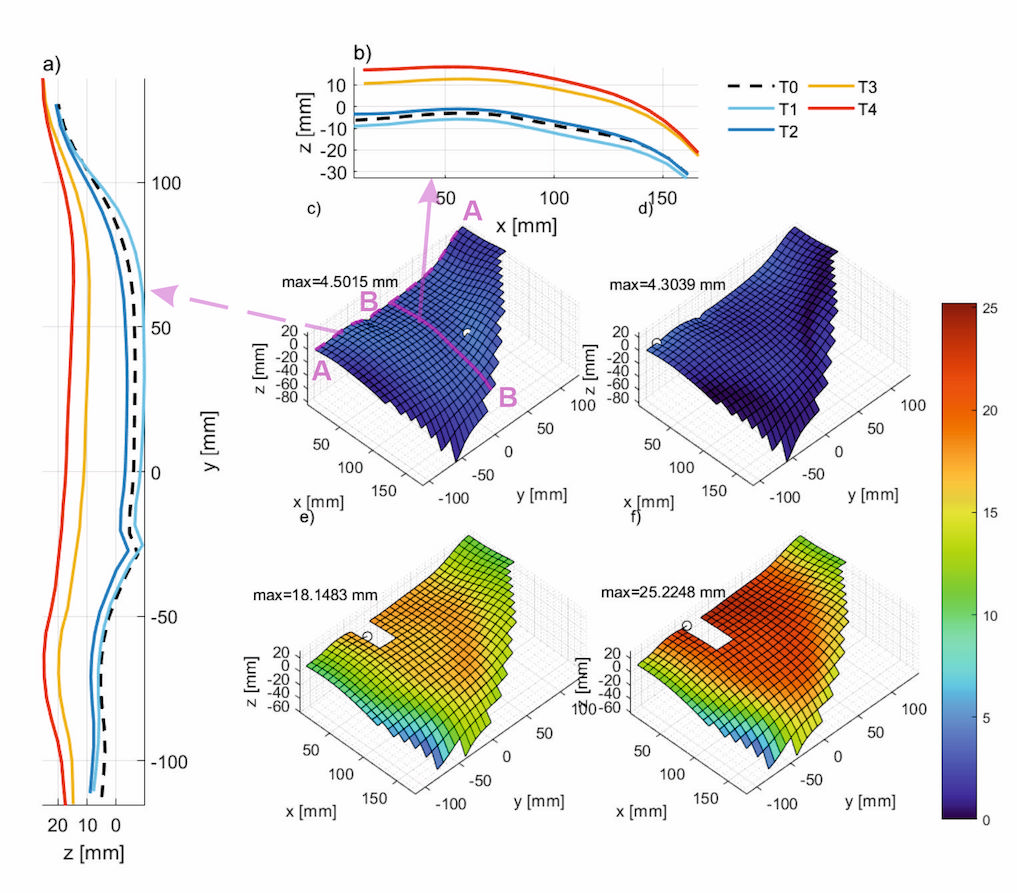}
   \caption{Shape and displacement of subject D5 in four stages T1--T4 and  reference one T0: a) profile of abdominal wall along mid-line A--A; b)  profile along transverse direction B--B; c--f) surfaces of abdominal wall with colour indicating total displacement [mm] in T1--T4, respectively with marked location of maximum displacement by a white circle; x is the mediolateral axis from right to left, y is craniocaudal axis from caudal to cranial, and z is anteriorposterior axis from anterior to posterior }
    \label{fig_dic5dis}
\end{figure}

\begin{figure}[ht!]\centering
    \includegraphics[width=\textwidth]{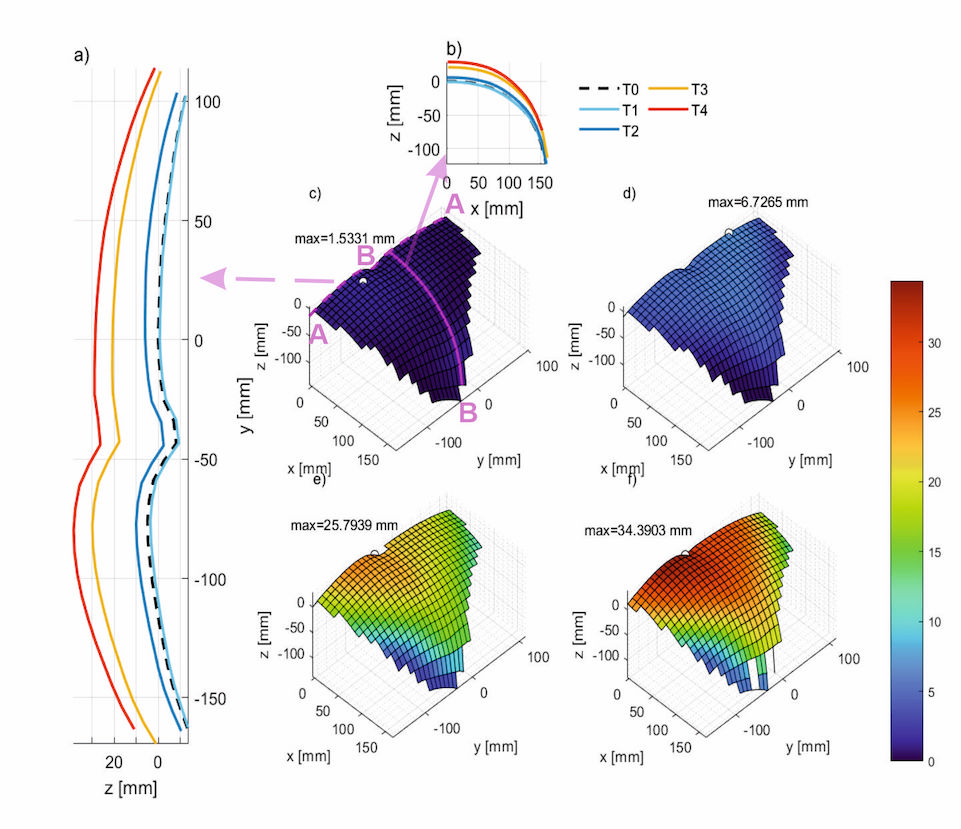}
   \caption{Shape and displacement of subject D6 in four stages T1--T4 and  reference one T0: a) profile of abdominal wall along mid-line A--A; b)  profile along transverse direction B--B; c--f) surfaces of abdominal wall with colour indicating total displacement [mm] in T1--T4, respectively with marked location of maximum displacement by a white circle; x is the mediolateral axis from right to left, y is craniocaudal axis from caudal to cranial, and z is anteriorposterior axis from anterior to posterior }
    \label{fig_dic6dis}
\end{figure}

\begin{figure}[ht!]\centering
    \includegraphics[width=\textwidth]{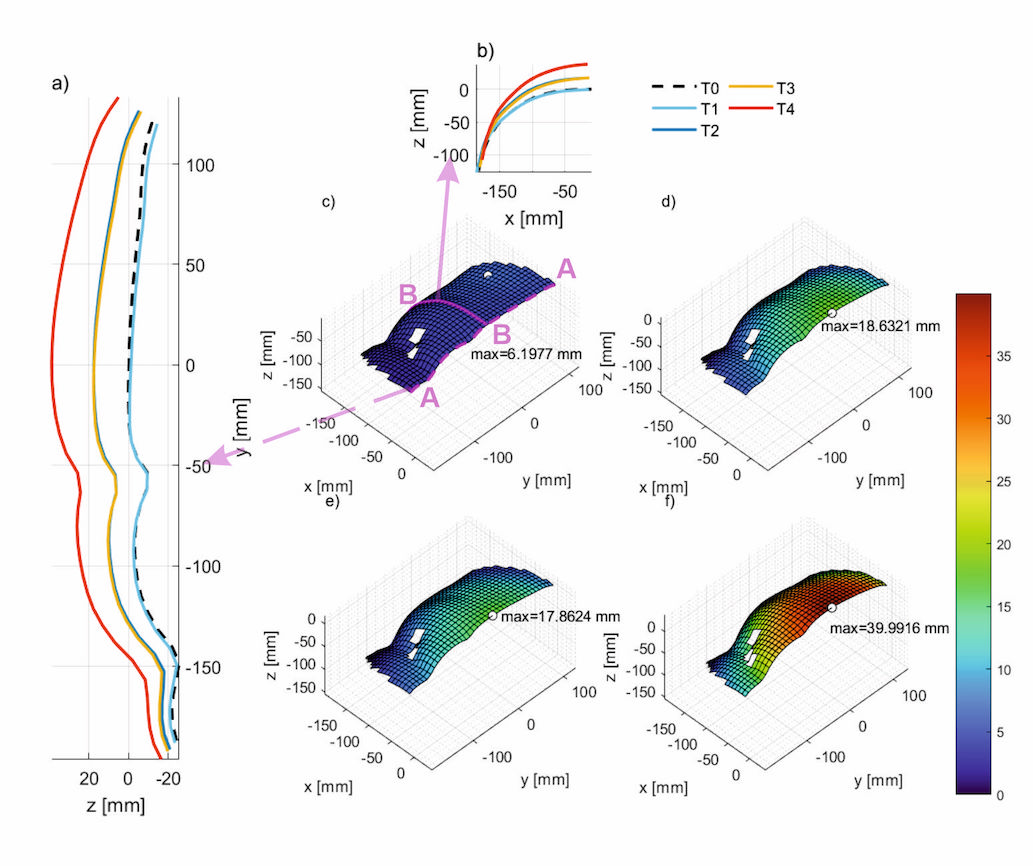}
   \caption{Shape and displacement of subject D7 in four stages T1--T4 and  reference one T0: a) profile of abdominal wall along mid-line A--A; b)  profile along transverse direction B--B; c--f) surfaces of abdominal wall with colour indicating total displacement [mm] in T1--T4, respectively with marked location of maximum displacement by a white circle; x is the mediolateral axis from right to left, y is craniocaudal axis from caudal to cranial, and z is anteriorposterior axis from anterior to posterior }
    \label{fig_dic7dis}
\end{figure}

\begin{figure}[ht!]\centering
    \includegraphics[width=\textwidth]{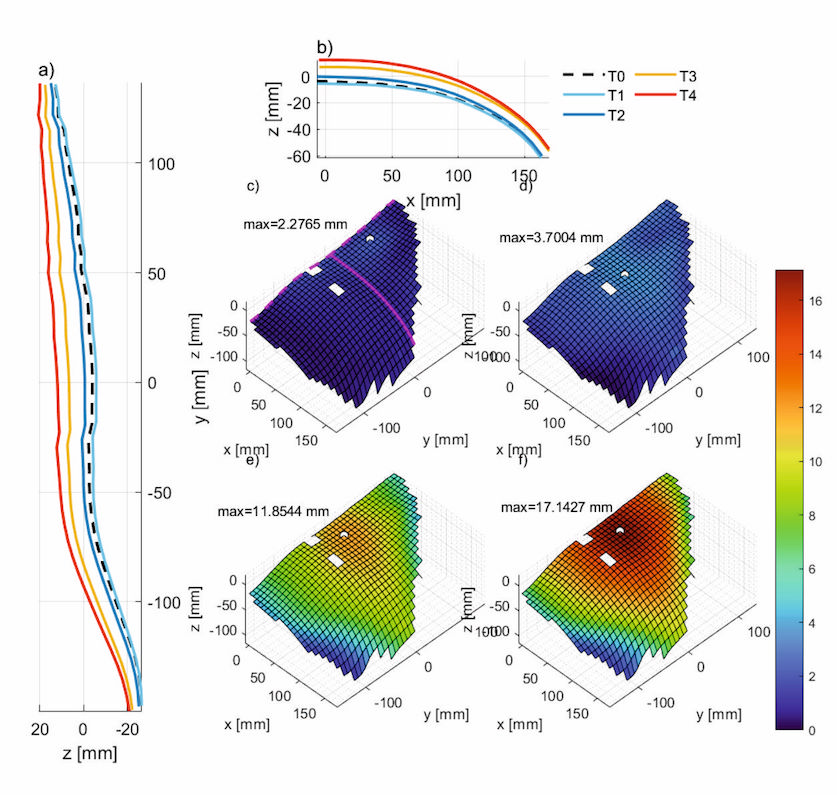}
   \caption{Shape and displacement of subject D9 in four stages T1--T4 and  reference one T0: a) profile of abdominal wall along mid-line A--A; b)  profile along transverse direction B--B; c--f) surfaces of abdominal wall with colour indicating total displacement [mm] in T1--T4, respectively with marked location of maximum displacement by a white circle; x is the mediolateral axis from right to left, y is craniocaudal axis from caudal to cranial, and z is anteriorposterior axis from anterior to posterior }
    \label{fig_dic9dis}
\end{figure}

\begin{figure}[ht!]\centering
    \includegraphics[width=\textwidth]{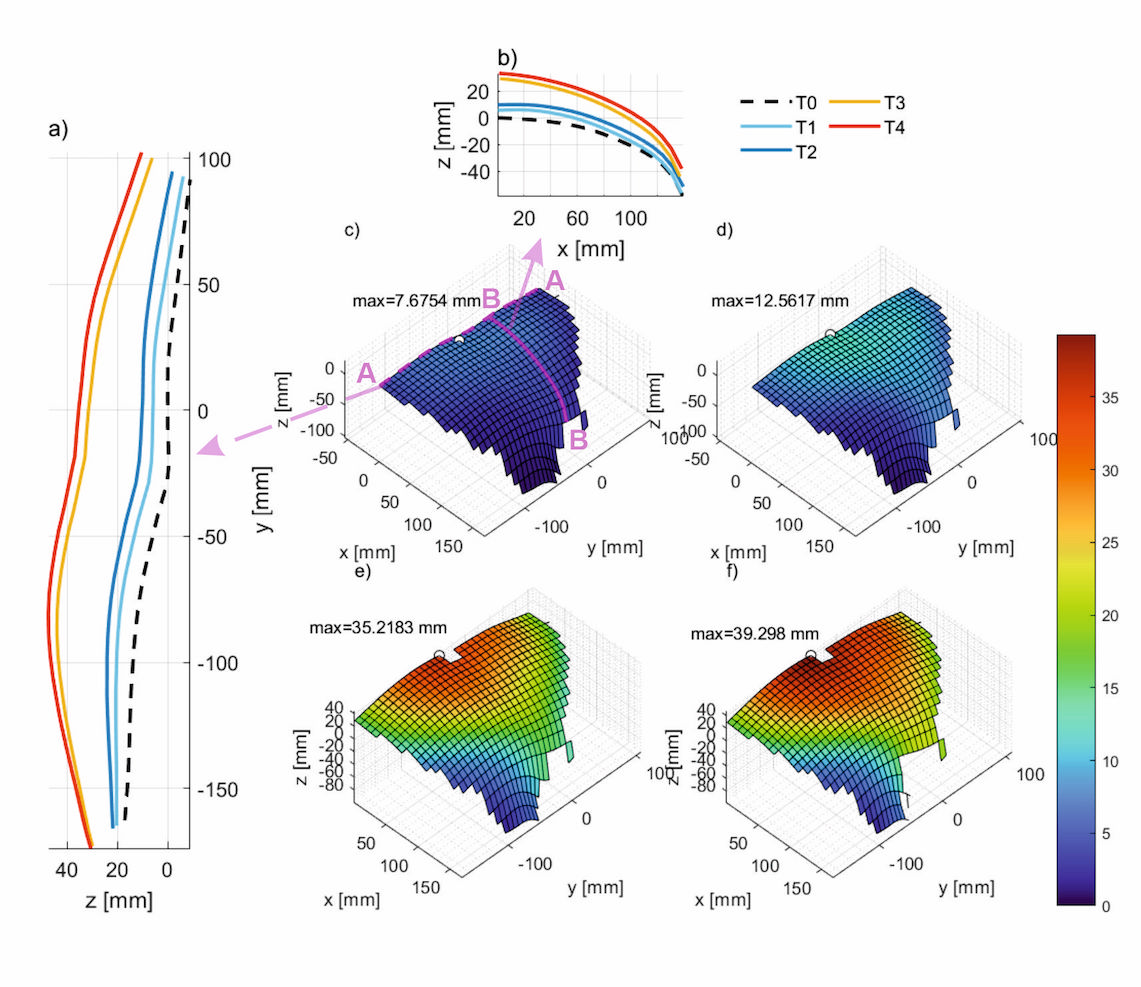}
   \caption{Shape and displacement of subject D10 in four stages T1--T4 and  reference one T0: a) profile of abdominal wall along mid-line A--A; b)  profile along transverse direction B--B; c--f) surfaces of abdominal wall with colour indicating total displacement [mm] in T1--T4, respectively with marked location of maximum displacement by a white circle; x is the mediolateral axis from right to left, y is craniocaudal axis from caudal to cranial, and z is anteriorposterior axis from anterior to posterior }
    \label{fig_dic10dis}
\end{figure}

\begin{figure}[ht!]\centering
    \includegraphics[width=\textwidth]{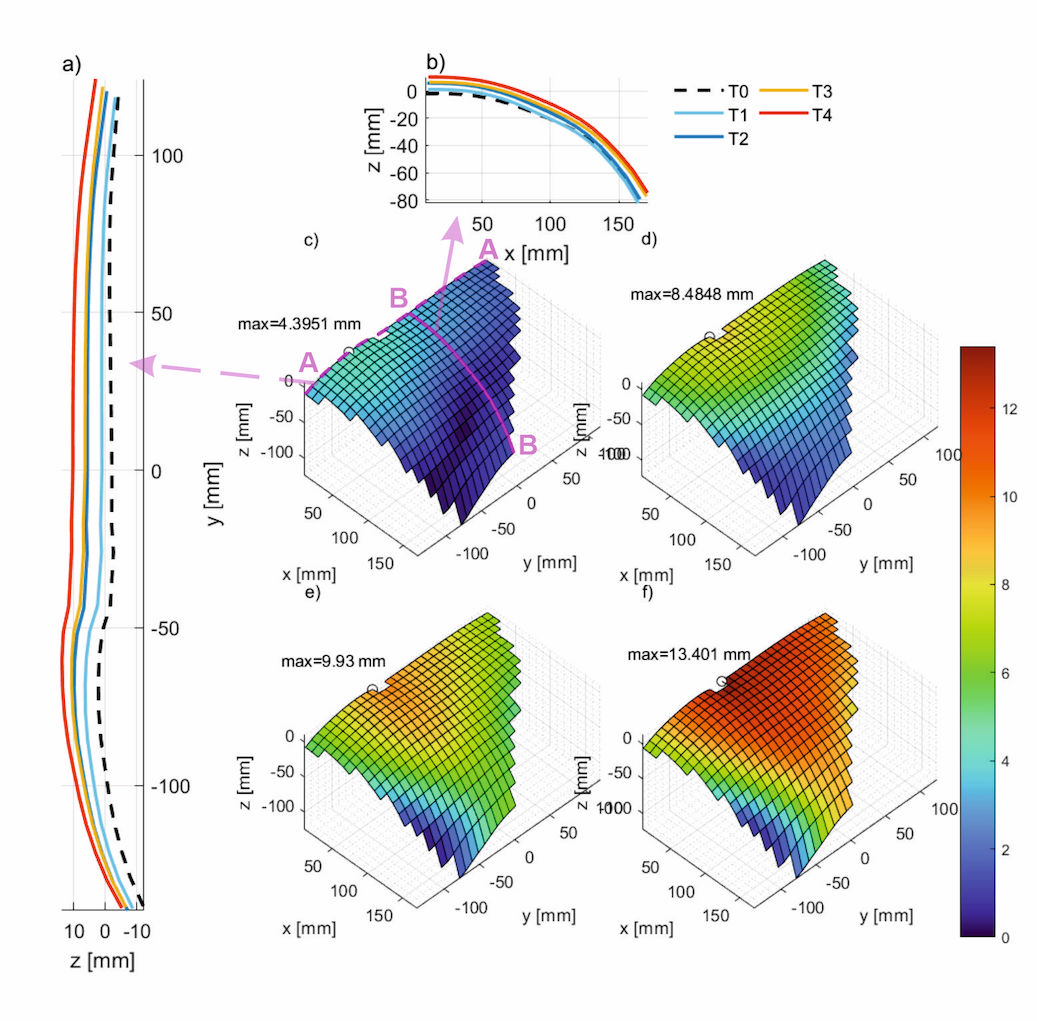}
   \caption{Shape and displacement of subject D11 in four stages T1--T4 and  reference one T0: a) profile of abdominal wall along mid-line A--A; b)  profile along transverse direction B--B; c--f) surfaces of abdominal wall with colour indicating total displacement [mm] in T1--T4, respectively with marked location of maximum displacement by a white circle; x is the mediolateral axis from right to left, y is craniocaudal axis from caudal to cranial, and z is anteriorposterior axis from anterior to posterior }
    \label{fig_dic11dis}
\end{figure}

\begin{figure}[ht!]\centering
    \includegraphics[width=\textwidth]{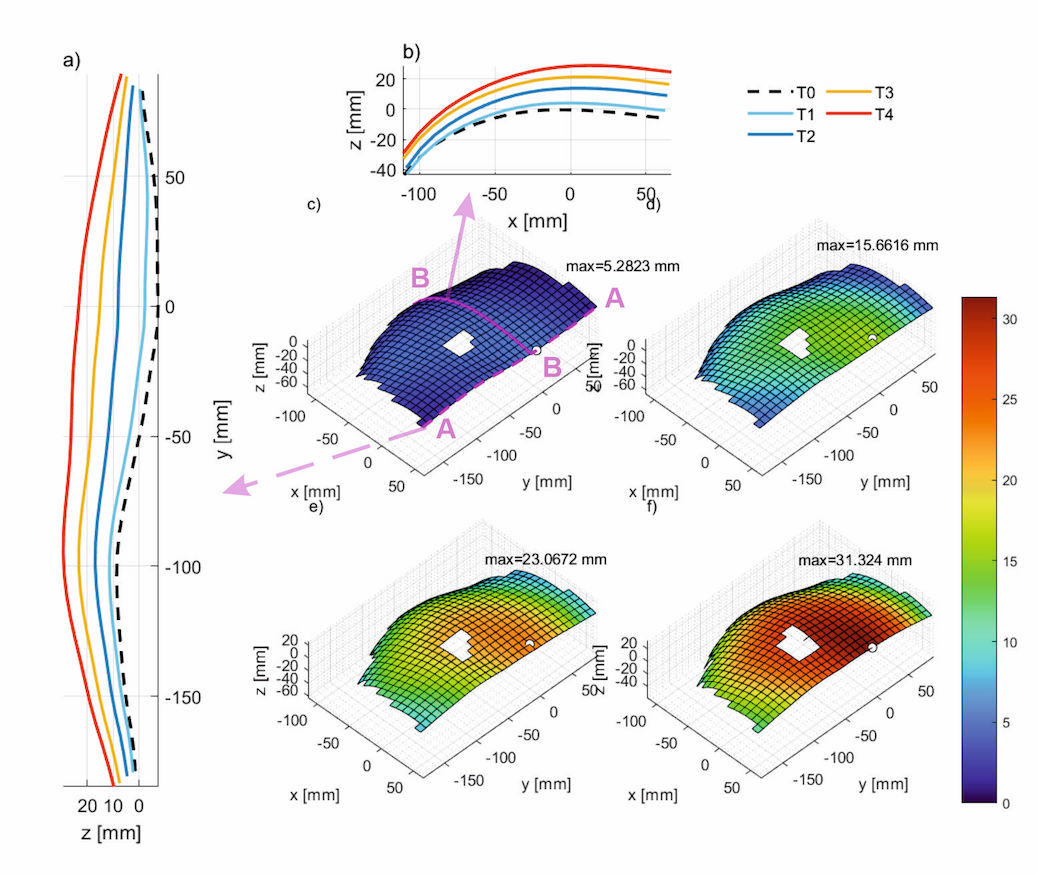}
   \caption{Shape and displacement of subject D12 in four stages T1--T4 and  reference one T0: a) profile of abdominal wall along mid-line A--A; b)  profile along transverse direction B--B; c--f) surfaces of abdominal wall with colour indicating total displacement [mm] in T1--T4, respectively with marked location of maximum displacement by a white circle; x is the mediolateral axis from right to left, y is craniocaudal axis from caudal to cranial, and z is anteriorposterior axis from anterior to posterior }
    \label{fig_dic12dis}
\end{figure}

Figures \ref{fig_dic1dis}--\ref{fig_dic12dis} show displacement of the abdominal wall in stages T1--T4 together with profile lines made along lines A--B and B--B (for subject D2, D4 and D8 see main file, Figures \ref{fig_dic2dis}--\ref{fig_dic8dis}).

\section{Strain maps}\label{appendix_strain}

\begin{figure}[ht!]\centering
    \includegraphics[width=1\textwidth]{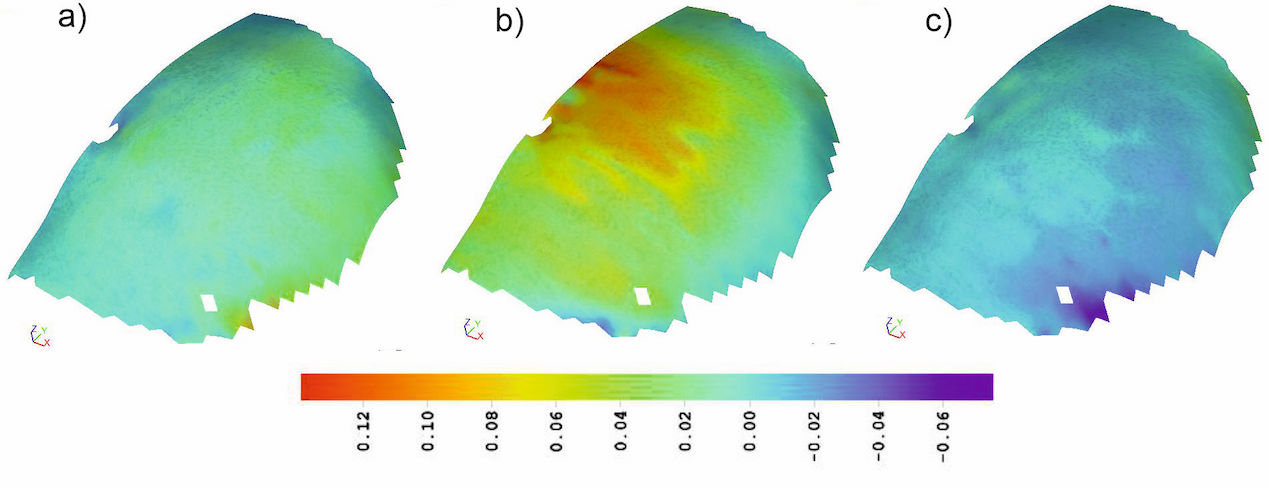}
    \caption{ Strain in T4 for subject D1, a) $\varepsilon_{xx}$, b) $\varepsilon_{yy}$ c) $\varepsilon_{xy}$}
    \label{fig_strainxy_dic1}
\end{figure}

\begin{figure}[ht!]\centering
    \includegraphics[width=1\textwidth]{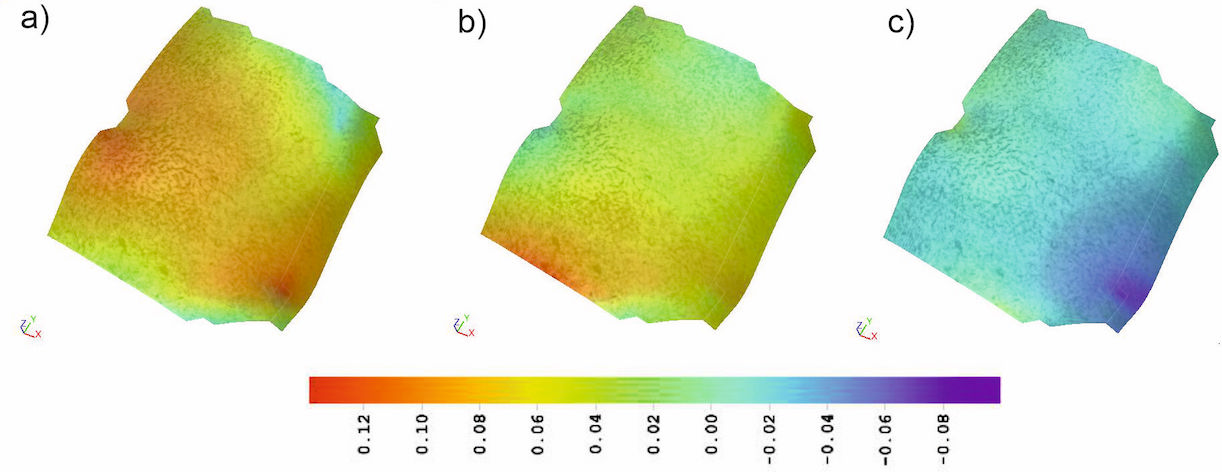}
    \caption{ Strain in T4 for subject D2, a) $\varepsilon_{xx}$, b) $\varepsilon_{yy}$ c) $\varepsilon_{xy}$}
    \label{fig_strainxy_dic2}
\end{figure}

\begin{figure}[ht!]\centering
    \includegraphics[width=1\textwidth]{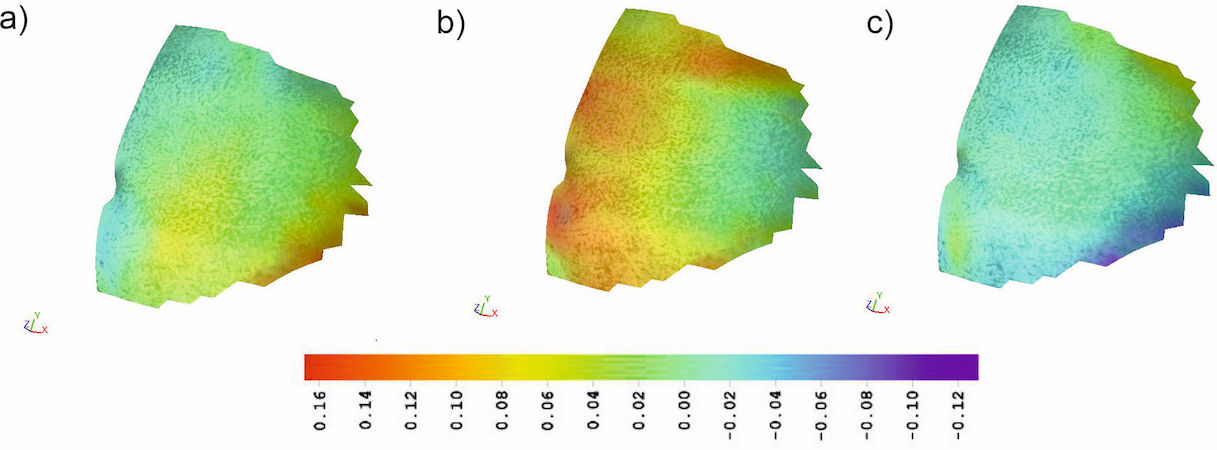}
    \caption{ Strain in T4 for subject D3, a) $\varepsilon_{xx}$, b) $\varepsilon_{yy}$ c) $\varepsilon_{xy}$}
    \label{fig_strainxy_dic3}
\end{figure}

\begin{figure}[ht!]\centering
    \includegraphics[width=1\textwidth]{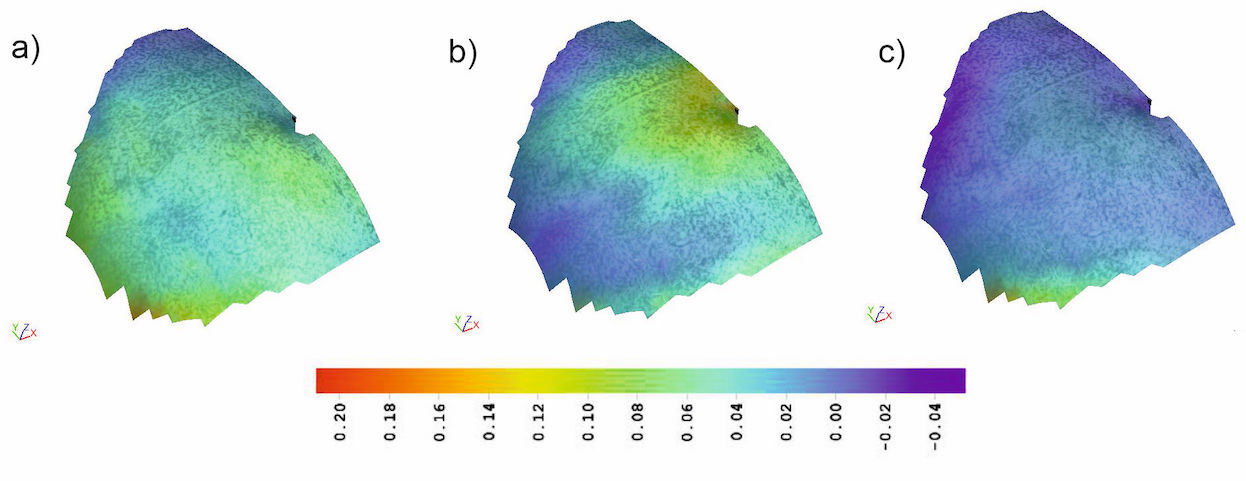}
    \caption{ Strain in T4 for subject D4, a) $\varepsilon_{xx}$, b) $\varepsilon_{yy}$ c) $\varepsilon_{xy}$}
    \label{fig_strainxy_dic4}
\end{figure}

\begin{figure}[ht!]\centering
    \includegraphics[width=1\textwidth]{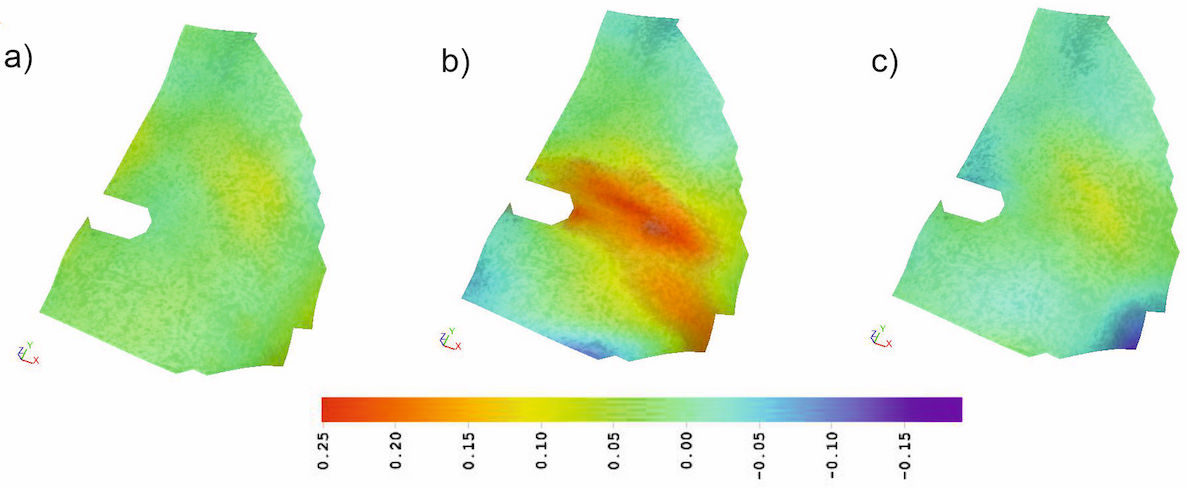}
    \caption{ Strain in T4 for subject D5, a) $\varepsilon_{xx}$, b) $\varepsilon_{yy}$ c) $\varepsilon_{xy}$}
    \label{fig_strainxy_dic5}
\end{figure}

\begin{figure}[ht!]\centering
    \includegraphics[width=1\textwidth]{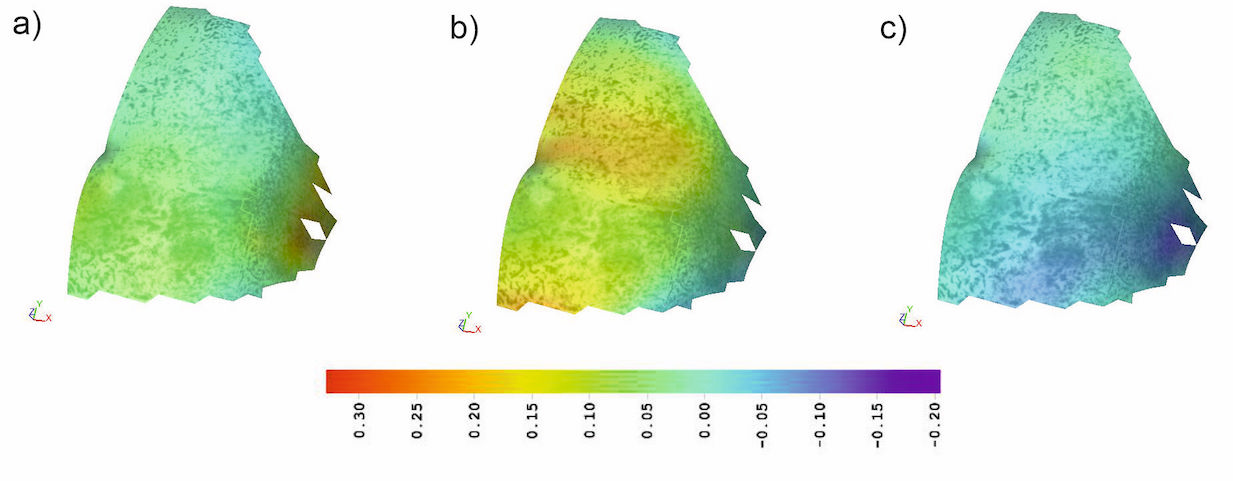}
    \caption{ Strain in T4 for subject D6, a) $\varepsilon_{xx}$, b) $\varepsilon_{yy}$ c) $\varepsilon_{xy}$}
    \label{fig_strainxy_dic6}
\end{figure}

\begin{figure}[ht!]\centering
    \includegraphics[width=1\textwidth]{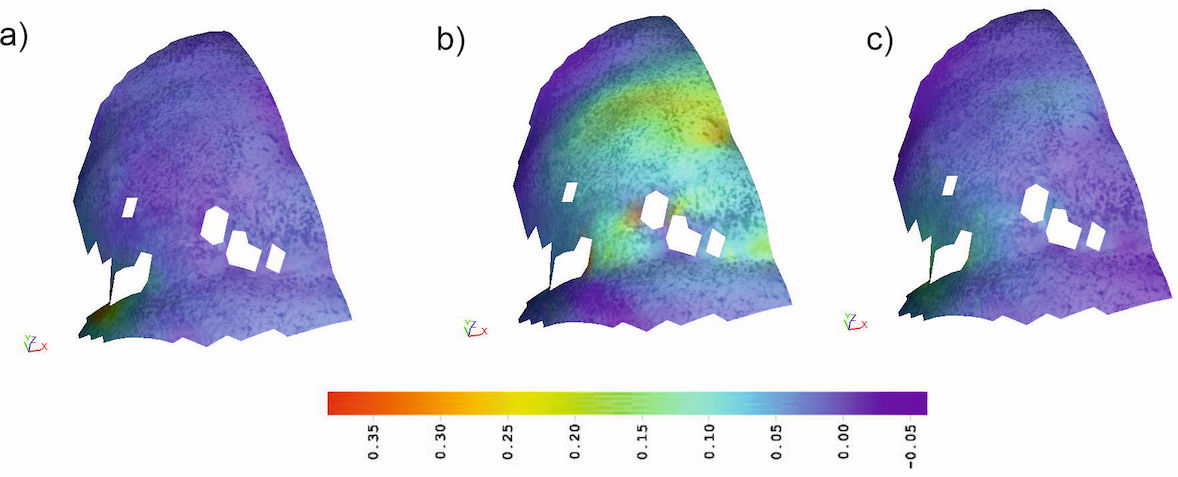}
    \caption{Strain in T4 for subject D7, a) $\varepsilon_{xx}$, b) $\varepsilon_{yy}$ c) $\varepsilon_{xy}$}
    \label{fig_strainxy_dic7}
\end{figure}

\begin{figure}[ht!]\centering
    \includegraphics[width=1\textwidth]{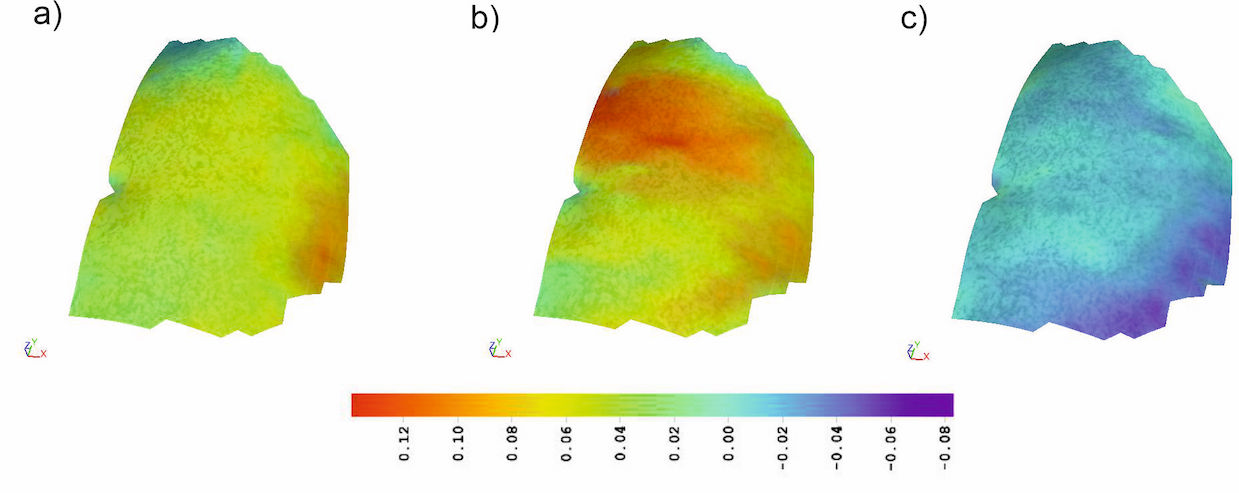}
    \caption{ Strain in T4 for subject D8, a) $\varepsilon_{xx}$, b) $\varepsilon_{yy}$ c) $\varepsilon_{xy}$}
    \label{fig_strainxy_dic8}
\end{figure}

\begin{figure}[ht!]\centering
    \includegraphics[width=1\textwidth]{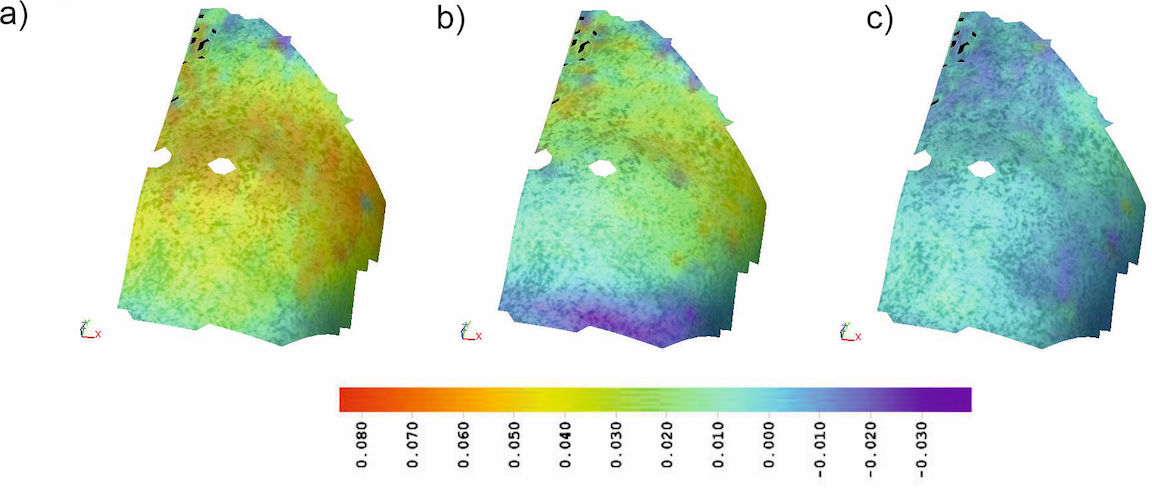}
    \caption{Strain in T4 for subject D9, a) $\varepsilon_{xx}$, b) $\varepsilon_{yy}$ c) $\varepsilon_{xy}$}
    \label{fig_strainxy_dic9}
\end{figure}

\begin{figure}[ht!]\centering
    \includegraphics[width=1\textwidth]{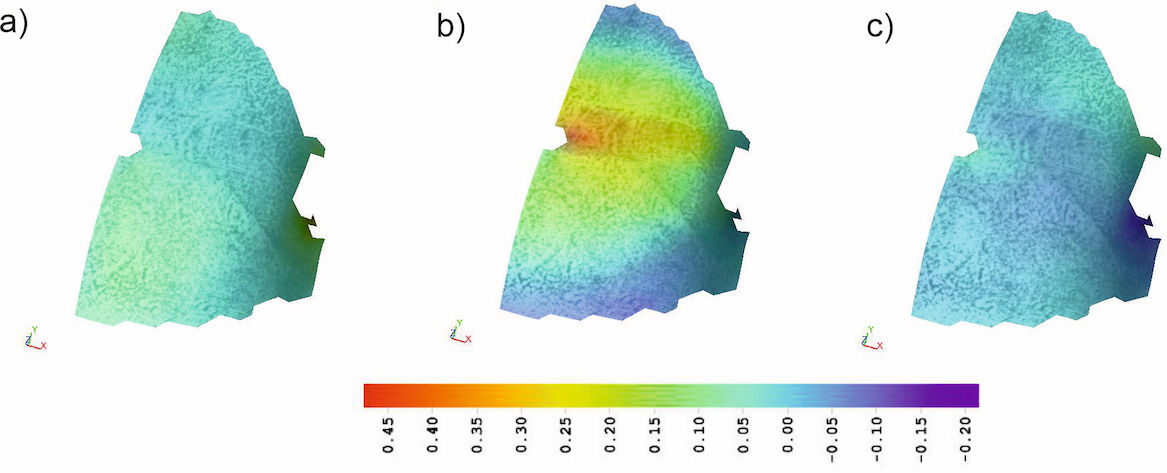}
    \caption{ Strain in T4 for subject D10, a) $\varepsilon_{xx}$, b) $\varepsilon_{yy}$ c) $\varepsilon_{xy}$}
    \label{fig_strainxy_dic10}
\end{figure}

\begin{figure}[ht!]\centering
    \includegraphics[width=1\textwidth]{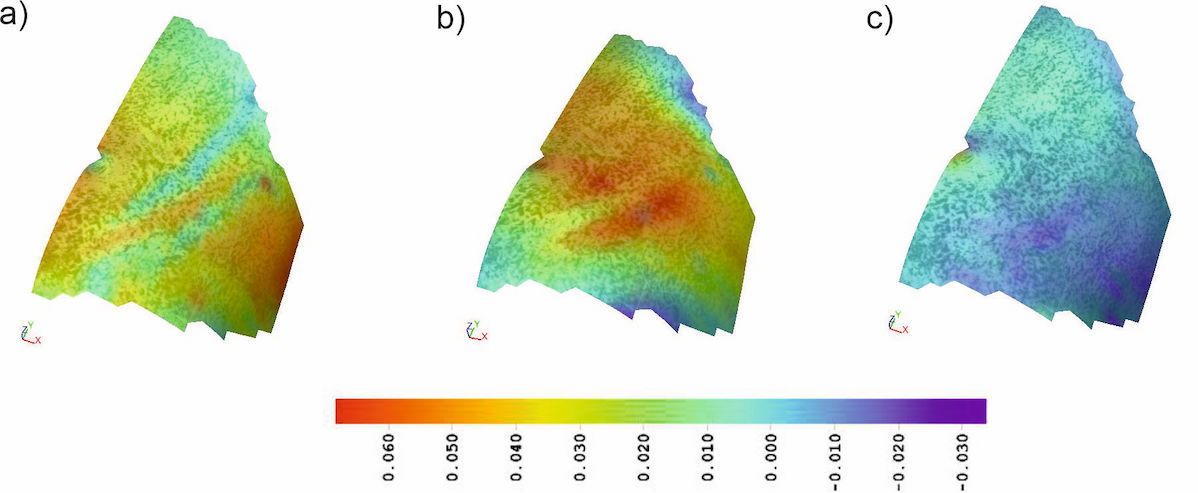}
    \caption{ Strain in T4 for subject D11, a) $\varepsilon_{xx}$, b) $\varepsilon_{yy}$ c) $\varepsilon_{xy}$}
    \label{fig_strainxy_dic11}
\end{figure}

\begin{figure}[ht!]\centering
    \includegraphics[width=1\textwidth]{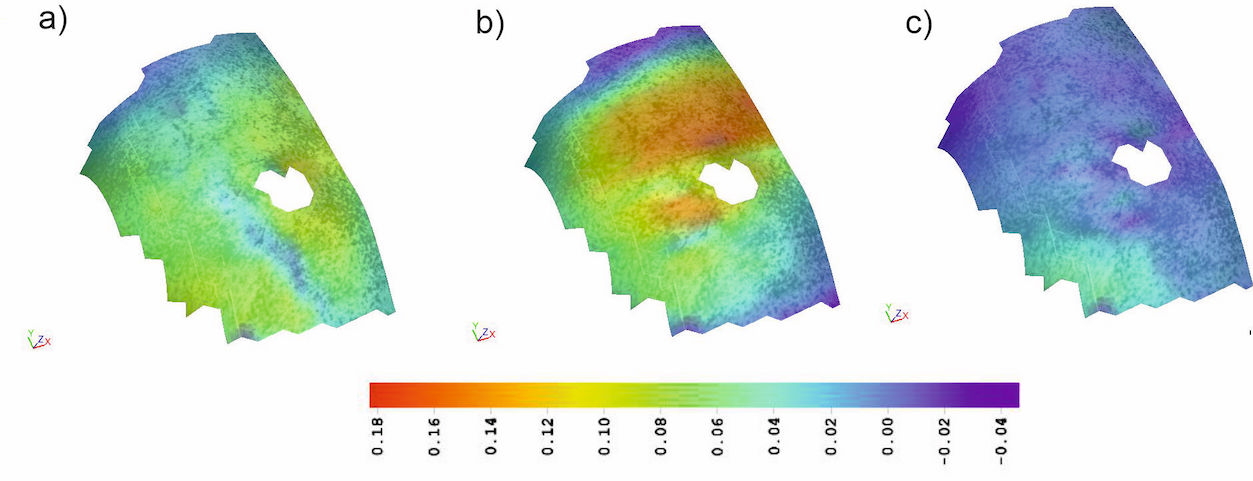}
    \caption{Strain in T4 for subject D12, a) $\varepsilon_{xx}$, b) $\varepsilon_{yy}$ c) $\varepsilon_{xy}$}
    \label{fig_strainxy_dic12}
\end{figure}

Figures \ref{fig_strainxy_dic1}--\ref{fig_strainxy_dic12} show strain maps of $\varepsilon_{xx}$,  $\varepsilon_{yy}$ and $\varepsilon_{xy}$.

\bibliography{mybibfile}

\end{document}